
%
%
%
%
%
%
%
\input harvmac
\def\figflag{I}
%
\global\newcount\figno \global\figno=1
\newwrite\ffile
\def\tfig#1{Fig.~\the\figno\xdef#1{Fig.~\the\figno}\global\advance\figno by1}
\def\figI{I}
%
\newdimen\tempszb \newdimen\tempszc \newdimen\tempszd
\newdimen\tempsze
\ifx\figflag\figI
\input epsf
\def\epsfsize#1#2{\expandafter\epsfxsize{
 \tempszb=#1 \tempszd=#2 \tempsze=\epsfxsize
     \tempszc=\tempszb \divide\tempszc\tempszd
     \tempsze=\epsfysize \multiply\tempsze\tempszc
     \multiply\tempszc\tempszd \advance\tempszb-\tempszc
     \tempszc=\epsfysize
     \loop \advance\tempszb\tempszb \divide\tempszc 2
     \ifnum\tempszc>0
        \ifnum\tempszb<\tempszd\else
           \advance\tempszb-\tempszd \advance\tempsze\tempszc \fi
     \repeat
\ifnum\tempsze>\hsize\global\epsfxsize=\hsize\global\epsfysize=0pt\else\fi}}
\epsfverbosetrue
\fi
%
%
\def\ifigure#1#2#3#4{
\midinsert
\vbox to #4truein{\ifx\figflag\figI
\vfil\centerline{\epsfysize=#4truein\epsfbox{#3.eps}}\fi}
\narrower\narrower\noindent{\bf #1:} #2
\endinsert
}
\def\ie{{\it i.e.}}
\def\eg{{\it e.g.}}

\def\CP{{\cal P}}
\def\CM{{\cal M}}
\def\CK{{\cal K}}

\def\ttstar{$tt^\ast$}
\def\emme#1{\textstyle{#1\over m}}

\def\Vsl#1{\,\raise.15ex\hbox{/}\mkern-13.5mu #1}

\def\BZ{{\bf Z}}
\def\BQ{{\bf Q}}
\def\BP{{\bf P}}
\def\BR{{\bf R}}
\def\CV{{\cal V}}
\def\BC{{\bf C}}
\def\CQ{{\cal Q}}
\def\BN{{\bf N}}
\def\CR{{\cal
R}}

\lref\gepner{D. Gepner, Com. Math. Phys. 141 (1991) 381.}
\lref\intri{K. Intriligator, Mod. Phys. Lett. 6A (1991) 3543.}
\lref\noisi{S. Cecotti and C. Vafa, {\it Ising Model and N=2
Supersymmetric
     Theories}, preprints Harvard HUTP--92/A044 and SISSA--167/92/EP
(1992).}
\lref\lgorv{C. Vafa, Mod. Phys. Lett. A 4
(1989)
1169.}
\lref\alvarez{L. Alvarez--Gaum\'e and P. Ginsparg, Comm. Math.
Phys. 102 (1985) 311.}
\lref\varc{A.N. Var\v cenko, Sov. Math. Dokl. 260 (1981) 272.}
\lref\delmix{P. Deligne, Pubbl. Math. I.H.E.S. 40 (1971) 5\semi
P. Deligne, Pubbl. Math. I.H.E.S. 44 (1973) 5.}
\lref\sch{W. Schmid, Invent. Math. 22 (1973) 211.}
\lref\ising{L. Onsager, Phys. Rev. 65 (1944) 117\semi   T.T. Wu and
B.M. McCoy,
    {\it The two dimensional Ising Model}, Harvard University Press,
Cambridge,
    Mass. 1973\semi   B. M. McCoy and T.T. Wu, Phys. Rev, Lett, 45
(1980) 675;
    B.M. McCoy, J.H.H. Perk and T.T. Wu, Phys. Rev. Lett. 46 (1981)
757.}
\lref\japanese{M. Sato, T. Miwa and M. Jimbo, Publ. R.I.M.S. 14
(1978) 223' 15
    (1979) 201; 577; 871; 16 (1980) 531; 17 (1981) 137\semi   M.
Jimbo
and T.
Miwa,
    {\it Aspects of holonomic quantum fields}, Lecture Notes in Phys.
vol.126,
    Springer 1980 p.429-491.\semi   M. Jimbo and T, Miwa,
    {\it Integrable Systems
    and Infinite Dimensional Lie Algebras}, in {\it Integrable
Systems in
    Statistical Mechanics}, Ed. G.M. D'Ariano, A. Montorsi, M.G.
Rasetti, World
    Scientific, Singapore, 1988\semi    M. Jimbo, Proceedings of
Symposia in Pure
    Mathematics, 49 (1989) 379.}
\lref\hodge{P. Griffiths {\it Topics in Transcendental Algebraic
Geometry},
    Annals of Mathematical Studies 106, Princeton University Press,
    Princeton, 1984.}
\lref\disorder{L.P. Kadanoff and H. Ceva, Phys. Rev. B3 (1971)
3918\semi
    E.C. Marino and J.A. Swieca, Nucl. Phys. B170 [FS1] (1980) 175.}
\lref\hitchen{N. Hitchin, Adv. in Math. 14 (1974) 1.}
\lref\WKD{E. Witten, Nucl. Phys. B258 (1985) 75.}
\lref\compact{S. Cecotti,L. Girardello and A. Pasquinucci, Int. J.
Mod. Phys.
    A6 (1991) 2427.}
\lref\residueapp{S. Cecotti, Int. J. Mod. Phys. A6 (1991) 1749.}
\lref\deligne{P. Deligne, {\it Equations differentielles a points
singuliers
    reguliers}, Lectures Notes in Math. 163, Springer--Verlag 1970.}
\lref\topatop{S.Cecotti and C. Vafa, Nucl. Phys. B367 (1991) 359}
\lref\TFT{E. Witten, Comm. Math. Phys. 118 (1988) 411\semi
    E. Witten, Nucl. Phys. B340 (1990) 281\semi
    T. Eguchi and S.K. Yang, Mod. Phys. Lett. A5 (1990) 1693\semi
    C. Vafa, Mod. Phys. Lett. A6 (1991) 337\semi   R. Dijkgraaf, E.
Verlinde, and
    H. Verlinde, Nucl. Phys. B352 (1991) 59.}
\lref\newindex{S. Cecotti, P. Fendley, K. Intriligator and C. Vafa,
{\it A New
    Supersymmetric Index}, preprint Harvard HUTP-92/A021, SISSA
68/92/EP,
    BUHEP-92-14, (1992). }
\lref\sigmamodels{S. Cecotti and C. Vafa, Phys. Rev. Lett. 68 (1992)
903\semi
    S. Cecotti and C. Vafa, Mod. Phys. Lett. A7 (1992) 1715. }
\lref\polymers{P. Fendley and H. Saleur, Boston and Yale preprint
    BUHEP-92-15, YCTP-P13-1992.}
\lref\veryold{S. Cecotti, Int. J. Mod. Phys. A6 (1991) 1749\semi
    S. Cecotti, Nucl. Phys. B355 (1991) 755.}
\lref\principles{P.A.M. Dirac, {\it The Principles of Quantum
Mechanics},
    4th Edition, Oxford University Press, Oxford 1958.}
\lref\chiralring{W. Lerche, C. Vafa and N. Warner, Nucl. Phys. B324
(1989)
427.}
\lref\zamolo{A.B. Zamolodchikov, JETP Lett. 43 (1986) 730.}
\lref\flatcoordinates{R. Dijkgraaf, E. Verlinde and H. Verlinde,
    Nucl. Phys. B352 (1991) 59\semi
    B. Blok and A. Var\v cenko, {\it Topological Conformal Field
Theories
    and the Flat Coordinates}, preprint IASSNS--HEP--91/5, January
    1991\semi   M. Saito Publ. RIMS, Kyoto Univ. 19 (1983) 1231\semi
    M. Saito, Ann. Inst. Fourier (Grenoble) 39 (1989) 27.}
\lref\FMVW{P. Fendley, S.D. Mathur, C. Vafa and N.P. Warner, Phys.
Lett. B243
    (1990) 257.}
\lref\cancoordinates{B. Dubrovin, {\it Integrable systems in
topological field
    theory} preprint Napoli INFN-NA-IV-91/26, DSF-T-91/26 (1991).}
\lref\solttstar{B. Dubrovin, {\it Geometry and integrability of
topological
    anti--topological fusion}, Napoli preprint INFN-8/92-DSF.}
\lref\landgins{E. Martinec, Phys. Lett. 217B (1989) 431\semi   C.
Vafa
and N.P.
    Warner, Phys. Lett. 43 (1989) 730.}
\lref\axioms{K. Osterwalder and R. Schrader, Comm. Math. Phys. 31
(1973)
    83\semi   K. Osterwalder and R. Schrader, Comm. Math. Phys. 42
(1975)
    281\semi   B. Simon, {\it  The $P(\phi)_2$ Euclidean (Quantum)
Field Theory},
    Princeton University Press 1974.}
\lref\specgeom{S. Ferrara and A. Strominger, {\it N=2 spacetime
supersymmetry
    and Calabi--Yau moduli space},
    presented at Texas A \&\ M University, String'
    89 Workshop\semi   S. Cecotti, Commun. Math. Phys. 131 (1990)
517\semi   A.
    Strominger, Commun. Math. Phys. 133 (1990) 163\semi   P. Candelas
and X.C. de
    la Ossa, {\it Moduli Space of Calabi--Yau Manifolds}, University
of Texas
    Report, UTTG--07--90\semi   R. D'Auria, L. Castellani and S.
Ferrara, Class.
    Quant. Grav. 1 (1990) 1767.}
\lref\isingFF{J. Cardy and G. Mussardo, Nucl. Phys. B340 (1990)
387\semi
    V.P. Yurov and Al.B. Zamolodchikov, Int. Mod. Phys.A6 (1991)
3419.}
\lref\raysinger{D.B. Ray and I.M. Singer, Adv. Math. 7 (1971)
145\semi
    D.B. Ray and I.M. Singer, Ann. Math. 98 (1973) 154.}
\lref\susyQM{S. Elitzur and A. Schwimmer, Nucl. Phys. B226 (1983)
109\semi
    M. Claudson and M.B. Halpern, Nucl. Phys. B250 (1985) 689.}
\lref\asforms{E. Witten, J. Diff. Geom. 17 (1982) 661.}
\lref\pathint{E.C. Marino, B. Schroer and J.A. Swieca, Nucl. Phys
B200 (1982)
    473.}
\lref\marino{E.C. Marino, Nucl. Phys. B217 (1983) 413\semi   E.C.
Marino, Nucl.
    Phys. B230 (1984) 149.}
\lref\french{O. Babelon, {\it From form factors to correlation
functions: the
    Ising model}, preprint Saclay SPhT-92-062; LPTHE-92-20.}
\lref\highermod{B. Dubrovin, {\it Differential Geometry of Moduli
Spaces
    and its Application to Soliton Equation and to Topological
Conformal
    Field Theory}, Preprint 117 of Scuola Normale Superiore, Pisa,
    November 1991.}
\lref\riemann{H. Flaschka and A.C. Newell, Commun. Math. Phys. 76
(1980) 67.}
\lref\oneloop{L.J. Dixon, V.S. Kaplunovsky and J. Louis, Nucl. Phys.
B355
(1991)
    649\semi   I. Antoniadis, E. Gava and K.S. Narain, preprints
IC/92/50 and
    IC/92/51
    \semi   S.Ferrara, C.Kounnas, D.L\"ust and F.Zwirner,
    preprint CERN-TH.6090/91.}
\lref\bill{W. Leaf--Herrmann, Harvard preprint HUTP-91-A061, and to
appear.}
\lref\witindex{E. Witten, Nucl. Phys. B202 (1982).}
\lref\markoff{A.A. Markoff, Math. Ann. 15 (1879) 381\semi
    A. Hurwitz, Archiv. der Math. und Phys. 3 (14) (1907) 185\semi
    L.J. Mordell, J. Lond. Math. Soc. 28 (1953) 500\semi
    H. Schwartz and H.T. Muhly, J. Lond. Math. Soc. 32 (1957) 379.}
\lref\mordell{L.J. Mordell, {\it Diophantine Equations},
    Academic Press, London 1969.}
\lref\lambdamat{F.R. Gantmacher, {\it The Theory of Matrices},
    Chelsea, 1960.}
\lref\monlemma{F. Lazzeri, Some remarks on the Picard--Lefschetz
monodromy,
     in {\it Quelques journ\'ees singuli\`eres}, Centre de
Mathematique de
     l'Ecole Polytechnique, Paris 1974.}
\lref\krona{L. Kronecker, {\it Zwei S\"atze \"uber Gleichungen mit
     ganzzahligen Coefficients}, Crelle 1857, Oeuvres 105.}
\lref\cotech{F.M. Goodman, P. de la Harpe, V.F.R. Jones, {\it Coxeter
     Graphs and Tower of Algebras}, Mathematical Sciences Research
     Institute Publications 14, Springer--Verlag, 1989.}
\lref\gross{B.H. Gross, Inv. Math. 45 (1978) 193.}
\lref\weilA{A. Weil, {\it Introdution a les Vari\'et\'es
K\"ahl\'eriennes},
     Hermann, Paris, 1958.}
\lref\griHarris{P. Griffiths and J. Harris, {\it Principles of
Algebraic
     Geometry}, Wiley--Interscience, New York, 1978.}
\lref\singt {V.I. Arnold, S.M. Gusein--Zade and A.N. Var\v cenko,
     {\it Singularities of Differentiable Maps}, Vol.II,
     Birkh\"ausser, Boston 1988.}
\lref\Strominger{A. Strominger, Commun. Math. Phys. 133 (1990) 163.}
\lref\pathrep{S. Cecotti and L. Girardello, Phys. Lett. B110 (1982)
39.}
\lref\acampo{N. A'Campo, Indag. Math. 76 (1973) 113.}
\lref\cyclopoly{K. Ireland and M. Rosen, {\it A Classical
Introduction
     to Modern Number Theory}, (Springer--Verlag, Berlin, 1982).}
\Title{ HUTP-92/A064, SISSA-203/92/EP }{\vbox{\centerline{On
Classification of
N=2  Supersymmetric Theories } }}

\bigskip
\bigskip

\centerline{Sergio Cecotti}
\medskip\centerline{International School for Advanced Studies,
SISSA-ISAS}
\centerline{Trieste and I.N.F.N., sez. di Trieste}
\centerline{Trieste, Italy}
\bigskip\bigskip

\centerline{Cumrun Vafa}
\medskip\centerline{Lyman Laboratory of Physics}
\centerline{Harvard University}
\centerline{Cambridge, MA 02138, USA}
\bigskip\bigskip

\vskip .3in

\noindent
We find a relation between the spectrum of solitons of massive $N=2$ quantum
field theories in $d=2$ and the scaling dimensions of chiral fields
at the conformal point.  The condition that the scaling dimensions be
real imposes restrictions on the soliton numbers and leads to
a classification program for symmetric $N=2$ conformal theories and their
massive deformations in terms of a suitable generalization of Dynkin diagrams
(which coincides with the A--D--E Dynkin diagrams for minimal
models).  The Landau-Ginzburg theories
are a proper subset of this classification.  In the particular
case of LG theories we relate the soliton numbers with
intersection of vanishing cycles of the corresponding
singularity; the relation between soliton numbers and the scaling dimensions
in this particular case is a well known application of
Picard-Lefschetz theory.

\Date{ 11/92}

\newsec{Introduction}
Quantum field theories in two dimensions have been under
intensive investigation recently in part due
to their importance in string theory and in part
serving as exactly soluble toy models for quantum field
theories in higher dimensions.  The interest in studying
them for string theory has mostly focused on conformal
field theories, i.e., the ones with traceless energy
momentum tensor (with only massless excitations).
On the other hand, as examples of interesting exactly soluble
quantum field theories with interesting $S$-matrices,
the massive ones have been under
investigation \ref\Zam{A.B. Zamolodchikov and
Al.B. Zamolodchikov, Ann. Phys. 120 (1980) 253.}.  In view of the
fact that
 massive QFT's can be viewed as deformation of the conformal
theories, it is natural to ask if there is any way to understand
properties of conformal theories, by studying the massive
analogs.  This program has been followed with a spectacular
degree of success originating with the work
of Zamolodchikov's \ref\Zam2{A.B.Zamolodchikov, JETP Lett. 46 (1987)
160.}\ref\TBA{Al.B. Zamolodchikov, Nucl. Phys. B342
(1990)695.}.
The method to relate properties
of integrable massive theories to the conformal ones uses
thermodynamical
Bethe ansatz (TBA).  In this way, just by studying the S-matrices
of the massive integrable theories one can deduce for example
the central charge of the conformal theory.

An interesting class of conformal theories for superstrings
is the class with $N=2$ superconformal symmetry.  These can be used
to construct string vacua.  For instance, $\sigma$--models
on Calabi-Yau manifolds provide examples of such theories.
In view of their importance in constructing string
vacua, it is natural to ask if one can classify all $N=2$
theories.  Progress in this direction was made \landgins\
when it was realized that $N=2$ Landau-Ginzburg theories is
an effective way of classifying some of them.
In particular all
the minimal $N=2$ models were found to have a simple
Landau-Ginzburg description which fitted with the known
classification of simple singularities \singt .
This program had the following limitation:  It is known
that not all the $N=2$ conformal theories can be
realized as a LG theory.  So this program leads to a partial
classification.

Massive integrable deformation of $N=2$ superconformal theories
has also been considered \FMVW
\ref\fenint{P. Fendley and K. Intriligator,
Nucl. Phys. B372 (1992) 533\semi BUHEP-92-5,
HUTP-91/A067.}\ref\lerwar{
W. Lerche and N.P. Warner, Nucl. Phys. B358 (1991) 571.}\ref\Lecl{A.
LeClair,
D. Nemeschansky and N.P. Warner,
1992 preprint, CLNS 92/1148, USC-92/010.}.
  Furthermore
the TBA has been applied to these theories (and in particular
the central charge and the charge of
chiral primary fields at the
conformal point has been
recovered in this way).  In this paper we will consider
massive perturbations of $N=2$ theories in 2 dimensions and
show that there is a very simple relation between the $U(1)$
charges of chiral fields at the conformal point
(the highest of which is equal to the central change) and the degeneracy
of solitons which saturate the Bogomolonyi bound in the massive
theory.   This relation exists {\it whether
or not the theory is integrable}.
Turning this around, we end up with the
following classification program:  Start with $n$-vacua, and
impose having a certain number of solitons between each pair.
Then deduce the structure of chiral ring at the conformal point.
In particular in this way we can compute the charges
of primary fields at the conformal point.
It turns out that the condition that the charges of chiral
fields be {\it real} puts a strong restriction on
the number of solitons allowed. For instance, we
show that for a minimal model, defined by the
condition that all chiral fields are relevant perturbations,
there are at most $1$ soliton allowed between vacua.  Using
the soliton numbers, we can associate a bilinear form
(with $2$'s on the diagonal)
to each massive $N=2$ theory.
We also find a relation
between the signature of the bilinear form and the
charges of chiral fields.
 We show that for
minimal models this
bilinear form is positive definite, which with the above
restriction leads to the well known ADE classification
of the minimal models.  This method explains in the most natural way
why the A--D--E classification arises
while classifying minimal models.
For theories with higher central charge more general
types of `Dynkin diagrams' arise, which encode the soliton
structure of the theory.

The organization of this paper is as follows:  In section 2
we describe the soliton structure of the $N=2$ LG models
and relate it to intersection theory of the Homology
cycles (as in singularity theory \singt ).
We will also show how, in this subclass, one can obtain the charges
of chiral fields  from
the number of solitons.  In section 3 we discuss
how to formulate these results generally {\it independently
} of whether they come from a LG theory.  In section 4
we give a proof of the general reformulation.  The proof
uses the topological-anti-topological equations ($tt^*$)
formulated in \topatop\
which has been reformulated as an isomonodromy deformation
of a linear system of equations by Dubrovin \solttstar .
We show that the phase of the eigenvalues of the monodromy of these
equations
are simply the chiral charges.  Relating the monodromy operator
to the soliton numbers gives the desired relation between the charges
and the soliton numbers.
In sect. 5 we discuss a criterion to select which massive
models have a non--degenerate UV limit.
 In section 6 we show how these
ideas lead to a classification program for massive $N=2$ theories
(up to addition of $D$-terms), or by
taking the UV limit to the classification of {\it conformal}
$N=2$ theories which admit a massive deformation (the
$D$-term being fixed by the conformal condition).
These ideas may be useful in classifying $c_1>0$ K\"ahler
manifolds (with diagonal Hodge numbers), as to each such manifold
(which admit
massive deformation) one can associate a particular
bilinear form by considering sigma models on them.
We give a number of examples where we can
use these techniques.  In particular in sect.7 we rederive
the A--D--E classification of minimal models, as
well as its `affine' counterpart (including orbifolds
of $S^2$)
and (in sect. 8)
supersymmetric sigma models on ${\bf C}P^n$ and Grassmanians.
Moreover we spell out
the classification of theories with up to 3 vacua as well
as that of models with a
$\BZ_n$ symmetry.
In section 9 we present our conclusions and
suggest some directions for future research.
In appendix A some aspects of the Grassmanian $\sigma$--models
are worked out.
In appendices B,C some further properties of the classification
program are discussed.

We would like to make a historical remark:  The order
we have decided to present our results does not reflect the order in
which
we discovered them, but rather the order in which
it can be understood most easily.  In particular
a time ordered sequence of our understanding is
roughly sections 4,6,3,2,5,7,8.

\newsec{Landau-Ginzburg Solitons and Monodromy}

An interesting subclass of $N=2$ QFT's in two dimensions
is given by Landau-Ginzburg theories (see e.g. \chiralring
\compact\
for the definition).  These theories are characterized
by a superpotential $W(x^i)$ which is a holomorphic
function of $n$ chiral superfields $x^i$, up to variation
in $D$-terms which is represented by a positive function
$K(x^i,\bar x^i)$.  The bosonic part of the LG action is
given by
$$S=\int d^2z \quad G_{i\bar j} \partial_\mu x^i \partial_\mu x^{\bar
j} +
 G^{i\bar j}\partial_i W \overline{\partial_j W}$$
where $G_{i\bar j}=\partial_i{\bar \partial}_j K$
(which is positive definite for a unitary theory).  The
scalar potential is
minimized at $x^j=a^j$ such that
$$\left.{\partial W\over \partial x^i}\right|_{a^j}=0 \qquad for \
all \quad
i$$
which thus correspond to vacua of this theory.
Let us assume that the vacua are non-degenerate, in the sense
that near each of them $W$ is quadratic.
This can always be arranged, if necessary, by perturbing
$W$.  Let us find
the number of solitons in this theory.
Our argument is a simple generalization of that given
in \FMVW\ from one variable case
to higher $n$.

Solitons are configurations
of fields as a function of space, where on the left $x^i(-\infty )
=a^i$ and on the right
$x^i(+\infty )=b^i$ where $a,b$ label two
distinct critical points of $W$.  Stable solitons are the
ones satisfying the above boundary condition which minimize
the energy.  Let us denote the space variable by $\sigma  $.
The energy of the soliton configuration is given by
$$E_{ab}=\int d\sigma  \left| \partial_\sigma   x -\alpha {\overline
{\partial
W}}\right|^2
+2{\rm Re}(\alpha^* \Delta W)$$
where $\Delta W =W(b)-W(a)$, and $\alpha$
is some arbitrary phase, and we have hidden all the indices
and raising and lowering of indices is done with $G_{i\bar j}$.  It
is easy to
see that there is a lower
bound for the energy:  choose $\alpha =  {\Delta W}/|\Delta W|$,
then we see from the above representation of $E$ that
$$E_{ab}\geq 2\left| \Delta W \right| $$
 Since $W$ is not renormalized
in the quantum theory (due to the existence of topological
ring which characterizes it) this is precisely the same as the
Bogomolnyi bound
in the quantum field theory.  So the number of solitons
which {\it saturate} the Bogomolnyi bound are given by
solving the equation with
\eqn\emot{\partial_\sigma   x^i= \alpha G^{i\bar j}
\overline{\partial_j
W}}
Note that for any such solution the image of the soliton
configuration in the $W$--plane is a straight line
$$\partial_\sigma   W=\partial_iW \cdot \partial_\sigma   x^i =\alpha
\left| \partial W \right|^2$$
In other words the image is a straight line connecting $W(a)$ to
$W(b)$.  Now we come to asking how many solutions are there to
\emot ?  For simplicity, and with no loss
of generality we take $W(a)=0$ and $W(b)$
to be a positive real number, which means taking $\alpha =1$.
First let us analyze solutions to \emot\ near $a$.
Again with no loss of generality we take $a$ to correspond to
 $x^i=0$ where near it we take $W=\sum_i (x^i)^2$ and $G_{i\bar
j}=\delta_{i\bar j}$.  Then the equation
for soliton \emot\ near the critical point becomes
\eqn\sph{\partial_\sigma   x^i= \bar{x^i}}
So the solution which at $\sigma   =-\infty$ is at the critical
point is given by
\eqn\solu {x^i=\alpha^i e^{\sigma  } \qquad with \quad
\alpha^i=(\alpha^i)^*}
Of course it is not clear if for all $\alpha^i$ we get a solution,
i.e.,
if this trajectory ends up on another critical point.  In order
to analyze how many of these initial conditions would correspond
to an actual soliton, we should look at the totality of allowed
solutions {\it near} each critical
point, and try to match them with solutions near others.
Let us consider the points $\Delta_a$
of the totality of all possible solutions \solu \ with a given value
of $W=r^2$
(where r is a small real number) near the critical point $a$. In
other
words let's look at the intersection of $W^{-1}(r^2)$ with all the
potential solutions originating from $a$.  This intersection is given
by the
condition
$$\Delta_a:\qquad  \sum_{i=1}^{n} (x^i)^2=r^2$$
where from \solu\  the only restriction on $x_i$ is that it be {\it
real}.
So the `wave front' of all possible solutions originating
from a critical point with a given value of $W$ is an
$n-1$--dimensional
sphere.  Note that this sphere vanishes as $r\rightarrow 0$.  This is
precisely
the definition of a {\it vanishing cycle} in singularity theory
\singt .
In fact near each critical point we get a vanishing cycle which
is diffeomorphic to $S^{n-1}$.
Now suppose we consider the vanishing cycle $\Delta_b$ near the
critical point $b$.  Those points will represent the points
which by \emot\ can flow from
the critical point $b$ (along the negative real
axis), where we need to set $\alpha =-1$ in \emot .  Now consider
going on a straight line in the $W$--plane connecting
the two critical values $W(a)=0$ and $W(b)$.  Let us fix a point $p$
on this line, say $W=W(b)/2$.  The wave front originating
from $a$ over $p$ continues to be an $n-1$ dimensional
cycle in $W^{-1}(p)$.  It gets deformed from the original
shape but it is still an $n-1$ dimensional sphere (as the flow
with the vector field given by \emot\ is just a diffeomorphism).
Let us still denote this cycle by $\Delta_a$.  Also consider the
intersection of wave front originating from the point $b$
with $W^{-1}(p)$ and denote the cycle by $\Delta_b$.  These
two cycles intersect at a discrete number of points (note
that each one is {\it half} the dimension of $W^{-1}(p))$.
{\it For each  point of their intersection we get a soliton}.
This is almost obvious:  For each point that they intersect
the flow of the vector field from $a$ which reaches that point
continues to flow to the critical point $b$.  Here
it is crucial that \emot\ is a first order equation.  So we get
a solution to \emot\ with the boundary condition
that $x(-\infty)=a$ and $x(+\infty)=b$.  Moreover the points
on $\Delta_a$ that do not intersect any point of $\Delta_b$
will not flow to $b$ when evolved with \emot\ as $\Delta_b$
is the totality of all such points that flow to $b$.  Therefore the
number
of solitons is exactly the number of points that $\Delta_a$ and
$\Delta_b$ intersect.
This is not necessarily the intersection number of these two
cycles, because the intersection number counts each intersection
point with $\pm 1$ depending on the orientations.
However the intersection number appears naturally for us as follows:
The solitons come in pairs, as they
are Bogomolnyi saturated states. We have been focusing
on the bosonic piece of the soliton, there will also
be a fermionic partner obtained by acting on this state
with $Q^-$ the supersymmetry charge (which
decreases the fermion number by 1).
In weighing the solitons with phases the natural thing
to consider is $(-1)^F$. However this would cancel
for pairs of solitons. Instead as in \newindex\ we
consider weighing the soliton pairs with $(-1)^F F$ which is
the same as weighing the bosonic components with $(-1)^F$.
What we will now
show is that the number of bosonic solitons weighed with $(-1)^F$, is
just this intersection number, i.e.,
\eqn\intn{\left| \mu_{ab}\right| = \left| \sum_{ab \ \rm
solitons}(-1)^F
F
\right|=
\left| \sum_{ab \ \rm bosonic\ solitons}(-1)^F\right| =\left|
\Delta_a \circ \Delta_b\right| }
{}From now on whenever we talk about soliton numbers we mean
this {\it weighted} soliton number.
Generically all the
solitons have the same fermion number and so this is just
the counting of the soliton.  At any rate this weighted
soliton number is more useful for our purposes than
the actual soliton number, in case they are not the same.
Also we show that the
absolute value signs can be taken out of the above equation
in the following sense:  First note that the fermion number
of any state in the $ab$ sector is
$f_{ab}+k$ where $f_{ab}$ is in general fractional
and can be written as a difference $f_a -f_b$ (see  \newindex )
and $k$ is an integer.  In fact in a Landau-Ginzburg theory $f_{ab}$
is given\foot{There is a general topological proof of this fact which
holds for any N=2 theory not just for LG models. In the general
case one has
$$\exp[2\pi if_{ab}]={\rm phase}\big[\eta_a/\eta_b],$$
where $\eta_a$ is defined by the corresponding TFT metric
as $\eta_{ab}=\delta_{ab}\eta_a$. The proof of this formula follows
from comparing three known facts: 1. $f_{ab}$ is defined (mod. $1$)
by
$Q_{ab}=\pm \exp[i\pi f_{ab}]|Q_{ab}|$ where $Q_{ab}$ is the new
index
of ref.\newindex\ computed in the spectral--flow `point basis'
$|e_a\rangle$ (see
\newindex). 2. In the canonical basis $|f_a\rangle$ $Q_{ab}$ is
real. See \noisi. 3. By definition
\noisi\ $|f_a\rangle=(\eta_a)^{-1/2}|e_a\rangle$. The statement in
the text
is obtained by replacing $\eta_a$ with its explicit expression for a
LG model.}
by  \fenint
$$e^{2\pi i f_{ab}}={\rm phase} \left[\det\ H(b)\over \det\ H(a)
\right]$$
where $H_{ij}=\partial_i\partial_j W$.
In the LG case $f_a$ and $f_b$ can be identified
with the phases of the determinant of Hessian at the respective
critical
points.
So $\mu_{ab}$ will in general carry a phase $\pm \exp(2 i\pi
f_{ab})$.
Viewing $\mu_{ab}$ as a matrix we see that we can get rid of phases
up to $\pm $ signs by a redefinition of the basis using $f_i$
as in \newindex .  Note
also from the definition \intn\ that $\mu$ is an anti-symmetric
matrix in this basis.

To remove the absolute value signs in \intn\ it is more
convenient to consider the case when we have an even number of
LG fields, i.e., $n$ is even.
This can be done with no loss of generality by simply
adding, if necessary, a field with $x^2$ contribution to
superpotential.
  In order to have a consistent definition
on the right hand side the intersection matrix should be
anti-symmetric
which is the case when $n$ is even, because vanishing cycles are
odd dimensional.  We will show that with this choice in
a suitable basis we have\foot{If we had chosen $n$ to be odd
we would need to order the vacua and use one definition
of sign when $a>b$ and the other when $b<a$.}
\eqn\fund{\mu_{ab}=\Delta_a \circ \Delta_b}
Note also the fact that there is no soliton from one vacuum
to itself $\mu_{aa}=0$ is automatic because an odd dimensional sphere
has
zero self intersection (the Euler character is zero).
We still have the freedom of redefining
the basis by multiplications by $\pm$.  So the invariant
quantities are obtained when we consider `cycles' which means
that if we consider $\mu_{i_1i_2}\mu_{i_2i_3}...\mu_{i_ri_1}$ it is
independent
of conventions.

Now we come to showing \fund\ which requires a rather
long and delicate analysis.  Suppose we have two different
soliton trajectories from $a$ to $b$, and we wish to show
that their relative contribution to the left and right hand
side of the above equation is the same.  In order to show
that we need to show that if these two trajectories
correspond to the same sign for intersection between cycles
they also have the same fermion number mod 2, and if they
have opposite intersection number their fermion number
differs by 1 mod 2.

Let us consider a given solution to
\emot\ (with $\alpha =1$) and consider the family of solutions
which is near this solution.  If we write the perturbation
as $x\rightarrow x+\delta$ the equation we get for $\delta$ is given
by
\eqn\smvar {{\partial_\sigma   \delta  } =H^* \delta^*}
Note that an obvious solution to this variational
equation is the `velocity vector' of the soliton
trajectory $v^i=\partial_\sigma x^i$.
Near a critical point $H$ is a constant, and the above
equation can be solved by finding solutions to
\eqn\eigc{H^*\delta_k^* =\lambda_k \delta_k}
where $\lambda_k >0$ for solitons which at $\sigma   =-\infty$
start at the critical point.
In fact the above equation has $n$ independent
solutions.  Indeed if we pair $(\delta , \delta ^*)$ as a $2n$
dimensional vector and consider
$${\cal H}=\left(\matrix{0&H^*\cr H&0 \cr }\right)$$
as a Hermitian hamiltonian, then its eigenvalues come
in pairs with opposite signs.  If $(\delta_i , \delta_i^*)$
has eigenvalue $\lambda_i>0$ then $(i\delta_i , -i \delta_i^*)$ has
eigenvalue $-\lambda_i$.
   The vectors tangent to the vanishing cycle $\Delta_a$
near the critical point $a$ are {\it real} linear
combinations of the vectors \eigc\ with positive
eigenvalues subject to the additional
constraint that
they are at the preimage of a fixed value of $W$, i.e.,
$$dW=\partial  W \cdot \delta  =0$$
Using the equation of motion \emot\ this means that
$$a_k\delta _k \cdot  {\bar v} =0$$
(where the field indices are implicit). Thus the vectors
tangent to vanishing cycle are always orthogonal to $v$ and $iv$.
This means that tangents
to vanishing cycle {\it near the critical point}
span positive eigenspace of $\cal H$ orthogonal to
$v$ (which itself belongs to this subspace). Note that near
$b$ the vanishing cycle $\Delta_b$ is spanned by the {\it negative}
eigenspace of \eigc\ (because $\alpha =-1$) which are orthogonal
to $v$ and $iv$.  Note that the $v$, which near $a$ belonged
to the positive subspace of \eigc , near $b$ belongs to the
negative subspace of \eigc , while $iv$ which
near $a$ belonged to the negative subspace
near $b$ belongs to the positive subspace.  The tangents to the
vanishing cycle
$\Delta_a$ near the critical point $b$ must however belong again
to the positive eigenspace of \eigc\ near the critical point $b$,
as they are orthogonal both to $v,iv$ and the tangents to $\Delta_b$.

We wish to compute the fermion number of this
trajectory.  This is the $N=2$ version of a similar problem which
arose
in Witten's considerations of Morse theory \ref\witmor{E. Witten,
J. Diff. Geom.  17 (1982) 661.}.  We have two options
in finding this sign: either
find the number of times the phase of ${\rm det} H$ wraps
around the origin modulo 2 as we go along soliton trajectories,
or more directly relate the sign of the amplitude by
relating fermions to the tangent vectors of the vanishing cycle.
We will use the second option.
First we note that the object we are computing is an index, and
thus can be computed by reducing the theory from 2 dimensions
to 1, and the question of determining the sign in this
set up is the same as determining the sign for an overlap
of the vacuum evolved from critical point $a$ transported
along the soliton trajectory, with the vacuum at point $b$.
It is important
to note that the equation \smvar\ is the same equation which
evolves the {\it fermions} of the theory (this follows from
supersymmetry
transformation), and each
vacuum will correspond for us to an $n$-form, when we identify
the fermions with the tangent vectors (or forms via the metric $G$).
We identify the fermionic degree of
freedom for the state evolving from $a$ with the
ordered set of vectors  $\delta_1,...,\delta_{n-1},v$
where  $\delta_1,...,\delta_{n-1}$ forms a basis for $\Delta_a$ near
$a$
and in such a way that the orientation of it
is compatible with that of $\Delta_a$, which means that
$\delta_1,i\delta_1,...,\delta_{n-1},i\delta_{n-1},v,iv$ give
the standard orientation of $x^i$ which is $\BC^n$.
The fermionic degree of freedom of the state evolving from $b$
will be identified with vector $\gamma_1,\gamma_2,...
,\gamma_{n-1},v$ where $\gamma_i$ form the tangent to the vanishing
cycle at $b$, ordered in the canonical way.  In taking the overlap
between
these two states, the $F$ insertion (in the definition of $\mu$)
removes the zero mode
we would have obtained and so
using the definition of Grassmann integration of fermions we are just
left with the standard definition of the intersection number
$\Delta_a
\circ \Delta_b$ for the contribution of this path integral.
Therefore
the sign of this amplitude is the same as the sign of the
intersection number.

This completes what was to be shown as
far as relative contribution of two trajectories
beginning at a critical point $a$ and ending at another
critical point $b$ is concerned.  By repeating this argument
by including another critical point $c$ it is easy to see that
the relative sign in the $ac$, $cb$ and $ab$ sectors
are correlated with $\Delta_a\circ\Delta_c$,
$\Delta_c\circ\Delta_b$ and
$\Delta_a\circ\Delta_b$.  This finally shows \fund\ is
true for a suitable choice of basis for the vacua.

We now consider what happens to the soliton numbers when
we perturb the superpotential $W$. As we perturb
$W$ the critical values move in the $W$--plane.  As long
as no critical value crosses the straight line connecting
two other critical values, the stability of intersection
numbers under continuous deformations guarantee that the
soliton numbers do not change.  But suppose the
critical value of a vacuum $j$ lies exactly on the straight
line from critical value of vacuum $i$ to $k$ (see \tfig\FiglabelA ).
Then
the wave front originating from $i$ cannot be continued
passed the point $j$, as some trajectories originating
from $i$ may get absorbed by $j$ or some new
trajectories may open up. So in this way
the soliton number of the $ik$ sector changes,
as some new solitons may appear which previously used
to go through $j$ and were not  {\it primitive}, or some primitive
solitons
in the $ik$ sector may become composite solitons $ij$,$jk$.  Can we
compute
this change in soliton number?  We should be able
to as it just involves understanding what happens to the
vanishing cycles as the vacua pass through an aligned
configuration.

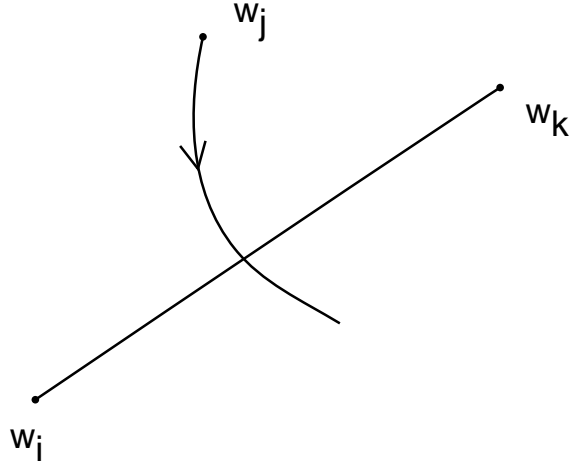
\ifigure\FiglabelA{Vacuum $j$ labeled by its critical value
in the $W$--plane will pass, by perturbing the theory, through
a straight line connecting two other vacua $i,k$.}{FigA}{2.5}

In order to discuss this it is useful to recall some
facts about vanishing cycles.  If we fix a non-critical value $t$
in  the $W$ plane and look at the preimage of that point,
we get an $n-1$ complex dimensional space.  The (compact) homology
cycles are in real dimension $n-1$ and they can be
described as follows \singt :
Connect the point $t$ to the critical values $w_1,..,w_n$ along
some cyclically ordered paths $\gamma_i$ which do
not cross any critical values (see \tfig\FiglabelB ), and consider
the
$n-1$ cycles that vanish as we go along these paths (by homotopy
lifting) to each of the critical values.
These form a basis for $n-1$ cycles.  Let us denote the $i$-th
vanishing cycle,
the one that vanishes along $\gamma_i$, by $\Delta_i$. The question
we
wish to address now is how to use the intersection between these
cycles
to find the soliton numbers $\mu_{ij}$.  If the point $t$ is along
the straight line connecting $i$ and $j$ and $\gamma_i$ and
$\gamma_j$
are the straight lines connecting $t$ to $i$ and $j$ respectively
then it is clear that $\mu_{ij}=\Delta_i\circ\Delta_j$.  Since
the intersection numbers are rigid under continuous deformations this
means that as long as we can deform $t$ and $\gamma$'s continuously
to the
above situation {\it without having the paths} $\gamma_i$ {\it cross
critical
values} $w_r$ we can still use these intersections to count the
corresponding
soliton numbers.
However sometimes this cannot be done with a particular choice of
the paths $\gamma_i$, and we will have to choose a different set of
paths connecting $t$ to critical values, and this will give us a
different
basis for the vanishing cycles.  Equivalently for any given choice of
paths
$\gamma_i$, by deforming the critical values by perturbation of the
theory,
we can arrange so that
 the intersection  numbers of the corresponding cycles  do count the
soliton
numbers of the perturbed theory.  So there is a one to one
correspondence
between the set of paths and the set of possible perturbations of the
critical
values.

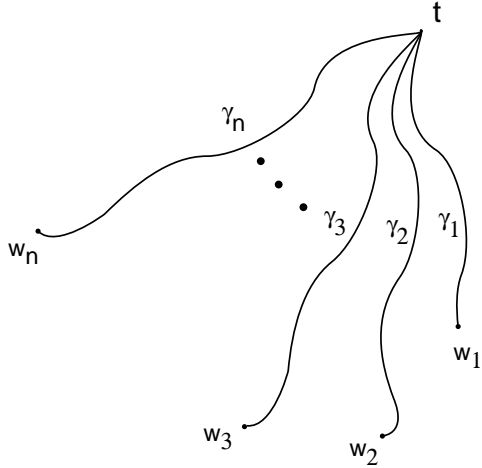
\ifigure\FiglabelB{The critical points in the $W$--plane
are connected to a point $t$ along some cyclically ordered
paths $\gamma_1,...,\gamma_n$.}{FigB}{2.5}

The theory of how the vanishing cycles change by choosing a different
basis of $\gamma_i$ is known as the Picard-Lefschetz theory \singt .
Suppose we wish to change a particular cycle $\gamma_i$ to
a path $\gamma_i'$ by passing
it through the critical value $w_j$ (see \tfig\FiglabelC ) .  If we
know how
the cycles
change under this particular change of basis, since we can get
an arbitrary basis by just repeating such steps over arbitrary
critical values we would be done.  The Picard-Lefschetz theorem
implies
that the new vanishing cycle $\Delta_i'$ is given by
\eqn\plef{\Delta_i'=\Delta_i \pm (\Delta_i\circ\Delta_j) \Delta_j}
where the $\pm$ sign corresponds respectively to whether the circle
$\gamma_i (\gamma_i')^{-1}$ is clockwise or counter-clockwise in the
$W$ plane (in the case of Fig.3
it is $+$ sign).  This formula is very much like the formula for
`Weyl
reflection'
and it is indeed exactly that for the example of minimal models that
we will
discuss later.

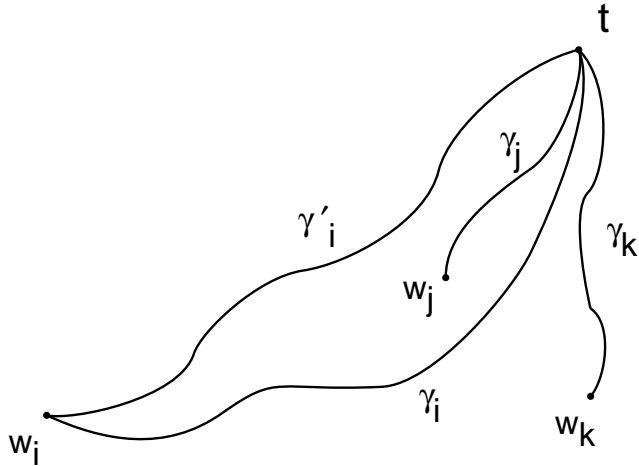
\ifigure\FiglabelC{The vanishing cycles change if we
choose a different set of paths.  In this case $\Delta_i$ changes
to $\Delta_i'$ as
we have deformed path $\gamma_i$ by passing it through $w_j$ to
a new path $\gamma_i'$.}{FigC}{2.5}

Now we are set to compute the change of the soliton number as
critical
values pass through configurations in which three critical values get
aligned.
As discussed above as the $j$-th vacuum crosses the $ik$ line  this can
be equivalently described by a change of basis of path
by changing $\gamma_i$ (see \FiglabelA\ and \
\FiglabelC ).  So the new soliton number
$\mu_{ik}'$ is
given by
\eqn\solnc {\mu_{ik}' =\Delta_i'\circ\Delta_j =(\Delta_i \pm
(\Delta_i\circ\Delta_j)
\Delta_j)\circ\Delta_k=
\mu_{ik}\pm \mu_{ij}\cdot \mu_{jk}}
where the $\pm$ will correspond respectively to whether the
right-hand
rule applied to the triangle $ijk$
{\it before} the $j$-th vacuum crosses $ik$ line is into or out of
the $W$
plane.  The formula
\solnc \ can be intuitively understood by noting that we get
new solitons (or lose solitons) by the fact that composite solitons
in
the $ik$ sector (composed
of two solitons in the $ij$ and $jk$ sectors) become primitive
solitons (or
vice-versa).
We will show how to derive equation \solnc \ from purely physical
reasoning
in the next section for arbitrary $N=2$ models, thus generalizing
this result for the Landau-Ginzburg theory.

Having set up all the machinery we now come to proving a surprising
relation
between the monodromy of vanishing cycles and the intersection
numbers.
Suppose we pick a point $t$ on the $W$ plane very far from the
critical
points.  Furthermore let us choose the paths $\gamma_i$ to be
straight lines
connecting $t$ to the critical values.  As $t$ goes around a large
circle in a
clockwise
direction with $|t|$ fixed (see \tfig\FiglabelD ), the vanishing
cycles undergo
a monodromy.  We can compute
what this monodromy is just by using \plef\ which tells us what
happens when $\gamma_i$ cross any of the vacua.  In other words,
consider
$n(n-1)$ half lines passing through the vacua in pair and originating
from one of the vacua.
Let us denote the half-line originating at the $j$-th and passing
through the
$i$-th vacuum
 by $l_{ij}$.  Then as $t$ cross the line $l_{ij}$ the basis for the
vanishing
cycles change, using \plef\ by multiplication with the matrix
$$M_{ij}=1-A_{ij}$$
where $1$ denotes the identity matrix and $A_{ij}$ is a matrix whose
only
non-vanishing entry is the
$ij$ entry and that is equal to
$A_{ij}=\mu_{ij}=\Delta_i\circ\Delta_j$.

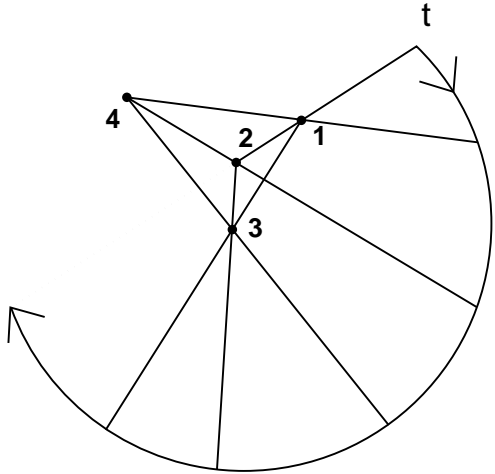
\ifigure\FiglabelD{The monodromy of the vanishing cycles
can be computed by taking $t$ on a large circle in the
$W$--plane connected by straight lines to the vacua.
As the straight lines overlap lines joining pairs of vacua
we pick up contributions to the monodromy.}{FigD}{2.5}

Note that
\eqn\usid{M_{ji}=1-A_{ji}=M_{ij}^{-t}}
where we used the fact that $A_{ji}=-A_{ij}^t$ and that as a matrix
$(A_{ij})^k=0$ for $k>1$.  Let $S$ denote the ordered product
(ordered according to which $l_{ij}$ line crosses the circle first)
of
matrices
$M_{ij}$
as we go half the way around the large circle
\eqn\defs{S=\prod_{l_{ij}\  cross\  half \ circle
}^{\rightarrow}M_{ij}}
Then as we go around the full circle, because of the identity \usid ,
and
because the order in which the lines cross the second half circle
is the same as the order in which they cross the first half circle
modulo replacing $l_{ij}$ with $l_{ji}$ we get the full
monodromy matrix $M$ to be
$$M=S^{-t}S$$

We will be interested in the eigenvalues of $M$.
We will compute the eigenvalues of the monodromy matrix in another
way:
We first note that the matrix $M$ is independent of finite
deformations of the
vacua.  So in the limit in which all the critical values become
equal, i.e.
the conformal case in which $W$ is quasi-homogenous, the eigenvalues
of the monodromy matrix $M$ can
be computed by a suitable choice of $n-1$ forms, which form a basis
for the
dual space to the vanishing
$n-1$ cycles. Let $\phi_k$ be a monomial basis for the chiral ring
${\cal R}={C[x_i]\over dW}$. Let $q_k$
be its degree (charge). Consider the $n$ form
$$\omega_k =\phi_k dx^1.... dx^n$$
Since $t$ is not a critical value of $W$ we define an
$n-1$-form $\alpha_k$ defined on the preimage of $W=t$ by
$$\omega_k=\alpha_k \wedge dW .$$
Then it is known that $\alpha_k$ form a basis for the dual
to the vanishing cycles \singt .
  Now consider deforming $t\rightarrow e^{2\pi i} t$.  This can
be undone, since $W$ has charge $1$ by letting
$$x^i\rightarrow e^{2\pi i q_i} x^i$$
So using the above formula for $\alpha_k$, we see that it transforms
by
$$\alpha_k \rightarrow (-1)^n e^{2\pi i (q_k -{\hat c \over
2})}\alpha_k$$
where $\hat c =\sum_i(1-2q_i)$ is the (normalized) central
charge of the conformal theory (we put $(-1)^n$ in the above to
cancel
the term involving $\sum_i 1/2$ in the definition of $\hat c /2$).
Now if we
take
even number of variables, as we have done, the $(-1)^n$ disappears
and we get
\eqn\fueq{{\rm Eigenvalues}(S^{-t}S)=e^{2\pi i q^R_k}}
where $q^R_k=q_k -{\hat c\over 2}$ denotes the charge of the
ground states of the Ramond sector.  This is the relation we were
after, which connects the information about the soliton spectrum on
the left, a property of the massive theory, to the spectrum
of the charges of chiral fields of the conformal theory on the right,
a property of the massless theory\foot{The reader
may wonder how reliable is the argument in the text
since we claim to be able to compute UV quantities
in the semi--classical limit (which is the IR limit
for the LG models). The point is that $\tr M^m$
is equal to $\Tr(-1)^F g^m$ where $g$ is the operator generating
a $U(1)$ transformation by $2\pi$. Such objects are
susy indices and so can be reliably computed. But then the
eigenvalues of $M$ can be computed in any regime we please.}.
 This theorem for the LG case was
known to the mathematicians (in the mathematical way
of thinking it is a relation between the intersection numbers
of vanishing cycles with the Milnor monodromy of the singularity)
\singt .
Note that equation \fueq\ gives the Ramond charges only mod integers.
We will find a method, which applies to an arbitrary $N=2$ QFT in
later sections, which also gives the {\it integral} part of the
charges.

The matrix $S$ can be simplified further if we choose a particular
deformation of the theory.  This certainly should not affect the
monodromy
as the monodromy is independent of the perturbation. We deform the
critical
values so that the polygon $w_1,w_2,...,w_n,w_1$ is convex\foot{ It
is
not clear that we can always do this, because not
all the chiral fields are relevant perturbations, and so generally
we cannot add all of them to the action.  Nevertheless
this does not modify the relation we derived for the monodromy,
as choosing such configurations correspond to conjugating $SS^{-t}$
by some matrix and does not affect the relation between charges and
the
monodromy.  Therefore it is useful to assume that at least formally
we can
choose such a configuration.}.  Moreover we assume that the polygon
is such that $l_{ij}$ crosses the half circle for all $i<j$.  This
configuration of vacua we call {\it standard} configuration.  Then
the matrix
$S$ given by \defs\ simplifies because  the products of $A$'s vanish
for
this convex geometry and we get
$$S=1-A$$
where
$$A=\sum_{i<j}A_{ij}$$
Note that $A$ is strictly an {\it upper triangular} matrix, and thus
in this deformed version $S$ is just upper triangular, with $1$'s on
the
diagonal
and $-\mu_{ij}$, i.e. minus the $ij$ soliton number, on the $ij$
entry with
$i<j$.

For a given $N=2$ theory there are many `standard' configurations.
Going
from one such configuration to another will give a new matrix $A$,
as the number of solitons will change.  So even after we restrict
to upper triangular matrices we will end up with many matrices $A$
which are {\it equivalent} modulo perturbations of the original
theory.  Indeed there is an action of the Braid group on $S$
which corresponds to this equivalence:
Consider ordering the vacua according to decreasing value of
${\rm Re}(W)$ from $1,...,n$.  Let us further assume that the vacua
are in the form of a convex polygon.
Order the $w_k$'s  so that ${\rm Re}\,w_k>{\rm
Re}\, w_j$ for $k<j$ and choose the imaginary parts
so that they form a `standard' convex polygon where $S=1-A $.
Let us deform the theory.  It is easy to see from
the definition \defs\ and \solnc\ that a deformation which leaves
 invariant the
real parts of the $w_k$'s does not change $S$ (
as long as $l_{ij}$ crosses the half circle for all $i<j$ ), although
in
general,
the soliton numbers change since some vacua get aligned.
Next let us perturb the
theory so that the $i+1$-th and $i$-th vacua exchange their positions
as shown in \tfig\FiglabelE .
We deform the $w_{i+1}$ coupling along a clockwise path making an
half
turn around $w_{i}$ in such a way that we end up with the `standard'
configuration but now with $w_i$ in the $(i+1)$--th place.
In doing this, $w_{i+1}$ crosses once all soliton lines
emanating from the point $w_{i}$ (except, of course, the line through
$w_i$ itself).
The effect on $S$, using our discussion of
how soliton numbers change, is
\eqn\efons{S\rightarrow P S P^t \qquad P=(1+A^t_{i,i+1})P_{i,i+1}}
where $P_{i,i+1}$ is the matrix permuting $i$ and $i+1$.
 It is clear
from this geometrical description that repeating this
operation for all $i$ forms a braid group.
Note that the above transformation on $S$ acts as
$$S^{-t}S\rightarrow P^{-t}(S^{-t}S)P^{t}$$
and thus does not change the eigenvalues of the monodromy matrix
$SS^{-t}$ as expected.

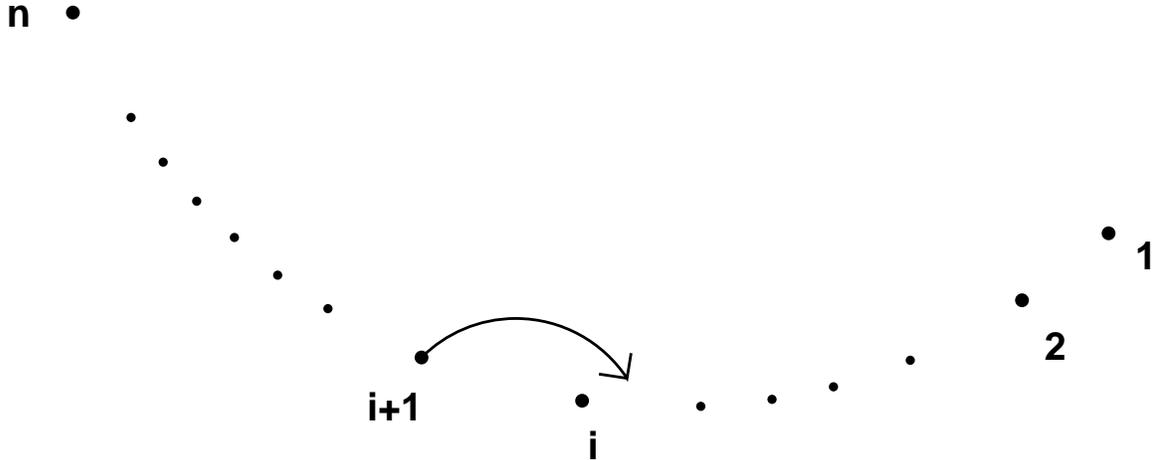
\ifigure\FiglabelE{The exchange of the $i+1$--th vacuum
with the $i$--th vacuum generate a Braid group.}{FigE}{2.5}

\newsec{Generalizations}

In this section we discuss how the results of the previous
section can be {\it stated} for any $N=2$ massive
quantum field theory in two dimensions.   This is not
automatic as even the definition of some of the objects in the
previous sections seemed to depend on the fact that we were
describing the Landau-Ginzburg models.  We will
show that this is not an obstacle.  We prove some of the general
statements
that we make, but the
proof of the main statement relating the $S$ matrix with
the $U(1)$ phases is left for the next section.

The first thing to discuss is what we
mean by a {\it massive} $N=2$ theory.  We mean one which has
a mass gap with non-degenerate
vacua.  In particular this means that each of the vacua support local
massive
excitations.  Let us
label the vacua by $i=1,...,n$.  In an abstract definition, this
`point basis' can be defined by the condition of diagonalizing
the chiral ring, i.e., we can choose representatives of the chiral
ring labeled by $\Phi_j$ such that
$$\Phi_j |i\rangle =\delta_j^i|i\rangle. $$
Note that the condition of having non-degenerate vacua
which is needed for a massive theory cannot be satisfied
for $N=2$ theories which have elements in the chiral ring
with non-vanishing fermion number $F$.  In particular since
fermion number is conserved by the $N=2$ algebra (even
for a massive theory) we will end up having degenerate
vacua.  So a necessary condition
for a conformal theory to admit a non-degenerate massive deformation
is that {\it it have vanishing fermion number for chiral ring
elements}\foot{It would be very interesting to study massive
theories which do not satisfy this constraint, an example
of which is provided by K\"ahler manifolds with positive
$c_1$ with non-vanishing off-diagonal Hodge numbers.}.

A crucial ingredient in our discussion of
the LG case was the understanding of the solitons in
the theory.  The definition of solitons of interest is as easy
in the general case:  We consider the $ij$
sector defined by the condition that we start with a vacuum $i$
on spatial infinity at left and end up with vacuum $j$ at spatial
infinity to the right. The solitons of interest to us are
the ones that saturate the Bogomolnyi bound.  What this
means is the following:  The $N=2$ algebra in the $ij$ sector
has a central extension which we denote by $w_{ij}$ and appears in
\eqn\cent{\{ Q^+,\bar Q^+\}=2w_{ij}}
It is easy to show, using the rest of the $N=2$ algebra,
that the mass $m$ of any state in the $ij$ sector satisfies
$$m\geq 2|w_{ij}|$$
As discussed in \newindex\ $\Tr(-1)^F F $ counts
the number of Bogomolnyi solitons.  So at least this  part
of the definition which we used in the LG case exists quite
naturally in the general set up.

In the previous section we also saw that the critical values
in the $W$--plane played a crucial role
in the change of soliton numbers as we perturb
the theory.  In particular when
three vacua passed through a configuration in which they were
 aligned in the $W$--plane the number of solitons
changed.  So if we wish to understand how soliton numbers
change we first need to see if we can define the notion of a critical
value of a vacuum.  This can be done as follows:  The central term
in the supersymmetry algebra \cent\ is additive, i.e.,
$$w_{ik}=w_{ij}+w_{jk}$$
This together with the fact that $w_{ii}=0$, implies that we can
assign to each vacuum $i$ a critical value $w_i$, unique up
to an overall shift, such that
$$w_{ij}=w_i-w_j$$
So the notion of critical value is also universal and not
restricted to LG theories.  So we now ask if the number of solitons
change as in the LG case when three vacua pass through an aligned
configuration. The answer is exactly as in the LG case, but
the proof will be different; after all in the general case we
do not have the analog of Picard-Lefschetz theory which gave
the formula in the LG case.  What we have instead is the fact
that the $tt^*$ equations (topological--anti--topological equations)
\topatop\ have continuous solutions.  In particular the new
supersymmetry
index defined in \newindex\ which computes $Q=\Tr(-1)^F F\,
\exp(-\beta
H)$
is a continuous function of moduli of the theory.  Now the leading
contribution, up to two soliton terms, to this index was computed in
\newindex .  Using the results of that paper, it is clear that there
will be a jump in the contribution of two particle solitons to
$Q$ in the $ik$ sector precisely as the $j$-th critical value
passes through the straight line connecting $w_i$ to $w_k$
(see equation 4.14 of \newindex ).  This
jump is unphysical, as $Q$ should be continuous.  Indeed the jump
in two soliton contribution is of the same form as the one soliton
contribution in the $ik$ sector.  So to compensate that jump the
number of solitons in the $ik$ sector must have jumped precisely
by
$$\mu_{ik}\rightarrow \mu_{ik}\pm \mu_{ij}\mu_{jk}$$
where the $\pm$ sign depends again on the orientations of the
$j$-th critical value crossing the $ik$ line
(as follows from equation 4.14 of \newindex ).  This is exactly
the same answer as in the LG case and so we have recovered it
without using Picard-Lefschetz theory.  This suggests that
{\it in the general case} $tt^*$ {\it equations are sufficiently
powerful to replace the  Picard-Lefschetz theory}.
Indeed we will find that not only this is true, but in
some sense it is even {\it stronger} than Picard-Lefschetz theory.
In particular we will use $tt^*$ equations to derive results
which were not known to mathematicians (as far as we know) using
Picard-Lefschetz theory.

Since we have translated the number of solitons and the geometry
of change of soliton numbers to the abstract `$W$--plane' even when
we are not dealing with LG, it is clear that all the rest of the
discussion about the LG case would lead to a natural guess about the
relation between the soliton numbers and the chiral charges at
the conformal point.  Namely the eigenvalues of $SS^{-t}$,
where $S$ is as defined in the previous section, should be related
to $\exp(2 \pi i q_i)$ where $q_i$ are the (left) charges of Ramond
vacua at the conformal point.  Also, the choice of a simple
vacuum geometry, i.e., the `standard configuration' for critical
values simplifies the formula for $S$ to be $S=1-A$ where $A$ is
strictly upper triangular
and counts the soliton numbers.  Also the
discussion about the action of the Braid group on $S$ at the end
of the previous section is equally applicable in the general set up.
In other words, we do not need the notion of {\it vanishing cycles}
which does not exist in any obvious sense in the general set up
to formulate the main results of the previous section.

We will indeed go one step further in the general set up, which was
not done in the LG case:  Note that from $S$ it seemed that we have
only a way of fixing the chiral charges $q_i$ modulo addition of
integers.
We will show in the next section that we can also fix its integral
part.  The idea is to consider
$$S(t)=1-A(t)$$
where $A(t)$ is a continuous function of $t$ and is a real
strictly upper triangular matrix interpolating from $0$ to $A$
as $t$ runs from $0$ to $1$.  We then consider the eigenvalues of
$SS^{-t}$ as a function of $t$.  Note that the eigenvalues are
never zero and so we can consider how many times a given eigenvalue
wraps around the origin as $t$ goes from $0$ to $1$.  This will be
the integral part of $q_i$\foot{The reader may worry about collision
of eigenvalues, but this can be avoided by considering a slight
perturbation of $A$ .}.  As far
as we know this is a new result even for the singularity theory\foot{
To make it fully rigorous we need one assumption which we have not
been able to rigorously prove.  See section 4.}.

\def\ie{{\it i.e.}}
\def\eg{{\it e.g.}}
\def\CP{{\cal P}}
\def\CM{{\cal M}}
\def\CK{{\cal K}}
\def\ttstar{$tt^\ast$}
\def\BZ{{\bf Z}}
\def\BQ{{\bf Q}}
\def\BP{{\bf P}}
\def\BR{{\bf R}}
\def\BC{{\bf C}}
\def\ipi{{1\over 2\pi i}}
\def\CE{{\cal E}}
\def\prodarrow{\mathop{\overrightarrow\prod}}



\lref\CCS{W. Barth, C. Peters, and A. van de Ven, {\it Compact
Complex
      Surfaces}, (Springer--Verlag, Berlin, 1984).}

\lref\graph{D. Cvetkovi\' c, M. Doob, H. Sachs, {\it
Spectra of Graphs. Theory and Applications}, Academic Press, New
York,
1980.}
\lref\moregraph{D. Cvetkovi\' c, M. Doob, I. Gutman and A. Torga\v
sev, {\it Recent Results in the Theory of Graph Spectra},
North--Holland,
Amsterdam 1988.}
\lref\surprise{M. Doob,
{\it A Surprising Property of the Least Eigenvalue of a Graph},
Linear
Alg. and Appl. 46 (1982) 1.}
\lref\bourbaki{N. Boubaki, {\it Groupes et alg\'ebres de Lie},
Hermann, Paris 1968.}
\lref\coxter{H.S.M.
Coxeter, {\it Regular Polytopes},  Macmillan, New York 1948.}
\lref\coxt{H.S.M. Coxeter, Duke Math. J. 18 (1951) 765.}
\lref\ising{L. Onsager, Phys. Rev. 65 (1944) 117\semi   T.T. Wu and
     B.M. McCoy,
    {\it The two dimensional Ising Model}, Harvard University Press,
Cambridge,
    Mass. 1973.}
\lref\mccoy{B. M. McCoy and T.T. Wu, Phys. Rev, Lett, 45 (1980) 675;
    B.M. McCoy, J.H.H. Perk and T.T. Wu, Phys. Rev. Lett. 46 (1981)
757.}
\lref\japanese{M. Sato, T. Miwa and M. Jimbo, Publ. R.I.M.S. 14
(1978) 223 15
    (1979) 201; 577; 871; 16 (1980) 531; 17 (1981) 137\semi
    M. Jimbo and T. Miwa,
    {\it Aspects of holonomic quantum fields}, Lecture Notes in Phys.
vol.126,
    Springer 1980 p.429-491.\semi   M. Jimbo and T, Miwa,
    {\it Integragrable Systems
    and Infinite Dimensional Lie Algebras}, in {\it Integrable
Systems in
    Statistical Mechanics}, Ed. G.M. D'Ariano, A. Montorsi, M.G.
Rasetti, World
    Scientific, Singapore, 1988\semi    M. Jimbo, Proceedings of
Symposia in
Pure
    Mathematics, 49 (1989) 379.}
  \lref\chiralring{W. Lerche, C. Vafa and N. Warner, Nucl. Phys. B324
(1989)
427.}
\lref\piiimath{B. McCoy, C.A. Tracy, and T.T. Wu, J. Math. Phys. 18
(1977)
    1058\semi   T.T. Wu, B.M. McCoy, C.A. Tracy, and E. Barouch,
Phys. Rev. B13
    (1976) 316. }
\lref\aits{A.R. Its and V.Yu. Novokshenov, {\it The Isomonodromic
    Deformation Method in the Theory of Painlev\'e Equations},
Lectures
    Notes in Mathematics 1191, Springer 1986.}
\lref\highermod{B. Dubrovin, {\it Differential Geometry of Moduli
Spaces
    and its Application to Soliton Equation and to Topological
Conformal
    Field Theory}, Preprint 117 of Scuola Normale Superiore, Pisa,
    November 1991.}
\lref\riemann{H. Flaschka and A.C. Newell, Commun. Math. Phys. 76
(1980) 67.}
\lref\markoff{A.A. Markoff, Math. Ann. 15 (1879) 381\semi
    A. Hurwitz, Archiv. der Math. und Phys. 3 (14) (1907) 185\semi
    L.J. Mordell, J. Lond. Math. Soc. 28 (1953) 500\semi
    H. Schwartz and H.T. Muhly, J. Lond. Math. Soc. 32 (1957) 379.}
\lref\mordell{L.J. Mordell, {\it Diophantine Equations},
    Academic Press, London 1969.}
\lref\lambdamat{F.R. Gantmacher, {\it The Theory of Matrices},
    Chelsea, 1960.}
\lref\monlemma{F. Lazzeri, Some remarks on the Picard--Lefschetz
monodromy,
     in {\it Quelques journ\'ees singuli\`eres}, Centre de
Mathematique de
     l'Ecole Polytechnique, Paris 1974.}
\lref\krona{L. Kronecker, {\it Zwei S\"atze \"uber Gleichungen mit
     ganzzahligen Coefficients}, Crelle 1857, Oeuvres 105.}
\lref\cotech{F.M. Goodman, P. de la Harpe, V.F.R. Jones, {\it Coxeter
     Graphs and Tower of Algebras}, Mathematical Sciences Research
     Institute Publications 14, Springer--Verlag, 1989.}
\lref\gross{B.H. Gross, Inv. Math. 45 (1978) 193.}
\lref\weilA{A. Weil, {\it Introdution a les Vari\'et\'es
K\"ahl\'eriennes},
     Hermann, Paris, 1958.}
\lref\griHarris{P. Griffiths and J. Harris, {\it Principles of
Algebraic
     Geometry}, Wiley--Interscience, New York, 1978.}
\lref\acampo{N. A'Campo, Indag. Math. 76 (1973) 113.}
\lref\cyclopoly{K. Ireland and M. Rosen, {\it A Classical
Introduction
     to Modern Number Theory}, (Springer--Verlag, Berlin, 1982)\semi
     R. Sivaramakrishnan, {\it Classical Theory of Arithmetic
Functions},
     (Dekkler, New York, 1989).}
\lref\Stokesph{W. Wasow, {\it Asymptotic Expansions for Ordinary
    Differential Equations}, Dover, New York, 1965\semi   for a
physicist's
     discussion see \eg\ G. Moore, Comm. Math. Phys. 133 (1990) 261.}
\lref\riehilb{N.I. Muskhelishvili, {\it Singular Integral Equations},
    Noordhoff, Groninger, 1953.}
\lref\saddlep{L. Sirovich, {\it Techniques of Asymptotic Analysis},
    Springer--Verlag, New York, 1971.}
\lref\piiideg{A.V. Kitaev, {\it The Method of Isomonodromic
Deformations for
     Degenerate Third Painlev\'e Equation} in {\it Questions of
Quantum Field
     Theory and Statistical Physics} 8 (Zap. Nauch. Semin. LOMI
v.161) ed. V.N.
     Popov and P.P. Kulish (Nauka, Leningrad) (in russian).}

\lref\massGap{E. Abdalla, M. Forger and A. Lima Santos, Nuclear Phys.
B256
     (1985) 145\semi   E. Abdalla and A. Lima Santos, Phys. Rev. D29
(1984)
     1851\semi   V. Kurak and R. Koberle, Phys. Rev. D 36 (1987)
627.}

\newsec{Isomonodromic Deformations and the General Solution of
\ttstar}

\subsec{The \ttstar\ Equations}

In this section we give a general proof of \fueq\ which does
not depend on a particular Lagrangian formulation of the theory,
Landau--Ginzburg or otherwise. The idea is to use
the differential equations
which describe the ground state geometry
 (\ttstar\ equations \topatop)
to connect the leading IR behaviour
(encoded in the soliton spectrum) to the UV one which is specified by
the $U(1)$ charges of the Ramond vacua $q_k$.
The basic quantity of interest is the `new index' \newindex\
\ie\ the matrix
\eqn\barba{Q_{ij}=
 \lim_{L\rightarrow \infty} {i\beta\over 2L}\Tr_{(i,j)}\Big[(-1)^F F
e^{-\beta H}\Big].  }
Here $\Tr_{(i,j)}$ means the trace over the sector $(i,j)$ of the
(infinite
volume) Hilbert space. This sector is specified by requiring that
as $x\rightarrow +\infty$ (resp. $-\infty$) the field configuration
approaches
the $j$--th (resp. $i$--th) vacuum.  By definition $Q$ is related
to the axial $U(1)$ charge of the vacua.  At the conformal point (UV
limit)
$Q$ is the same as the left (or right) charges of Ramond
ground states, and at the IR the leading contribution to
\barba\ counts the number of solitons. So this object
knows about both sides of \fueq\ and is thus no surprise
that studying it would lead to proving \fueq .

For a completely massive theory there is a natural system of
coordinates in
coupling constant space, \ie\  the canonical coordinates\foot{For
simplicity,
we  assume we are in a {\it generic} situation, \ie\
 $w_i\not=w_j$ for $i\not=j$.} $w_k$
($k=1,\dots,n$) \refs{\cancoordinates,\noisi,\solttstar}.
They are defined as follows. Let $Z=\{Q^+,\bar Q^+\}$ be the central
charge
in the N=2 algebra. Then as discussed before we can set
$$ w_i-w_j=\half Z\Big|_{(i,j)}.$$

In the canonical coordinates one
has \noisi\
\eqn\Qg{Q=-\sum_k w_k g\partial_k
 g^{-1}=-\half g \beta\partial_\beta g^{-1},}
where $g_{i\bar j}=\langle\bar j|i\rangle$, is the ground
state--metric in
the canonical basis\foot{The canonical basis $|i\rangle$ is the
topological
basis (see \topatop)
such that: 1) the chiral ring $\CR$ is diagonal, and
2) the topological metric is
normalized to $1$. The canonical basis is unique up to sign.}.
 $g$ satisfies the differential equations (\ttstar\ {\it equations})
\eqn\diif{\eqalign{&
\bar\partial_i\big(g\partial_j g^{-1}\big)=[C_j,\bar
C_i],\cr
&  [g\partial_j g^{-1},C_k]=[g\partial_k g^{-1},C_j],\cr}}
where $C_k$ are the matrices representing in $\CR$ the multiplication
by the chiral primary operator $\phi_k$ such that
$$\delta S= \sum_k \delta w_k \int \phi_k^{(2)}.$$
By definition, in the canonical basis we have
\eqn\whatC{(C_k)_i^{\ j}=\delta_{ki}\delta_i^{\ j}.}
Then, in this basis, the \ttstar\ equations take a universal form
\refs{\noisi,\solttstar} which is
nothing else than the equations for the Ising
$n$--point functions (see \noisi\ for details).
Different models differ {\it only in the boundary conditions}
satisfied by solutions of $tt^*$ equations.  Thus, finding
a universal way to describe the boundary
conditions will lead to a classification of different models.

{}From the thermodynamical interpretation of $Q_{ij}$ \newindex\
it is clear that  the
general solution to \ttstar\ can be
written in the form of a soliton expansion, and that
the specific boundary conditions for \diif\ are
encoded in the soliton spectrum. More precisely,
we have $n(n-1)/2$
soliton `fugacities'
$\mu_{ij}=-\mu_{ji}$ corresponding to the $n(n-1)/2$ possible
soliton masses  \FMVW
$$m_{ij}=|Z|\,\Big|_{(i,j)}=2|w_i-w_j|.$$
 The `fugacities' are defined by the
asymptotics\foot{The convention--dependent phases are chosen so that
$\mu_{ij}$ is real. Here and below $K_\nu(\cdot)$ are modified
Bessel functions.} \newindex\
\eqn\fugadef{Q_{ij}\Big|_{\beta\rightarrow\infty}
\simeq -{i\over 2\pi}\mu_{ij}
m_{ij}\beta\,  K_1(m_{ij}\beta), }
or, in terms of $g$, (\topatop\ App. B)
\eqn\gAsymp{g_{i\bar j}\simeq \delta_{ij}-{i\over
\pi}\mu_{ij}K_0(m_{ij}\beta).}
Although regular solutions to \diif\ exist for real\foot{Only for
$\mu_{ij}$
small enough the solutions are expected to be regular. This reflects
the fact
that there is an upper bound for the UV central
charge $\hat c$ of a unitary `massive' theory with a given Witten
index $n$.
For instance, for $n=2$ we get $\hat c\leq 1$. Stated differently,
let us
order
the eigenvalues $q_i$ of the $Q$--matrix in increasing order. Then
the gaps
$(q_{i+1}-q_i)$ cannot be too big.} $\mu_{ij}$, in  the physical case
 $|\mu_{ij}|$ is an integer counting the number of soliton
species connecting the $i$--th and $j$--th vacua.
On physical grounds one expects that varying $\mu_{ij}$ one gets all
possible solutions to the \ttstar\
equations.
The UV asymptotics is
\eqn\smallbeta{Q_{ij}\Big|_{\beta\rightarrow 0}=q_{ij}. }
where $q_{ij}$ are the $U(1)$ charges of the Ramond vacua at the UV
fixed
point \topatop. Since the solution depends on the boundary data
$\mu_{ij}$, the
\ttstar\ equations may be seen as a map from the soliton spectrum
$\mu_{ij}$
to the possible values of the $U(1)$ charges. Below we show that this
map
is precisely the one predicted by eq. \fueq .

 Since
$Q_{ij}$ can be computed from the
ground--state metric, we should be able
 to read the  $n$ soliton contribution to
the $Q$ matrix from the \ttstar\ equations.
Indeed the general solution to the \ttstar\
equation (for a massive model) has
naturally the form of a grand--canonical
sum over $n$--soliton sectors. For the case of two
vacua (corresponding to PIII) this has been shown in \refs{\noisi,
\newindex}.
This case is particularly easy since
there is only one soliton of mass
$2|w_1-w_2|$. In the soliton expansion of \barba,
 the $n$--soliton
sector contributes a term of
order
$$\exp\big(-2\beta|w_1-w_2|n\big)
\qquad {\rm for}\ \beta\ {\rm large}.$$
For PIII the soliton expansion (first obtained in Ref.\piiimath) is
in terms of Ising form--factors.
By the remark after \whatC\ this is true in general.

\subsec{The Integral Formulation of \ttstar\ \rm \solttstar}

In principle to
get the general soliton expansions we
could start from the Ising form
factors or, equivalently, from
the known series for the Ising correlation functions \mccoy.
However it is more convenient
to take advantage of the analysis of the \ttstar\ equations
due to
Dubrovin \solttstar. He was able to reformulate the (massive)
\ttstar\ equations  as a
Riemann--Hilbert problem\foot{For a review of this problem,
see \eg\ \riehilb.} having a very
convenient expression in terms of
linear integral equations.
Here we recall the aspects of his work we need in the following.

Introducing the covariant derivatives
\eqn\covadev{\eqalign{& \nabla_i=\partial_i+(g\partial_i g^{-1})-x
C_i,\cr
&
\overline{\nabla}_{\bar i}=\bar\partial_{\bar i}-x^{-1}\bar C_{\bar
i},\cr}}
where $x$ is a spectral parameter,
we can rewrite the equations \diif\
 as the consistency (zero--curvature) conditions for the system of
linear differential equations,
\eqn\lindiff{\nabla_i \Psi(x,w_k)=\overline{\nabla}_{\bar
i}\Psi(x,w_k)=0,}
where $\Psi(x,w_k)$ is an $n\times n$ matrix.
  Clearly, in order to solve the
\ttstar\ equations it is enough to compute\foot{To save print we
usually omit the dependence of $\Psi$ on the couplings $w_k$ and
$\bar w_k$.} $\Psi(x)$. To completely specify the $tt^*$ geometry
one needs to impose the condition that $g$ is independent of an
overall rotation in the value of $w_i$.  In order to incorporate
this condition naturally, let
 us consider the dependence of $\Psi$ on the
overall scale $\beta$ and the overall chiral angle $\theta$. This
amounts
to a redefinition of the canonical coordinates as
\eqn\overall{w_k\rightarrow \beta e^{i\theta} w_k.}
{}From \covadev, \lindiff\ we get (after the identification
$x=e^{i\theta}$, natural in view of \covadev)
\eqn\diffx{
x{\partial\phantom{x}\over \partial x}\Psi=
\big(\beta x C + Q-\beta x^{-1}\bar
C\big)\Psi}
\eqn\diffbeta{\beta{\partial\phantom{\beta}\over \partial \beta}\Psi=
\big(\beta x C + Q+\beta x^{-1}\bar
C\big)\Psi,  }
where
$$C=\sum_k w_k C_k,\qquad \bar C= g C^\dagger g^{-1}.$$
Notice that $Q$, $C$ and $\bar C$ are independent of $x$. Indeed, the
$w_k$'s overall phase can be absorbed in the phase of the fermions.
The introduction of a spectral parameter $x$ allows us to extend
$\Psi(x)$,
which originally was defined only for $|x|=1$, to
a {\it piecewise analytic} function in the whole $x$ plane $\Psi(x)$,
whose dependence on $x$ is governed by \diffx. In fact the nice thing
about
\diffx\ is that the compatibility of this equation with \lindiff\
automatically implies that the solution to $tt^*$ are independent
of $\theta$ (are `self similar'), a condition which was previously
imposed by
hand.
So the compatibility of the above linear system of equations
completely
captures the $tt^*$ geometry.

The differential
equation \diffx\ has two {\it irregular}
singular points for $x=0$ and $\infty$. Then $\Psi(x)$ presents the
Stokes
phenomenon \Stokesph. This means that
 $\Psi(x)$ is well defined only in certain
angular sectors centered at the origin. In the present case we need
 (at least) two angular sectors. For convenience we choose
these two sectors to be two suitable angular
neighborhoods of the upper and lower half--plane, respectively.
This means that $\Psi(x)$ should be replaced
by a couple of $n\times n$ matrices
$(\Psi_+(x),\Psi_-(x))$ which are analytic
in the half--planes ${\rm Im}\,x>0$ and ${\rm Im}\,x<0$
respectively.
In the overlap between the two angular sectors,
$\Psi_+$ and $\Psi_-$, being both solutions to the
linear equation \diffx, should satisfy a
 relation $\Psi_-=\Psi_+M$ for some {\it constant} matrix $M$.
More precisely, along the real axis they satisfy the following
Riemann
boundary condition (here $y>0$)
\eqn\rieproblem{\eqalign{& \Psi_-(y)=\Psi_+(y)S\cr
&\Psi_-(-y)=\Psi_+(-y)S^t.\cr}  }
General Stokes theory gives constraints\foot{We choose the overall
phase of the $w_k$'s so that ${\rm Re}(w_i-w_j)\not=0$ for $i\not=j$.
Of course, this can be done only locally in coupling space. To get
the
global solution one has to glue all the local solutions so obtained.}
 on
the  matrix $S$
\eqn\Wstokes{\eqalign{& S_{ii}=1\cr
& S_{ij}=0\quad {\rm for}\ {\rm Re}\,(w_i-w_j)<0.\cr}  }
Moreover, PCT requires $S$ to be real.

The matrix $\Psi(x)$ satisfies the following boundary condition
\eqn\beainfty{\lim_{x\rightarrow\infty}
\Psi(x)\exp[\beta(x C+ x^{-1}C^\dagger)]=1.  }
Using this boundary condition and the well--known identity ($P$ means
principal part)
\eqn\disidentity{{1\over x-y\mp i\epsilon}
=P{1\over x-y}\pm i\pi \delta(x-y),}
one rewrites the above Riemann boundary problem as
 the integral equation
\eqn\dubrovin{\eqalign{\Phi(x)_{ij}=\delta_{ij}+
&\ipi \sum_k\int\limits_0^\infty {dy\over y-(x+i\epsilon)}
\Phi(y)_{ik}A_{kj}e^{-\beta(y
\Delta_{kj}+y^{-1}\bar\Delta_{kj})}\cr
& +\ipi  \sum_k\int\limits_{-\infty}^0 {dy\over y-(x+i\epsilon)}
\Phi(y)_{ik}A^t_{kj}e^{-\beta(y
\Delta_{kj}+y^{-1}\bar\Delta_{kj})}\cr} }
where
$$\Delta_{kj}=w_k-w_j\equiv \half m_{ij} e^{i\phi_{ij}},$$
and
$$\Psi_+(x)=\Phi(x)\exp[-\beta(x C+x^{-1} C^\dagger)],$$
In terms of $A$ the Stokes matrix reads
\eqn\Stokes{S=1-A,}
so, in particular, \Wstokes\ gives
$$A_{kj}\not=0\qquad {\rm only\ if\ }\ {\rm Re}\, \Delta_{kj}>0,$$
which is nothing else than the condition
needed in order to make sense out of
 the integrals in  \dubrovin.

The solution to \dubrovin\ is automatically a solution to all the
equations \diif. Indeed, the matrix $S$ encodes (with respect to the
chosen
angular sectors) the
monodromy properties of the linear differential
equations with rational coefficients \diffx.
In particular the monodromy around the singular point $x=0$ is given
by
$H=S(S^t)^{-1}$.
 {\it A priori}  the
monodromy data depend on the coefficients in eq.\diffx, i.e. on
$w_k$,
$Q$ and the ground--state metric $g$. However the equations \diif\
just represent the isomonodromic deformations of eq.\diffx, that is
they describe the
variations of the coefficients in \diffx\ which {\it
do not change} its monodromy data.
Said differently, the fact that $\Psi$ is a solution to
\covadev\ and \diffx\  means that the matrix $S$ is a constant
independent of both $x$ {\it and} $w_i$.
 In fact, from the
general theory of isomonodromic deformations \aits\ we know that
the condition for having isomonodromic deformations is just
the zero--curvature
condition above\foot{
Recall that the massive \ttstar\ equations are  those for
the Ising correlations. It is well known that these equations
describe
isomonodromic deformation.
In fact this is precisely what the Kyoto school mean when they
talk of {\it `holonomic field theory'} \japanese.}, \ie\ the \ttstar\
equations themselves.

Now, the solution to \dubrovin\ for a given (fixed) $S$
 is certainly a
family of isomonodromic solutions to \diffx\ parametrized by the
couplings
$w_k$. Indeed the monodromy data $S$ is
a constant by construction. Then it
must be also a solution to \diif. (Mathematically  oriented people
may find
complete proofs in \solttstar; for the special  $n=2$ case see also
\refs{\aits,\riemann}). This `monodromic' viewpoint
also explains how the Stokes parameters $A_{ij}$ encode the
boundary conditions needed to specify a solution of \diif.

For small temperatures, $\beta\rightarrow\infty$, the kernel in
\dubrovin\
is exponentially suppressed. Hence for small enough temperature we
have
a unique solution with given monodromy data $A_{ij}$. Whether this
can be extended to a regular solutions for all $\beta$'s depends on
the
particular $A_{ij}$. This should happen for the physical values of
the
Stokes parameters.
{}From the Riemann problem \rieproblem\ and the uniqueness of the
solution we
infer that the piecewise analytic function
$\Psi\equiv(\Psi_+,\Psi_-)$
satisfies \solttstar\
\eqn\propsi{\eqalign{& \Psi(x)\Psi^t(-x)=1\cr
& \overline{\Psi(1/\bar x)}=g^{-1}\Psi(x),\cr} }
where the second equation is nothing else than the statement that
complex
conjugation acts on the vacuum wave function\foot{Indeed for LG
models
$\Psi(x)$
is related by a linear integral transform to the usual SQM
wave function.} as the ground
state metric $g$ \topatop.

{}From \beainfty\ and \propsi\ we get
\eqn\gmetric{g_{i\bar j}\equiv \lim_{x\rightarrow 0}\Phi(x)_{ij}.}

\subsec{The Ultra--Violet Limit: The $Q$--matrix}

Now we study the large temperature asymptotics \smallbeta\
of the solutions to
\ttstar.
This would give us the conformal dimensions of the chiral primary
operators at the UV fixed point as a function of the Stokes
parameters
$A_{ij}$.

To get
the eigenvalues $q_i(A)$
of the matrix $q_{ij}(A)$ we exploit its  physical meaning.
 As $\beta\rightarrow 0$ the
$U(1)$ invariance is restored and  $q_i(A)$ are just
the vacuum values of the corresponding conserved charge. Therefore
when  we increase $\theta$ by  $2\pi$  in eq.\overall\ the
wave functions $\Psi$ pick up phases $\exp[2\pi i
q_j(A)]$.
This can also be seen from the differential equation \diffx\
satisfied
by $\Psi(x)$.
As $\beta\rightarrow 0$, and as long as we restrict ourselves to
the region
\eqn\region{\beta \ll |x| \ll \beta^{-1},}
we can approximate  eq.\diffx\
 by one with constant coefficients, namely\foot{
Recall that $q_{ij}=\lim_{\beta\rightarrow 0} Q_{ij}$.}
$${d\phantom{\theta}\over d\theta}\Psi_i\approx q_{ij}\Psi_j.$$
Hence in the region \region\ we have
\eqn\falsesol{\Psi(\theta+2\pi i)_i\approx
\big( e^{2\pi i q}\big)_{ij}\Psi(\theta)_j.}
On the other hand from \rieproblem\ we see that
$$\Psi(\theta+2\pi i)= \Psi(\theta) S (S^{-1})^t.$$
Comparing the last two equations we get
\eqn\qiswhat{\exp[2\pi i q_j]={\rm Eigenvalues}[S(S^{-1})^t].}
This is the equation expressing the UV charges $q_j$ in terms
of the Stokes parameters we look for, modulo showing the relation
between $A$ defined here and the soliton numbers $\mu_{ij}$ which
needs
a detailed analysis which we postpone to the next two subsection.

We will now see that we can use $tt^*$ to also fix the integral
part of $q_i$.  To do this note that even though the physical
values for the matrix $A$ are integer, as we will relate
it to soliton numbers, as far as the $tt^*$ equations are concerned
we can take them to be arbitrary.  Consider $A\rightarrow A(t)$
with $A(0)=0$ and $A(1)=A$.  Then
at $t=0$ we get the trivial theory with the charges equal to zero.
As we vary $t$ from $0$ to $1$, we can trace the eigenvalues
of $H(t)=S(t)(S(t))^{-t}$ on the complex plane.  Since these
eigenvalues do
correspond to $exp(2\pi i q)$ where $q$ is the solution of $tt^*$ at
the UV point\foot{See the discussion below on the requirement of
the existence of solution to $tt^*$.}(unphysical as they may be), by
continuing
the eigenvalues until we get to $t=1$ we can deduce
the integral part of the charges by the number of times they
have wrapped around the origin in the complex plane.
This clearly shows the power of $tt^*$ equations as they
can be used even in the unphysical regime
(non--integral soliton numbers) to give some
physical results (with integral soliton numbers).  This result
applies
in particular to the LG case, and as far as we know it was
not known to the mathematicians
how to fix the integral part of charges purely from the $S$
matrix.   In the singularity
language the trick we are using is like taking a `continuous real
intersection number' which is not easy to see how would one
interpret.

Using the idea of `building up the charge'
we can also learn something about the signature of the
matrix $B=S+S^t$.  Note that  this matrix is a symmetric integral
matrix with $2$'s on the diagonal.  It can be interpreted as
the bilinear form for an integral lattice.  It is useful to discuss
the signature of this form when we begin to classify $N=2$ quantum
field theories.  We know that at $t=0$ and $t=1$ the eigenvalues
of $H(t)=SS^{-t}$ are pure phases
(i.e. have norm 1).  Let us assume that by a proper
choice of $t$-dependence of $S$ which connects these points we go
{\it only} through phases.  Let us consider the signature of $B(t)$.
Clearly $B(0)$ is positive definite.  For its signature to change
we should come across a zero eigenvector of $B$, which means we must
have a vector $v$ with
$$Sv=-S^tv$$
which implies
$$H^tv=-v$$
In other words the signature changes precisely when one of the
eigenvalues
crosses the negative real axis.  Of course if that eigenvalue
crosses the negative real axis another time, it will change back
the signature.  Now noting the connection between the integral part
of
charges and the number of times an eigenvalue wraps around the origin
we see that the number of positive directions $r$
and negative directions $s$ of $B$ are given by
\eqn\sigb{r=\#(2n-{1\over 2}<q<2n+{1\over 2})\qquad s=\#(2n+{1\over
2}<q<2n+{3\over 2})}
When there are some charges equal to $1/2$ mod $1$ we also get some
null directions.  This result for the signature of $B$ agrees
with what is known to mathematicians in the context of singularity
theory.

We made the assumption that by continuously changing the parameters
of $S$ we can vary the eigenvalues of $H$ maintaining the condition
that they remain roots of unity.
Indeed if the eigenvalues
of $H$ end up having norm other than one, then $q$ becomes complex
and this implies that for the solution of $tt^*$ there is some
singularity because otherwise $q$ is given by the
eigenvalues of $-g\beta \partial_\beta g^{-1}/2$ and that is real for
a
solution of $tt^*$.
So as long as the regular solution space of $tt^*$ is connected, we
should
be able to go through phases only.
 In general it is easy to see
(as we will argue later) that for small $t$ this is the generic
case.  Indeed the eigenvalues of $H$ come in groups of four
$\lambda , \lambda ^*, \lambda^{-1} ,\lambda ^{*-1}$ and for
$t$ near zero ($S$ near one) it is easy to see that they
come in pairs because $\lambda$ is a root of unity.  In general
if we just take an arbitrary deformation of $S$ like letting
$A\rightarrow tA$ this condition will not be maintained for
larger values of $t$.  However, it is natural to expect
that by proper tuning of the coefficients of A
(with arbitrary real functions of $t$), we should be able
to maintain the condition that eigenvalues of $H$ be pure phases.  It
would be
nice to prove this highly plausible statement.  The fact
that in the LG case the result obtained in this way
agrees with what mathematicians had obtained lends further support
to this statement.

\vglue 10pt
\noindent\underbar{\raise 3pt\hbox{\it Subtleties with Asymptotic
Freedom}}
\vglue 2pt

At first sight one would also expect that the two matrices
$\exp[2\pi i q]$ and $H=S(S^t)^{-1}$ are similar.
However it is not so:
$H$ may have
non--trivial Jordan blocks. This possibility
arises because of UV sub--leading terms that we have
neglected in the above analysis.
Instead of discussing the (well--known) mathematics
of this phenomenon, let us explain
its deep physical meaning.
To make things as simple as possible, we consider
a specific model, namely
the supersymmetric
${\bf C}P^1$ $\sigma  $--model \sigmamodels.
This model has a mass--gap \massGap. Since it is asymptotically free,
its UV fixed point is just free field theory. At this UV fixed point
the (unique) non--trivial chiral primary field has
dimension $({1\over 2},{1\over 2})$. However it is not really
a marginal operator, otherwise the $\sigma  $--model would be
conformal for all $\beta$'s.
As it is well--known, this  state of affairs leads
to {\it logarithmic violations of scaling}.
The non--trivial Jordan blocks
are related to these violations.
For instance, for ${\bf C}P^1$
\eqn\cpmono{H=\left(\matrix{1& -2\cr 2 & -3\cr}\right)\buildrel
\scriptscriptstyle \rm similarity \over \longmapsto
\left(\matrix{-1& 1\cr 0 & -1\cr}\right),}
and so we expect logarithmic corrections to scaling\foot{These
corrections can be found by an exact computation, see
 ref.\sigmamodels.}.
The Jordan structure of \cpmono\ can be extracted directly from the
basic equations \diffx\diffbeta.
It is natural to look for a solution of the form
$$\Psi(x,\beta)=\exp\big[q\,  (\log x+\log\beta)\big]
\Phi(x,\beta).$$
In the limit $\beta\rightarrow 0$
the differential equation for $\Phi(x,\beta)$
reduces to
\eqn\zeze{x{d\phantom{x}\over dx}\Phi=[B-x^{-2}\bar B]\Phi+
O\left({1\over
\log\beta}\right),}
where
\eqn\BB{\eqalign{& B=\lim_{\beta\rightarrow 0} \beta
x\big[(x\beta)^{-q}
C(x\beta)^{q}\big]\cr
& \bar B=\lim_{\beta\rightarrow 0} \beta x \big[(x\beta)^{-q}\bar C
(x\beta)^q\big].\cr}  }
The matrix
$C$ represents in $\CR$ some chiral operator $\hat\phi\equiv
\sum_k w_k\phi_k$.
Let us decompose $\phi$
into a sum $\sum_{i\in I}\tilde\phi_i$
of operators having definite $U(1)$ charge
$q_i$ at the UV fixed point. Let $\bar q=\max_{i\in I}\{q_i\}$.
Then, for small $\beta$,
$(x\beta)^{-q}C(x\beta)^q$ is of order $\beta^{-\bar q}$.
Thus, if our perturbation $\phi$
has an UV dimension less than 1 (\ie\ it is
`super--renormalizable') $B=\bar B=0$ and there is no new subtlety.
Instead for
an `asymptotically free' (AF)
theory $\bar q$=1 and $B$ is finite\foot{If $\bar
q>1$ it is not clear how to make sense of the corresponding
perturbation.
Below we shall see that (typically) the non--renormalizable
interactions lead to singular solutions to \ttstar\ and so they are
`pathological'.}. For
instance, in the ${\bf C}P^{n-1}$ case we have (up to similarity)
\eqn\excpn{B=\left(\matrix{0 &
1 & 0 & 0 & \dots & 0\cr
                  0 & 0 & 1 & 0 & \dots & 0\cr
                  \dots & & \dots & & \dots & \cr
 0 & 0 & 0 & \dots & \dots & 1\cr}\right), }
and $B^n=0$. From \zeze\ we see that for $x$ large (but still $x\ll
\beta^{-1}$)
$$\Psi(x,\beta)\sim \exp[q\, (\log x+
\log \beta)]\exp[B (\log x+\log\beta)]\Phi_0,$$
from which it is manifest that the Jordan structure of $H$ is
that of
\eqn\Jstruct{\exp[2\pi i q]\exp[2\pi i B]. }
In particular, for the
${\bf C}P^{n-1}$ models $H$ should consist of a single block of
length $n$.

\subsec{Infra--Red Asymptotics}

To complete our proof of the formula relating $q_j$ to the soliton
matrix
$\mu_{ij}$ we have still to find the relation between the Stokes
parameters
$A_{ij}$ and the soliton numbers $\mu_{ij}$. In order to do this, we
have to find the asymptotic behaviour as $\beta\rightarrow\infty$
of the
\ttstar\ solutions. Here the integral formulation of \S 4.2 becomes
crucial.

We write symbolically eq.\dubrovin\ as
$$\Phi=1+\Phi\CK.$$
For $\beta$ large enough we can solve this equation by the method
of successive approximations.
In this way we get a {\it convergent} (for $\beta$ large enough)
series for the ground--state metric
\eqn\gensolu{g=1+\left. 1\cdot \sum_{m=1}^\infty
\CK^m\right|_{x=0}.}
The term $1\cdot \CK^m$ is of order $O(A^m)$. To begin with,
let us consider the first order contribution.
Using the formula
(valid for ${\rm Re}\, a>0$ and  ${\rm Re}\, b<0$)
\eqn\Bessel{\int_0^\infty x^{\nu-1} \exp[-a x- b x^{-1}]=
2\left({b\over
a}\right)^{\nu/2}  K_\nu(2\sqrt{ab}),  }
one gets
\eqn\leading{g_{i\bar j}=\delta_{ij}-i(A_{ij}-A_{ji}){1\over \pi}
K_0(2|w_i-w_j|\beta )+O(A^2).}
The first order contribution has precisely the form predicted
by the large $\beta$ asymptotics \gAsymp.
This  may suggest
that the first order saturates the one--soliton contribution and,
more generally,
that the $m$--th order term $1\cdot \CK^m$ corresponds to
 $m$ soliton processes.
This is almost {\it but not quite} true.
Explicitly $[1\cdot\CK^m]_{i\bar j}$ can be written
as a sum of terms, one for each sequence $\alpha(k)$ ($k=1,\dots,m$)
in $\{1,2,\dots, n\}$ with $\alpha(1)=i$, $\alpha(m)=j$. The sequence
$\alpha(k)$ specifies a particular chain of $m$ would--be
`solitons' connecting
the $i$--th vacuum to the $j$--th one. Then
\eqn\mthterm{1\cdot\CK^m= \sum_{ m-{\rm chains}}
G_\alpha(\beta, A, w_k),}
where $G_\alpha(\beta, A, w_k)$ has the general
form (here $\tilde A=A-A^t$)
$$G_\alpha(\beta,A,w_k)=\left(\prod_{k=1}^m
\tilde A_{\alpha(k)\, \alpha(k+1)}\right)
\int_0^\infty
\prod_k dx_k F_\alpha(x)\times$$
\eqn\ofform{\times \exp\bigg[-\beta \sum_{k=1}^m\sigma  _k\Big(
x_k (w_{\alpha(k)}-w_{\alpha(k+1)})+
x_k^{-1}(\bar w_{\alpha(k)}-\bar w_{\alpha(k+1)})\Big)\bigg],}
where $F_\alpha(x)$ is an
universal function independent of the parameters and
$$\sigma  _k={\rm sign}\big[{\rm
Re}(w_{\alpha(k)}-w_{\alpha(k+1)})\big].$$
Now, {\it were the kernel $\CK$ non--singular}, we could evaluate the
large $\beta$ asymptotics of \ofform\ by the usual
saddle--point method. The relevant
saddle point is at
$$x_k= \sigma  _k \sqrt{\bar w_{\alpha(k)}-\bar w_{\alpha(k+1)}\over
w_{\alpha(k)}-w_{\alpha(k+1)}}$$
and then we would have
\eqn\betalarge{G_\alpha(\beta, A,w_k)\approx \exp\bigg[-2\beta
\sum_{k=1}^m|w_{\alpha(k)}-w_{\alpha(k+1)}|\bigg], }
which is the expected result for a chain of $m$ solitons
having  masses \ $2|w_{\alpha(k)}-w_{\alpha(k+1)}|$.

However, since $\CK$ is singular, \betalarge\ is not necessarily
correct.
Indeed in order to use the saddle point technique
\saddlep\  we have to deform the
integration contour to pass through the saddle point. In this
process we may cross poles (resp. cuts) of the integrand and hence we
pick
up residue (resp. discontinuity) contributions to \ofform. From
\disidentity\ it is clear that these contributions have also the
general
structure \ofform\ but with a {\it smaller} $m$. Moreover the
presence of
these additional terms depends in a crucial way on the angles in
$W$--plane since the number and type of singularities encountered
while
deforming the path depends on the vacuum geometry in $W$--space.

Because of this mechanism, the $k$--soliton processes may get
contributions
from all terms in  \gensolu\ with $m\geq k$. This, in
particular, holds for the one soliton term which defines the soliton
matrix $\mu_{ij}$. So\foot{Note that our definition of $A$ in this
section defers from the one used in section 2.  There it was
defined in terms of soliton numbers.  Here it is defined by $S=1-A$.
In the `standard' configuration the two definitions are the same.},
\eqn\muA{\mu_{ij}=A_{ij}-A_{ji}+O(A^2).}
Under our genericity assumption, the rhs of \muA\ is a {\it finite
polynomial}. Indeed, without deforming the integration contour, we
get the weaker bound (for $\beta$ large)
$$G_\alpha(\beta,A,w_k)\leq C\exp\bigg[-2\beta\sum_{k=1}^m|{\rm Re}\,
(w_{\alpha(k)}-w_{\alpha(k+1)})|\bigg],$$
and thus only sequences satisfying
\eqn\wbound{\sum_{k=1}^m|{\rm Re}\,
(w_{\alpha(k)}-w_{\alpha(k+1)})|\leq |w_i-w_j|}
may contribute to $\mu_{ij}$. Clearly, there are only finitely many
such
sequences. Then to compute $\mu_{ij}$ we can truncate the
expansion after a {\it finite} number of terms. However, this method
is
rather impractical since the number of terms needed varies very much
from
model to model. For this
reason, we shall adopt a different strategy based on the known
properties
of the solution \propsi\ rather than on the integral equation itself.
In order to do this, we need more details on the analytic
properties of the functions which appear in the expansion \gensolu.
We pause a while to digress on this more technical material.
The reader may wish to jump directly to \S  4.5.

\vglue 10pt
\noindent\underbar{\raise 3pt\hbox{\it Some Useful Functions}}
\vglue 2pt

The purpose of this digression is to describe the functions one gets
when the integrals \ofform\ are computed along a contour for which
the
saddle--point analysis is correct. As discussed above, the functions
appearing in the expansion \gensolu\ can be expressed in terms of
these
ones, the precise relation being determined by
 their analytic properties as
well as the vacuum geometry.

We introduce a function $\CF[z,\zeta]$, where $z$ is a
real positive
variable and $\zeta$ is a variable taking value in the complex
plane cut along the positive real axis, by
\eqn\effe{\CF[z,\zeta]={1\over 2\pi}\int\limits_0^\infty {ds \over
s-\zeta}
e^{-z(s+s^{-1})}  }
For $\zeta$ real
positive,  $\CF[z,\zeta]$ is defined to be
$\CF[z,\zeta+i\epsilon]$. The discontinuity at the cut along the
positive real
axis is given by
\eqn\discontinuity{\CF[z,x+i\epsilon]-\CF[z,x-i\epsilon]
=i e^{-z(x+x^{-1})}. }
As $z\rightarrow\infty$ one has the asymptotic expansion
(for $\zeta$ {\it not real positive})
\eqn\asympex{\CF[z,\zeta]\approx {1\over 2\pi}
 \sqrt{\pi\over z} e^{-2z}\sum_{k=0}{g_k(\zeta)\over
(2z)^k}, }
where
\eqn\defig{g_k(\zeta)=
{ \zeta^{k+{1\over 2}}\over k!}\left({d\phantom{z}\over d
\zeta}\right)^{2k}\left[{\zeta^{(k-{1\over 2})}
\over 1-\zeta}\right].}
These formulae are a consequence of
$$\CF[z,\zeta]= {1\over 2\pi}\sum_{k=0}^\infty
\zeta^k\int\limits_0^\infty {ds\over s^{k+1}} e^{-z(s+s^{-1})},$$
together with \Bessel.
Instead for $\zeta$  real positive
\eqn\realcase{\CF[z,x]={1\over 2\pi}P\int\limits_0^\infty
{ds\over s-x} e^{-z(s+s^{-1})}+{i\over 2}e^{-z(x+x^{-1})}   }
where $P\int$ means principal part and we used \disidentity.
As $z\rightarrow \infty$, the integral in \realcase\ has
the same asymptotic expansion as above, except for $x=1$ when it
vanishes more rapidly. In this case the leading term is the
second one.
Hence for $\zeta=1$ the asymptotics is
$$\CF[z,1]\sim {i\over 2} e^{-2z}.$$
Then, for $\zeta=1$ the large $z$ behaviour differs\foot{This
discontinuity in the low temperature behaviour can be understood
physically
as due to contact terms.} for a factor
$i\sqrt{z \pi}$ with respect that for $\zeta\not=1$.

Next we
define a function $\CF^{(2)}[z_1,z_2,\zeta,\phi]$, (where $z_i$ are
real
positive, $\zeta$ is as above, and $\phi$ is an angle
with $\phi\not=0$ mod. $2\pi$)
$$\CF^{(2)}[z_1,z_2,\zeta,\phi]={1\over 2\pi}\int\limits_0^\infty
{ds \over s-\zeta} e^{-z_1(s+s^{-1})}\CF[z_2, e^{i\phi}s].$$
If $\phi=0$ we define this function as the limit as $\phi\downarrow
0$.
Here again we have a discontinuity for $\zeta$ real as well as for
$\phi=0$. In particular,
\eqn\gdisco{\CF^{(2)}[z_1,z_2,\zeta,\epsilon]
-\CF^{(2)}[z_1,z_2,\zeta,-\epsilon]=
i\CF[z_1+z_2,\zeta]}
\eqn\ddisco{\CF^{(2)}[z_1,z_2,x+i\epsilon,\phi]
-\CF^{(2)}[z_1,z_2,x-i\epsilon,\phi]=i
e^{-z_1(x+x^{-1})}\CF[z_2,e^{i\phi}x].  }
So one has\foot{Taken at the face value, this  says that two
soliton represented by two exactly
aligned segments give a contribution which looks like a `half'
soliton of mass
$m_1+m_2$. This strange effect is effectively seen in explicitly
computable
models (see appendix B for an example).
It can also be understood from the $S$--matrix viewpoint.}
\eqn\halfsoli{\eqalign{&\CF^{(2)}[z_1,z_2,\zeta,0]=\cr
&={1\over (2\pi)^2}\int\limits_0^\infty {ds\over
s-\zeta} e^{-z_1(s+s^{-1})}
P\int\limits_0^\infty {dt\over t-s} e^{-z_2(t+t^{-1})}+
{i\over 2}\CF[z_1+z_2,\zeta].\cr} }
If $e^{i\phi}\not=1$ and $\zeta\not=1$, for $z_i\rightarrow \infty$
we have
\eqn\twoasy{\CF^{(2)}[z_1,z_2,\zeta,\phi]\sim {e^{-2(z_1+z_2)}\over
4\pi
\sqrt{z_1 z_2}} {1\over (1-\zeta)(1-e^{i\phi})}. }
If $e^{i\phi}=1$ one has instead
\eqn\twoasyspe{\sim {i\over 4\pi}\sqrt{\pi\over z_1}e^{-2(z_1+z_2)},}
so we have again the discontinuity in the large $z_i$ behaviour.
We have a similar result when $\zeta=1$ and when both variables are
equal
to 1. In this last case both factors $z_i^{-1/2}$ cancel.

Clearly, the above analysis may be generalized.
Let us define recursively the functions
$$\eqalign{\CF^{(k)}&[z_1,\dots, z_k,\zeta,\phi_1,\dots,\phi_{k-1}]
=\cr
&={1\over 2\pi}\int\limits_0^\infty {ds\over s-\zeta}
e^{-z_k(s+s^{-1})}
\CF^{(k-1)}[z_1,\dots,z_{k-1}, e^{i\phi_{k-1}}s,
\phi_1,\dots,\phi_{k-2}]\cr}$$
Here $z_i$ are real positive variables, $\zeta$ is a complex
parameter taking value in the plane cut along the real semi--axis
(for $\zeta$ real, by convention, we define the function
as its limit by above), and $\phi_i$ are angular variables in the
range
$0<\phi_i<2\pi$,
and for $\phi_i=0$ we take as definition the limit by above.
As $z_i\rightarrow\infty$ one has
$$\CF^{(k)}[z_1,\dots,z_k, \zeta,\phi_1,\dots,\phi_k]\sim e^{-2\sum_i
z_i},$$
up to a power of the $z_i$ which depends on $\zeta$ and the
$\phi_i$'s.

{}From their recursive definition, it is clear that the discontinuity
of
$\CF^{k}$ for $\zeta$ real positive (resp. for $\phi_i=0$) can be
expressed in terms of $\CF^{(h)}$ with $h<k$, possibly multiplied by
factors $\exp[-z_i(x+x^{-1})]$.

\vglue 10pt
\noindent\underbar{\raise 3pt\hbox{\it Sample Integrals}}
\vglue 2pt

As a preparation to
\S 4.5, and illustration of the above mechanism, we  compute some
sample
 integrals one gets  in \gensolu.
At the first order the typical integral is
\eqn\firstord{
\ipi\int\limits_0^\infty {dy\over y-(x+i\epsilon)}
A_{ij}e^{-\beta(y \Delta_{ij}+y^{-1}\bar\Delta_{ij})},}
where ${\rm Re}\,\Delta_{ij}>0$ and
$$\Delta_{ij}=\half m_{ij}\,  e^{i\phi_{ij}},\qquad {\rm with}\
-\half\pi < \phi < \half\pi.$$
Assume $0\leq \phi_{ij}<\half\pi$,
and let $C_R$ be the segment $y=te^{-i\phi_{ij}}$, $0\leq t \leq
R$,
and $\gamma_R$ the arc $y= R e^{i\theta}$, $-\phi_{ij}\leq \theta
\leq
0$. Denote by $F(y)$ the integrand in eq.\firstord.
$F(y)$ is holomorphic in the lower half--plane. Then we have
$$-\int\limits_0^R F(y) dy+ \int_{C_R}
F(y)dy+\int_{\gamma_R}F(y)dy=0,$$
As $R\rightarrow\infty$, the last integral vanishes exponentially.
Hence \firstord\ reduces to
\eqn\caseI{\int_{C_\infty} F(y) dy =
-iA_{ij}\CF\big[\half m_{ij}\beta, e^{i\phi_{ij}} x\big].}
Consider now the case $-\half\pi<\phi_{ij}<0$.
This time $C_R$ is in the upper half--plane. Since $F(y)$ has a pole
for
$y=x+i\epsilon$, we have
$$\int\limits_0^\infty F(y)dy +\int_{\gamma_R}F(y)dy-
\int_{C_R} F(y) dy =\vartheta(x)  {\rm Res}_{x+i\epsilon} F(y).$$
Taking $R\rightarrow\infty$ we get
$$\int\limits_0^\infty F(y)dy= \int_{C_\infty} F(y) dy+\vartheta(x)
A_{ij}
\exp\big\{-\half m_{ij}\beta[ e^{i\phi_{ij}}x+e^{-i\phi_{ij}}x^{-1}]
\big\}.$$
The integral in the rhs is given by \caseI; but in
 this second case there is also a contribution  from the residue.
Notice that this term is present only if the angle
$\phi_{ij}$ belongs to the IV quadrant.

A typical integral appearing in the next order is
\eqn\firstint{\ipi A_{ik}A_{kj}\int\limits_0^\infty {dy\over
y-(x+i\epsilon)}
\CF\big[\half m_{ik}\beta, e^{i\phi_{ik}}y\big]
\exp\big[-\half m_{kj}\beta(y e^{i\phi_{kj}}+ y^{-1}
e^{-i\phi_{kj}})\big],  }
where $-\half \pi <\phi_{ik},\phi_{kj}< \half\pi$.
Again, the idea is to deform the integration contour to the ray
$y=e^{-i\phi_{kj}}t$.
 When deforming the contour we can cross two kind
of singularities, \ie\ the pole at $y=x+i\epsilon$ and the cut of the
function $\CF[z,\zeta]$ for $\zeta= x+i\epsilon$,
$x$ real positive (\ie\ on the ray $y= e^{-i\phi_{ik}}t$).
There are
four distinct cases
\eqn\fourcases{\eqalign{&{\rm case\ 1}\qquad 0\leq\phi_{kj} <\half
\pi
\ {\rm and}\  \phi_{kj}<\phi_{ik}<2\pi\cr
&{\rm case\ 2}\qquad 0\leq\phi_{kj} <\half \pi
\ {\rm and}\  0\leq\phi_{ik}\leq\phi_{kj}\cr
&{\rm case\ 3}\qquad -\half\pi <\phi_{kj} < 0
\ {\rm and}\  0<\phi_{ik}<\phi_{kj}\cr
&{\rm case\ 4}\qquad -\half\pi <\phi_{kj} < 0
\ {\rm and}\  \phi_{kj}\leq\phi_{ik}\leq 0\cr}}
In case 1 we encounter no singularity
when deforming the contour from the real positive semi--axis to
the ray  $y=s e^{-i\phi_{kj}}$ $0\leq s\leq \infty$.
Instead in case 2 deforming the contour to the ray
$y=e^{-i\phi_{kj}}s$
we encounter a cut along the ray $y=e^{-i\phi_{ik}}t$.
Using the discontinuity \discontinuity, we find
$$\eqalign{&\ipi\int\limits_0^\infty {dy\over y-(x+i\epsilon)}
\CF\big[\half m_{ik}\beta, e^{i\phi_{ik}}y\big]
 \exp\big[-\half m_{kj}\beta(y e^{i\phi_{kj}}+ y^{-1}
e^{-i\phi_{kj}})\big]=\cr
&=-i\CF^{(2)}\big[\half m_{kj}\beta, \half m_{ik}\beta,
e^{i\phi_{kj}}x, \phi_{ik}-\phi_{kj}\big]
+\cr
&\int\limits_0^\infty {d t\over 2\pi (t- e^{i\phi_{ik}}x})
e^{-m_{ik}\beta(t+t^{-1})/2}
\exp\big[-\half m_{kj}\beta(e^{i(\phi_{kj}-\phi_{ik})}t
+ e^{i(\phi_{ik}-\phi_{kj})}
t^{-1})\big].\cr}$$
Consider the integral
\eqn\borint{{1\over 2\pi}\int\limits_0^\infty {d t\over t-
e^{i\phi_{ik}}x}
e^{-m_{ik}\beta(t+t^{-1})/2}
\exp\big[-\half m_{kj}\beta(e^{i(\phi_{kj}-\phi_{ik})}t
+ e^{i(\phi_{ik}-\phi_{kj})}
t^{-1})\big]. }
Using  the identity
\eqn\useide{\eqalign{&m_{ik} e^{i\phi_{ik}}+m_{kj} e^{i\phi_{kj}}=
m_{ij}
e^{i\phi_{ij}}\cr
& 0<\phi_{ik}<\phi_{ij}<\phi_{kj}<\half\pi\qquad \rm (case\ 2),\cr}
}
\borint\ becomes
$${1\over 2\pi}\int\limits_0^\infty {dt\over t- e^{i\phi_{ik}}x}
\exp\big\{-\half
m_{ij}\beta[e^{i(\phi_{ij}-\phi_{ik})}t+e^{i(\phi_{ik}
-\phi_{ij})}t^{-1}]\big\},$$
\ie\ the typical first order.
Again we deform the integration contour to the ray
$t=e^{-i(\phi_{ij}-
\phi_{ik})}s$. From \useide\ we see that
we encounter no singularity in this process. Hence
\borint\ is equal to
$$\CF\big[\half m_{ij}\beta, e^{i\phi_{ij}}x\big],$$
\ie\ \caseI, but for the third side of
 the triangle $(w_i,w_k,w_j)$.
Then, in case 2, \firstint\ is
\eqn\twocom{-iA_{ik}A_{kj}
\CF^{(2)}\big[\half m_{kj}\beta,\half m_{ik}\beta,
e^{i\phi_{kj}}x, \phi_{ik}-\phi_{kj}\big]
+A_{ik}A_{kj}\CF\big[\half m_{ij}\beta, e^{i\phi_{ij}}x\big], }
As $\beta\rightarrow \infty$, the first term in the rhs is of order
\twoasy, whereas the second one is of order
$$\exp[-2 m_{ij}\beta]\gg \exp[-2(m_{ik}+m_{kj})\beta],$$
{\it unless} the three points $w_i$, $w_j$ and $w_k$ are aligned,
in which case the two sides are roughly of the same order.
If no three points are aligned in $W$--space, the one--soliton
contributions are unambiguously determined to be the coefficient of
$\CF[m_{ij}\beta/2]$; the second term in \twocom\ is an explicit
example of an $O(A^2)$ contribution to $\mu_{ij}$.
In case 3. (resp. 4) we get the same result as in case 1. (resp. 2)
except that now when deforming the contour we pick up also a
contribution from the residue at $y=x+i\epsilon$.

\vglue 10pt
\noindent\underbar{\raise 3pt\hbox{\it Large $\beta$ Asymptotics of
$\Phi$}}
\vglue 2pt

We use the following short--hand
$$\eqalign{&\CF_{ij}(x)=\CF\big[\half m_{ij}\beta,
e^{i\phi_{ij}}x\big]\cr
&\CE_{ij}(x)=\exp\big[-\half m_{ij}
\beta(e^{i\phi_{ij}}x+e^{-i\phi_{ij}}x^{-1})\big].\cr}$$
Notice the identities
\eqn\otherid{\eqalign{&\CE_{ik}(x)\CE_{kj}(x)=\CE_{ij}(x)\qquad
{\rm not\ summed\ over}\ k
\cr
&\CF_{ij}(-x)=\CF_{ji}(x),\qquad \CE_{ij}(-x)=\CE_{ji}(x).\cr}.}
Moreover, ($t$ real positive)
\eqn\discont{\CF_{ij}(e^{-i\phi_{ij}+i\epsilon}t)-
\CF_{ij}(e^{-i\phi_{ij}-i\epsilon}t)=i\CE_{ij}(e^{-i\phi_{ij}}t)  }

The previous
discussion shows
that just above the real positive (resp. negative)
axis $\Phi(x)$ has the following
IR expansion
\eqn\IRexp{\eqalign{\Phi_{i\bar j}(x)& =
\delta_{ij}-i\mu_{ij}\CF_{ij}(x)+B_{ij}\CE_{ij}(x)-\cr
&\quad -i\sum_k D^k_{ij}\CF_{ik}(x)\CE_{kj}(x)+{\rm terms\
containing\ higher}\ \CF'{\rm s}\qquad (x>0)\cr
& =
\delta_{ij}-i\mu_{ij}\CF_{ij}(x)+\tilde B_{ij}\CE_{ij}(x)-\cr
&\quad -i\sum_k \tilde D^k_{ij}\CF_{ik}(x)\CE_{kj}(x)+{\rm terms\
containing\ higher}\ \CF'{\rm s}\qquad (x<0).\cr}  }
The various coefficients in this expansions are polynomials in the
Stokes
parameters $A_{ij}$.
As long as no three points $w_j$ are aligned, the omitted terms are
subleading in the IR limit. Notice that $B_{ij}$ (resp. $\tilde
B_{ij}$)
can be non--vanishing only if $\cos(\phi_{ij})>0$ (resp. $<0$). A
similar
condition holds for $D_{ij}^k$, $\tilde D_{ij}^k$. Then the third
and fourth terms in the rhs of
\IRexp\ vanish exponentially as $x\rightarrow 0$, and hence by
\gmetric\
do not contribute to $g$. The coefficient of $\CF_{ij}$ is fixed
by the asymptotics \gAsymp,\asympex\ to be $-i\mu_{ij}$.

Eq.\propsi\ gives strong restrictions on
the various coefficients in \IRexp. Indeed, (for $x$ real positive)
one has
$$\eqalign{
\delta_{ij}&=\Big(\Phi(x+i\epsilon)\Phi^t(-x-i\epsilon)\Big)_{ij}=
\cr
&=\sum_{k,l}\Phi_{ik}(x+i\epsilon)\big(\delta_{kl}-A_{kl}\CE_{kl}(x)\big)
\Phi_{jl}(-x+i\epsilon),\cr}$$
which, in view of \otherid, gives
\eqn\obvious{\mu^t+\mu=0,}
\eqn\cru{D_{ij}^k=\mu_{ik}B_{kj},
\qquad \tilde D_{ij}^k=\mu_{ik}\tilde B_{kj}, }
\eqn\Stokes{(1+\tilde B^t)(1+B)=(1-A)^{-1}\equiv S^{-1}.}
This last relation allows us to read the Stokes parameters
directly from the IR expansion of $\Phi$ near the real axis.

For simple situations, the relation between $\mu_{ij}$ and
$A_{ij}$ can be obtained by inserting this truncated expansion for
$\Phi$ in the integral equation \dubrovin.
However this does not work in general, since --- because of the
singularity of $\CK$ --- the integrals of
 terms ignored in \IRexp\ may
contribute to the coefficients $\mu$ and $B$.
A better approach is presented below.

\subsec{Multi--Sector Formulation}

The rays $t e^{-i\phi_{ij}}$ ($t$
real positive) divide the plane into $n(n-1)$ sectors\foot{
To simplify the discussion we assume that the angles $\phi_{ij}$ are
all distinct. Notice
that the relevant rays are not the soliton lines but their mirror
images with respect the real axis. Indeed $\Psi(x)$ is the
 momentum space wave function and
the above condition corresponds to alignment in position space.}.
We number
these sectors according to the anti--clockwise order starting from
the
one containing the real positive axis
 which is called sector $1$. The
ray separating the $\alpha$--th sector from the $(\alpha+1)$--th one
is called the $\alpha$--th ray. To each $\alpha$ there is associated
an angle $-\phi_{ij}$. The corresponding indices will be denoted by
$i(\alpha),j(\alpha)$, respectively.
The sector containing the negative real axis
is the $(m+1)$--th one, where
$m=\half n(n-1)$.
If $M=(M_{kl})$
is a $n\times n$ matrix, we denote by $M^{[\alpha]}$ the matrix
$$(M^{[\alpha]})_{kl}= \delta_{k\, i(\alpha)}\, \delta_{l\,
j(\alpha)}
M_{i(\alpha)\, j(\alpha)}.$$

The analysis of \S  4.4 with the Stokes axis rotated by suitable
angles
in the $x$--plane shows that
in (some angular neighborhood of) the $\alpha$--th sector the
function $\Phi(x)_{i\bar j}$ has an IR expansion of the form
\eqn\IRE{\eqalign{\Phi_{i\bar j}^{(\alpha)}(x)& =
\delta_{ij}-i\mu_{ij}\CF_{ij}(x)+B^{(\alpha)}_{ij}\CE_{ij}(x)+\cr
&-i\sum_k D^{(\alpha),k}_{ij}\CF_{ik}(x)\CE_{kj}(x)+{\rm
higher\ }\CF'{\rm s}.\cr}  }
Comparing with \IRexp, we have
\eqn\realaxis{\eqalign{& B^{(1)}_{ij}=B_{ij}\cr
& B^{(m+1)}_{ij}=\tilde B_{ij}.\cr}  }
As before,
$B^{(\alpha)}_{ij}$ may be non--vanishing only if
\eqn\vanishing{
{\rm Re}(e^{i\phi_{ij}}x)>0\qquad {\rm for}\ x\ {\rm in\
the}\ \alpha-{\rm th\ sector}.}

The crucial point is the identity
\eqn\crucial{D_{ij}^{(\alpha),k}=\mu_{ik}B^{(\alpha)}_{kj}.}
(cf. \cru).
This can be seen as before: Let $\tilde\alpha$ be the sector
opposite to $\alpha$ (i.e. $x$ belongs to the  $\tilde\alpha$--th
sector if $-x$ belongs to the $\alpha$--th one). Then
inserting \IRE\ into the identity
\eqn\symmetry{\Phi^{(\alpha)}(x)
\big[\Phi^{(\tilde\alpha)}(-x)\big]^t=1,}
we get
\eqn\second{B^{(\alpha)}=-[(1+B^{(\tilde\alpha)})^{-1}
B^{(\tilde\alpha)}]^t,}
\eqn\third{[D^{(\alpha),k}(1+B^{(\tilde\alpha)})^t]_{ij}=
-\mu_{ik}B^{(\tilde\alpha)}_{jk}.}
Plugging \second\ into \third\ yields \crucial.

The function $\Phi_{i\bar j}(x)$
is globally defined in the upper half--plane.
Then $\Phi^{(\alpha+1)}$ and
$\Phi^\alpha$ should
agree on the $\alpha$--th ray.
On the other hand, the single terms in \IRE\ are discontinuous as we
cross
the $\alpha$--ray because of \discont. Then the continuity of the sum
gives
relations between the coefficients in \IRE. Notice that (assuming no
three
vacua get aligned) the discontinuity of terms omitted in \IRE\
cannot
contribute to $B^{(\alpha+1)}$.
On the contrary, they do contribute to
$D^{(\alpha+1)}$. Luckily there is no need
to control these terms: Their net effect is
just to produce the right discontinuity so that eq.\crucial\ remains
true as we cross the $\alpha$--ray.
Eq.\discont\ yields
$$\CF_{ij}(x)\bigg|_{\alpha+1}-\CF_{ij}(x)\bigg|_\alpha=-i\delta_{i\,
i(\alpha)}\, \delta_{j\, j(\alpha)} \CE_{ij}(x).$$
Then, comparing the coefficients of $\CE_{ij}(x)$ in
$\Phi^{(\alpha+1)}$ and
$\Phi^{(\alpha)}$, with the help of \otherid\ we get
$$B^{(\alpha+1)}=-\mu^{[\alpha]}+B^{(\alpha)}-\mu^{[\alpha]}B^{(\alpha
)},$$
or
$$(1+B^{(\alpha+1)})=(1-\mu^{[\alpha]})(1+B^{(\alpha)}).$$
In view of \realaxis\ this implies
$$(1+\tilde B)=(1+B^{(m+1)})=
\prodarrow_{1\leq \alpha\leq m}\big(1-\mu^{[\alpha]}\big)
(1+B)$$
where the overarrow  means that the  product is taken in the
anti--clockwise order.

Finally, let
\eqn\finally{L:=(1+\tilde B)(1+B)^{-1}=\prodarrow_{1\leq \alpha\leq
m}
\big(1-\mu^{[\alpha]}\big).}

Using \Stokes, the monodromy reads
$$\eqalign{S(S^t)^{-1}&=(1+B)^{-1}(1+\tilde B^t)^{-1}(1+B^t)(1+B)\cr
&=(1+B)^{-1}(L^t)^{-1}L (1+B),\cr}$$
i.e. (up to a unimodular change of bases) the monodromy is given
by $(L^t)^{-1}L$, or explicitly\foot{This equation has been obtained
under the condition that all $\phi_{ij}$ are distinct. However it
is valid
even if this condition does not hold (provided no three vacua are
aligned).
Indeed if, say, $\phi_{ij}=\phi_{kl}$ then there is an $\alpha$ for
which the $\alpha$--th and $(\alpha+1)$--th rays coincide. Then the
order
of the corresponding matrices $(1-\mu^{[\alpha]})$ and
$(1-\mu^{[\alpha+1]})$ is
ambiguous . But (since the four points
$w_i,w_j,w_k,w_l$ are all distinct) these two matrices commute
and hence the order ambiguity is totally immaterial.}\foot{We can
reinterpret this equation in terms of the original soliton
lines $t e^{i\phi_{ij}}$. We have just to take the product in the
clockwise order.}
\eqn\monodromy{\prodarrow_{1\leq \alpha\leq
2m}\big(1-\mu^{[\alpha]}\big). }

Eq.\monodromy, together with
\qiswhat, is our  relation between the soliton numbers and the
UV $U(1)$ charges.

Let us consider \finally\ in more detail. We know that $B_{ij}$
(resp. $\tilde B_{ij}$) can be non--vanishing only if
$\cos(\phi_{ij})>0$
(resp. $<0$). In view of this remark, Lazzari's lemma \monlemma\
applied to \finally\ gives
$$\eqalign{ (1+B)^{-1}&=\prodarrow_{\rm I\
quadrant}(1-\mu^{[\alpha]})\cr
 (1+\tilde B)&=\prodarrow_{\rm II\ quadrant}(1-\mu^{[\alpha]})\cr}$$
where the ordered products are on
the $\alpha$'s whose corresponding angles
$-\phi_{ij}$ belong to the first (resp. second) quadrant. Then
$$(1+\tilde B^t)^{-1}=\prodarrow_{\rm IV\
quadrant}(1-\mu^{[\alpha]}).$$
Finally, from \Stokes\ we have
\eqn\Finally{S=(1+B)^{-1}(1+\tilde B)^{-1}
=\prodarrow_{\rm right\ half-plane}
(1-\mu^{[\alpha]}).}
This shows that the formula we derived for the relation
between $S$ and soliton numbers in the context of LG theories
in section 2 is generally valid for any massive $N=2$
quantum field theory.

\newsec{More on Degenerate UV Critical Theories}

\subsec{The `Strong'  Monodromy Theorem}

In this section we wish to study in slightly
more detail the critical theories one gets
as the ultra--violet limit of a given massive N=2 model.
In general one may get a degenerate superconformal theory, \ie\ a
model
with a continuous spectrum of dimensions.
For instance, in the $\BC P^n$ case the UV limit corresponds to free
field theory and this limit is reached up to logarithmic deviations.
Typically a degenerate limit looks like
a $\sigma$--model with a {\it
non--compact} target space. In this case $L_0$
 has a continuous spectrum and hence the states of definite dimension
are
not normalizable. In particular $|1\rangle$ is not a normalizable
state,
as it is obvious from classical geometry (harmonic forms in
non--compact
manifolds are usually non--normalizable).

One of the purposes of this section is to characterize the massive
theories
having `nice' UV limits.
If a model has a nice UV limit, we can find a basis $\CO_i$ of $\CR$
such that as $\beta \rightarrow 0$,
$${\langle \overline{\CO}_i|\CO_i\rangle \over \langle \bar
1|1\rangle}
\simeq C_i \beta^{-q_i}\Big(1+ O(\beta^a)\Big),\qquad a>0,$$
for some constants $C_i$. Equivalently,
$$Q_{ij}(\beta)=q_{ij}+O(\beta^a).$$
This is just the statement that the UV theory has a positive gap $a$
in the spectrum of dimensions. In particular, this implies the
normalizability of $|1\rangle$
$$\langle\bar 1|1\rangle <\infty,$$
where in the lhs we mean the state obtained by spectral--flow of $1$
in a special field
representation\foot{I.e. in a basis such that $\eta_{ij}$ is
constant. }. This no--degeneracy criterion fails, say, for the $\BC
P^1$
$\sigma$--model where \sigmamodels\
$$\langle\bar 1|1\rangle \simeq -4\log\beta, \qquad \beta\rightarrow
0.$$

The first remark is that the UV limit cannot be non--degenerate if
 the monodromy $H=SS^{-t}$ has non--trivial
Jordan blocks.
This was shown in \S.4.3, see eq.\Jstruct.
Then we have
the natural question: {\it is the triviality of the Jordan structure
of $H$
enough to ensure the non--degeneracy of the UV limit} (assuming
that the original massive theory is regular)?

To begin with, let us consider the Landau--Ginzburg
models with a polynomial superpotential. In this case the UV limit
is
 `nice' if and only if in the limit the superpotential $W(X_i)$
becomes
a quasi--homogeneous function. Indeed, this is precisely the
condition
needed in order $W(X_i)$ to be $U(1)$--invariant. If $W(X_i)$ is
quasi--homogeneous, \ie\
if there are rational numbers $q_i$ such that
$$W(\lambda^{q_i}X_i)=\lambda W(X_i)\qquad \forall \lambda\in\BC,$$
then
\eqn\euler{W(X_j)=\sum_i q_i X_i \partial_i W(X_j)
 \simeq 0 \quad {\rm in}\ \CR. }
Conversely, let $W(X_i)_{\rm uv}$ be the superpotential in the
UV limit, and
$C_{\rm uv}$ the matrix representing multiplication by the chiral
operator $W(X_i)_{\rm uv}$ in $\CR$. We claim that we can choose the
additive
constant in $W(X_i)$ so that all the eigenvalues of $C_{\rm uv}$
vanishes.
Indeed, for all $\beta\not=0$, multiplication by the superpotential
is
represented by the matrix $\beta C$ and hence\foot{Notice that this
argument does
not imply that $C_{\rm uv}$ is the matrix $0$. Indeed, the
transformation
relating the canonical basis of
$\CR$ to the operator one becomes singular as
$\beta\rightarrow 0$. Otherwise, the ring $\CR$ itself would
trivialize at
the UV fixed point, which is obviously not the case. The
characteristic
polynomial is invariant under changes of bases and hence we are
allowed
to take its  limit as $\beta\rightarrow 0$.}
$$ \det[z-C_{\rm uv}]=\lim_{\beta\rightarrow 0}\det[z-\beta C]=z^n.$$
Then $C_{\rm uv}$ is nilpotent and
therefore it is fully determined by the dimensions
of its Jordan blocks. The UV limit superpotential is
quasi--homogeneous
if and only if these blocks are all trivial. For $W(X_j)$ a
polynomial,
this is an easy consequence of \euler.
So in the LG case the UV limit is `nice' iff $C_{\rm uv}=0$.

The `strong' monodromy theorem of Singularity Theory
(first proven by Var\v cenko \varc) states that the Jordan structures
of
$C_{\rm uv}$ and of the Milnor monodromy $H$ are equal.
Then for LG models (with polynomial superpotentials) the answer to
our
question is {\it yes}.

Now let us go to the general case.
By analogy with the LG case,
to answer {\it yes} we have to show that: 1. the `strong' monodromy
theorem holds in general, and 2. that the UV limit is nice iff
$C_{\rm
uv}=0$. We have already shown 1.
Indeed from \BB\ we see that
$$C_{\rm uv}=B\Big|_{|x|=1},$$
and so the `strong' monodromy theorem is equivalent to the remark
just
after eq.\Jstruct.
Instead 2. is a well--known consequence of the \ttstar\ equations.
In fact, these equations imply (here the matrix
$C^\prime$ is $\beta C$
rewritten in the operator basis)
$$\bar\partial_i Q= [C^\prime, \bar C_i],$$
we see that $Q$ has a constant limit if and only if the UV limit of
the
rhs vanishes for all $i$'s, \ie\
\eqn\nnnj{[C_{\rm uv},g_{\rm uv}
 C^\dagger_i g_{\rm uv}^{-1}]=0\qquad \forall i. }
Assume
that the UV limit is a non--degenerate conformal theory.
Then the metric $g_{\rm
uv}$ is a non--singular positive--definite inner product on $\CR$.
In this case \nnnj\ implies\foot{Indeed a nilpotent
matrix which commutes with its own adjoint, vanishes.} $C_{\rm
uv}=0$.

The fact that a `strong' version of the monodromy theorem holds
allows us to borrow many results from Algebraic Geometry which
are consequences of this theorem. Some of these results were
developed in the context of the degeneration theory for complex
structures over algebraic manifolds and eventually evolved in
Deligne's theory of {\it mixed Hodge structures} \delmix.
Physically they are
related to `mirror symmetry'.
It is not appropriate to discuss further
these developments here, so we
limit ourselves to the simplest result in this direction (the
Schmid's orbit theorems \sch) that we need below.

The basic idea is that from the Jordan structure of $H$ we can cook
up
an $SU(2)$ action on $\CR$. In fact, given a nilpotent matrix $L$
acting on a vector space $\CV$,
we can always find
 (by Jacobson--Morosov)
an $sl(2)$ representation on $\CV$ such that
the generator $J_+$ is mapped into $L$. Applying this remark to
the nilpotent
matrix $B$ (acting on $\CR$),  we see that
we can use $SU(2)$ representation theory
to `measure' the degeneration of the UV critical theory.
The bigger the `angular momentum' the more degenerate the UV limit
is. In particular the
theory is non--degenerate if and only if the corresponding $SU(2)$
representation is trivial. More generally,
we get logarithmic corrections of the form $(\log\beta)^k$
where $k/2$ is the larger `spin' appearing in
the above $SU(2)$ representation.
We illustrate the physical applications of
this viewpoint in the special case of $\sigma$--models.

\subsec{AF $\sigma$--models}

We consider an AF
$\sigma$--model with action
$$S=\sum_{(1,1)\ \rm classes} t_a\int\omega_a^{(2)}+ D-\rm term.$$
Asymptotic freedom requires the Ricci tensor $R_{i\bar j}$ to be
(strictly) positive--definite. The corresponding $(1,1)$ form
$R$ can be decomposed as
$$R=\sum_{(1,1)\ \rm classes} s_a\omega_a^{(2)}.$$
By definition, the matrix $\beta C$ represents in $\CR$
multiplication
by the operator $\hat\phi$ such that
$$\delta_{\rm RG}S=\int d^2z d^2\theta \hat\phi+ \rm h.c\ +\
D-terms,$$
(here $\delta_{RG}$ is the infinitesimal Renormalization Group flow).
In the present case $\hat\phi$ is just the
chiral field associated to the
Ricci form, and thus $\beta C$ is
the matrix representing multiplication by the Ricci class in
the {\it quantum} cohomology ring.

As $\beta \rightarrow 0$, the target space metric $G_{i\bar j}$
flows towards
one cohomologous to a
K\"ahler--Einstein metric of infinite volume\foot{Indeed, by
\alvarez\
we have $G_{i\bar j}(\beta)\simeq
G_{i\bar j}(1)-R_{i\bar j}\log\beta$,
where $\simeq$ means equality in cohomology.}.
Moreover, this is the weak coupling ($=$ semiclassical) limit.
 In this
limit the chiral ring reduces to the classical cohomology ring.

Hence $B$ is proportional
to the matrix representing multiplication by the (asymptotic)
K\"ahler class
in the (classical) cohomology ring.
For instance, in the $\BC P^n$ case $B$
is given by eq.\excpn. From that equation
it is obvious that $B$ represents the multiplication by the K\"ahler
class
in the cohomology ring.

Then for an AF
$\sigma$--models having as target space a (compact
K\"ahler) manifold $\CM$ of complex dimension $d$,
the Jordan structure of
$H$ is completely specified in terms of the geometry of
$\CM$. Indeed the set of all
harmonic forms on $\CM$ can be decomposed into
irreducible representations of $SU(2)$ ({\it Lefschetz decomposition}
\griHarris). Comparing the hard Lefschetz theorem with
 the our construction above, we see that the Lefschetz $SU(2)$
coincides with the one measuring the degeneracy of the UV theory.
Let $\{s_j\}$ be the set of
`spins' appearing in the Lefschetz decomposition (counted with
multiplicity). Then the length $(k_j+1)$
of the $j$--th Jordan block
is equal to $(2s_j+1)$.
In particular in the (AF) $\sigma$--model case
$H$ has {\it one and only one} Jordan block
of length $d+1$, and no Jordan block has length $l>d+1$.
Moreover for all blocks $k_j\equiv d$ mod.$2$.
These geometrical facts are easily recovered from the
general classification of N=2 superconformal models
discussed in the present paper.

This example also `explains' in which sense the Jordan structure
measures
the failure of the UV fixed theory to be a nice superconformal
theory.
The Ricci tensor is the $\beta$--function of the model, and its
topological
class (\ie\ the first Chern class)
measures the obstruction to find a fixed
point \ie\ a point where the $\beta$--function really vanishes.
But $B$ encodes exactly this topological information.

{}From the above formulae one can also extract the leading UV
behaviour
for the ground--state metric $g$.
Again we illustrate this in the ${\bf C}P^{n-1}$
case. Since $B^n=0$, we have
$$\Phi\sim \sum_{r=0}^{n-1}{1\over r!} \big(\log\beta\big)^r
B^r\Phi_0.$$
Let $X$ be the chiral primary operator dual to the hyperplane
section. In the UV limit it
 acts on the ring as the matrix $c^{-1}B$, for some
normalization coefficient\foot{One has $c^{-1}=2n$.
This can be seen as follows. In our conventions,
$C$ represents on $\CR$ the chiral field
$2\mu\partial_\mu\omega$, where $\omega$
is the K\"ahler class and $\mu$ is the RG scale. For ${\bf C}P^{n-1}$
we have $\omega=-\log(\alpha/\mu^n)\, X$, and thus $C=2n X$.} $c$.
Then as $\beta\rightarrow
0$,
$$X^k\Phi= c^{-k} B^k\Phi \sim {1\over (n-1-k)!}
\big(c\log\beta\big)^{(n-1-k)}
{1\over c^{n-1}}B^{n-1}\Phi_0.$$
On the other hand, by definition\foot{In writing this equation we
used the
fact that $q$ commutes with $g$ for $\beta\sim 0$.}
$$X^k\Phi= \overline{X^{n-1-k}\Phi}\, \langle \bar X^k|X^k\rangle,$$
and thus
\eqn\guv{\langle \bar X^k|X^k\rangle\sim
{k!\over (n-1-k)!}\big(-|c|\log\beta\big)^{(n-1-2k)},}
in agreement with Ref.\sigmamodels.
Of course, this is just
the result predicted by classical geometry \weilA.

\newsec{The Classification Program}
We have seen in the previous sections that the number of vacua and
the number of solitons between them is enough to give the full
solution
to $tt^*$ equations.  This means that the geometry of ground
states of the supersymmetric theory are completely determined
by the IR data which is the counting of the soliton numbers.
Note that the geometry of the ground state is sensitive only
to $F$--term perturbations and are insensitive to $D$--terms.
Therefore
two theories which differ only by a variation of the $D$--term will
have the same ground state geometry and soliton numbers.
However as we have seen the soliton numbers fully capture the
$F$--term perturbations of the theory.  As an example, if we consider
$CP^1$ sigma model, the Kahler class of the metric is the information
contained in the $F$--term, whereas the precise form of the K\"ahler
metric is determined by the $D$--term.  In particular there are
infinitely
many ways to vary the $D$--term which is equivalent to the space
of all K\"ahler metrics with a fixed K\"ahler class.  So what we will
be able to do is therefore to begin classifying massive $N=2$ quantum
field
theories up to variation of $D$--terms.  Indeed this turns out
to be {\it equivalent} to classifying all $N=2$ CFT's which admit
a massive deformation.  The reason is that the condition of conformal
invariance automatically picks a $D$--term for a given $F$--term.
This can be proven rigorously in the SCFT by noting that the
only supersymmetric perturbations which preserve conformal
invariance is via chiral fields, which are $F$-term, i.e., there
is no continuous variation of the $D$--term which preserves
conformal invariance.  So the UV limit of any of the theories
we consider will automatically label a {\it conformal} theory,
the $D$--term of which is adjusted to make the theory conformal!
In this way we get a mapping between soliton numbers and $N=2$
superconformal models.  As we discussed before this will not give all
superconformal models, but only those which admit a
non-degenerate massive
deformation,
a precondition of which is that the left ($q_L$) and right ($q_R$)
charges of chiral fields be equal (i.e., chiral fields have
zero fermion number).  It is not clear that all conformal theories
satisfying $q_L=q_R$ automatically admit a massive deformation
but we know of no counterexample to such an expectation. Assuming
this is
generally
true, our method thus classifies
all the $N=2$ CFT's with left-right symmetric $U(1)$ charges for
Ramond ground
states.

Note that a particularly interesting class of conformal theories
for constructing string vacua, i.e., Calabi-Yau case, admit
no massive deformation (there are no relevant operators).  However
we know that one can obtain examples of Calabi-Yau by considering
orbifolds of LG models.  The same is true for the left-right
symmetric theories under consideration here\foot{The
reader should be careful to distinguish
the usage of `left-right symmetric  $N=2$
theories' here from that used in the context of RCFT's.}\ from
which we can obtain Calabi-Yau manifolds by taking orbifolds (it
is an interesting question to see if for every Calabi-Yau
there exists a point on moduli space
which is related to a symmetric
theory by orbifoldizing).

We may be interested in classifying non-degenerate (or `compact')
$N=2$ CFT's, in which case we have to impose the condition
that $H=SS^{-t}$ has a trivial Jordan block structure, as discussed
before.  As an example the $UV$ limit of $CP^1$ is $R^2$ which is
degenerate (in the sense that it has a continuous spectrum). In
this section we consider both degenerate and non-degenerate theories.

We may also be interested in uncovering the allowed perturbations of
our
theories which send some vacua to infinity.  This would for example
be interesting in understanding the RG-flows among the theories.
{}From the classification program all the perturbations which send
some vacua to infinity are allowed as long as we end up with {\it
real}
$U(1)$ charges for the theory with fewer vacua.  In other words
if the reduced $S$ matrix gives rise to real $U(1)$ charges
then it presumably is an allowed perturbation of the theory.

To begin with classifying the theories, we first fix the Witten
index of the theory to be $n$.  Then we take an arbitrary strictly
upper-triangular
integral $n \times n$ matrix $A$ which is taken to count the soliton
numbers
(taking into account $(-1)^F$) between these vacua (where we assume
the vacua
to be in a `standard'
configuration in the $W$--plane)\foot{As usual, we order the vacua
such that ${\rm Re}\,(w_i-w_j)>0$ for $i<j$.}.
For a general
triangular matrix $A$, the eigenvalues of
$H\equiv (1-A)(1-A^t)^{-1}$ need  not
have norm 1. However in the physical case
they should, since
$q$ is hermitian. This gives a severe restriction on the entries of
physically
allowed
Stokes parameters\foot{It is conceivable that the
condition that eigenvalue be a phase is already implied by the
regularity of
the solution for
all $\beta$'s.  Though the {\it integrality} of the matrix $A$ is not
guaranteed
by regularity alone, as there are counterexamples,
and should be viewed as an additional physical constraint.}
$A$ which count soliton numbers.
Thus  $H\in SL(n,\BZ)$ is a {\it modular} matrix.
{}From Lazzari's lemma \monlemma,  $A$ can be recovered uniquely
from $H$.

Then  the classification of  N=2 superconformal models
having a totally massive perturbation is reduced to
the following Diophantine problem\foot{Strictly speaking, those
discussed in
the text are only {\it necessary} conditions. However, experience
suggests
that these conditions are very close to being also sufficient.}.
Find
all {\it
integral} strictly upper--triangular $n\times n$ matrices $A$ such
that all the
eigenvalues $\lambda_i$ of the modular matrix  %
$$H=SS^{-t}=(1-A)(1-A^t)^{-1}$$
belong to the unit circle $|\lambda_i|=1$.
Two solutions
$A$ and $A^\prime$ are `equivalent' if
they are related by a braiding transformation and a change of sign in
the
canonical basis discussed in section 2.
The very same
number--theoretical problem arises in Algebraic--Geometry \hodge\ and
Singularity Theory \singt .
Unitarity gives  further restrictions on the physically allowed
solutions. In particular
 in any (irreducible) unitary theory we have only
one chiral primary with vanishing charge, \ie\ 1. Then the smallest
value of $q$ should be non degenerate.

Here we discuss some general facts about this classification
program.  In the following subsections we apply these methods
to obtain the complete
classification for the case of small Witten indices $n <4 $.  In the
next
section
we rederive the $ADE$ classification of minimal models using our
methods.
It turns out to be extremely simple to obtain this classification with
these
methods.

Standard number--theoretical argument (based on Kronecher's
theorem \refs{\krona\cotech}) shows that $H$ is
{\it quasi--idempotent} i.e. there exist
integers $m$ and $k$ such that
\eqn\emme{(H^m\pm 1)^{k+1}=0\qquad {\rm weak\ monodromy\ theorem},}
(here $m$, $k$ are
assumed to be the smallest integers for which \emme\ is
true; $k$ is known as the {\it index} of $H$).
In particular the
$q_k$'s are {\it rational numbers}.  This is our first
general conclusion.
In the geometrical case one has also a strong form of
the monodromy theorem stating that the index $k$
is always less or equal to the (complex) dimension $d$.
In the physical context $d$ should be replaced by
the UV central charge $\hat c$. The known N=2 theories satisfy
this stronger statement (in particular the $\sigma  $ models,
as we saw in sect.5). It is tempting to conjecture that the strong
form of the monodromy theorem is always
true.  Indeed, in the
LG case, the theorem is a simple
consequence of the `strong' monodromy theorem we
discussed in \S.5.2. Since this `strong' theorem holds in full
generality,
it is reasonable to expect that also the bound $k\leq \hat c$ is
always valid. Here we limit ourselves to a sketch of the proof for
the
general case, under the additional assumption that in
the (degenerate) UV
critical theory the only primary chiral field with vanishing charge
is the
identity operator $1$. In this case, all nilpotent chiral operators
are
linear combinations of fields of {\it positive} charge.
This remark, in particular, applies to the
field $\hat\phi$ corresponding to the matrix $B$. Consider then the
subset of operators $\phi_i^\prime$
($i=1,\dots,k+1$) belonging to a Jordan block of
$H$ of maximal index $k$,
and  let $q^\prime_i$ be their $U(1)$ charges.
By definition, $q^\prime_i=q^\prime_j$ {\it modulo
one}. On the other hand, the arguments of sect.5
imply
$$\phi_r^\prime\approx \hat\phi^{r-1}\phi^\prime_1.$$
Since $\hat\phi$ has positive charge, we have $q^\prime_r >
q^\prime_{r-1}$,
which implies $q_r^\prime \geq q_{r-1}^\prime+1$. Then
$q^\prime_r\geq
(r-1)+q^\prime_1$, which gives
\eqn\chatbound{\hat c=q_{\rm max}
-q_{\rm min}\geq q^\prime_{k+1}-q^\prime_1\geq k, }
which is the strong form of the monodromy theorem.

Consider the characteristic polynomial
$P(z)=\det[z-H]$. It satisfies $P(0)=(-1)^n$ and
$$P(z)=(-z)^n P(1/z).$$
Then $A$ is a solution to \emme\ if and
only if all roots of $P(z)$ are $m$--roots of
$1$ \ie\ if $P(z)$ has the form
\eqn\cycloprod{P(z)=\prod_{d|m}\Phi_d(z)^{k_d},}
where $\Phi_d(z)$ are the cyclotomic polynomials \cyclopoly.
Since ${\rm deg}\, P(z)=n$,
we get a relation between the Witten
index and the possible $U(1)$ charges
$q_k$ of a Ramond ground state\foot{
Some of the following restrictions
can also be derived using
integrality of the number of states in twisted
sectors of the orbifolds of the corresponding
conformal theory.}. Indeed let $q_k=r/s$ with $(r,s)=1$.
Then $\phi(s)\leq n$,
where $\phi(s)$ is Euler's totient function.
Moreover, if we have $n_s$ Ramond vacua
with charge $r/s$ mod.1 and $(r,s)=1$, corresponding to a set of
Jordan blocks of lengths $(k_{j_s}+1)$, then for all
$\l\in ({\bf Z}/s{\bf Z})^\times \simeq {\rm Gal}({\bf
Q}(e^{2\pi i/s})/{\bf Q})$ there are precisely
$n_s$ Ramond vacua with charge
\eqn\dimenum{q={l\over s}\quad {\rm mod.}\ 1,}
and they are
organized in Jordan blocks of the same lengths\foot{This follows from
the fact that the
minimal polynomial of $H$ belongs to $\BZ[z]$.}.

Not all products of cyclotomic polynomial can appear in
\cycloprod. Let
\eqn\decomposition{P(z)=\prod_{m\in\bf N}
\Big(\Phi_m(z)\Big)^{\nu(m)},}
where  $\nu(m)\in\bf N$ are
almost all vanishing. Then one has the following constraints on the
possible
$\nu(m)$'s (physically they are selection rules on the
allowed $U(1)$ charges)
$$\eqalign{& 1.\qquad \sum_m\nu(m)\phi(m)=n\cr
&2.\qquad \nu(1)=n\ {\rm mod.}\,  2\cr
&3.\qquad {\rm for}\ n\ {\rm even},\ either\ \nu(1)>0\ {\rm or}\
\sum_{k\geq 1}\nu(p^k)=0\ {\rm mod}.\, 2\
{\rm for\ all\ primes}\ p.\cr}$$
For instance, in
degree $2$ there are $6$ polynomials of the form \decomposition.
 Only three of these
satisfy the selection rules, namely
$\Phi_1^2$, $\Phi_2^2$, and
$\Phi_6$. In the same way, in degree $3$ only $5$ out
of $10$ possibilities are allowed, and in degree $4$ only $12$ out of
$24$
(\eg\ $\Phi_8$, $\Phi_5$ and $\Phi_3 \Phi_2^2$ cannot appear).

This is shown as follows: 1. is obtained by equating the degree
of both sides of \decomposition\ (recall that deg $\Phi_m=\phi(m)$).
2. Follows from the fact that $P(0)=\det[-H]=(-1)^n$, whereas
$\Phi_m(0)=1$
for all $m$'s but for $m=1$, where  $\Phi_1(0)=-1$.

To get 3., notice the identity
$$P(1)\equiv \det[1-H] = \det[S^t-S]\det[S^t]
=\Big({\rm pf}[S^t-S]\Big)^2,$$
where ${\rm pf}[\cdot]$ is the Pfaffian. Hence
\eqn\identity{
\prod_m\Big(\Phi_m(1)\Big)^{\nu(m)}=\Big({\rm pf}[S^t-S]\Big)^2.}
Moreover, one has
\eqn\valuesone{\Phi_m(1)=\cases{0\qquad {\rm if}\ m=1\cr
p\qquad {\rm if}\ m=p^k,\ p\ {\rm prime},\  k\geq 1\cr
1\qquad {\rm otherwise}.\cr} }
Now, if $n$ is odd, ${\rm pf}[S^t-S]\equiv 0$
in agreement with 2. Instead, if $n$ is even,
either ${\rm pf}[S^t-S]=0$ or,  by \identity,
$\prod_m(\Phi_m(1))^{\nu(m)}$ is a
non--trivial square, and hence its order at each prime should be
even.
The order at the various primes is easily computed with the help of
\valuesone. This gives 3.

For small $n$ it is easy to solve the above
Diophantine problem thus getting a complete classification.
Here we limit ourselves to $n=1$, $2$, $3$.

For $n=1$ there are no solitons, and then we have only the trivial
solution
to our Diophantine problem.
This solution corresponds
to the free massive model. This model has an unbroken $U(1)$ symmetry
whose
charge $q$ counts the number of Bose particles. On the vacuum
$Q=q=0$, as
required by PCT.

\subsec{Complete Solution for $n=2$}

For $n=2$ we have
$$S=\left(\matrix{1 & a\cr
0 & 1\cr}\right).$$
Notice that the sign of $a$ is physically irrelevant since
it can be flipped
by redefining the canonical basis. Then we can assume $a\leq 0$.
Moreover, $|a|$ is equal to the number $|\mu|$ of solitons connecting
the two vacua.
The characteristic polynomial of $H$ reads
\eqn\chtwo{P(z)=z^2+(a^2-2)z+1.}
There are three values of $a$ for which $P(z)$ has
the form \decomposition.
These correspond to the three possibilities allowed by the selection
rule.
They are
\item{1.}{$a=0$. This gives $P(z)=\Phi_1(z)^2$, \ie\ $q=0$. This case
corresponds to the trivial model (no solitons at all).
\item{2.}{$a=-1$. This gives $P(z)=\Phi_6(z)$, \ie\ }
$$q_i=(-\textstyle{1\over 6},\textstyle{1\over 6})\  .$$
\item{}{The integral part of the charge is fixed here and in what
follows by taking $a\rightarrow t a$
and letting $t$ go from $0$ to $1$, as discussed
before.  This solution
corresponds to the Landau--Ginzburg model with superpotential}
$$W(X)=X^3-X.$$
\item{}{The uniqueness of solution for \dubrovin\ also
 fixes the integral part of $q$. Thus $|a|=1$ implies $q=\pm 1/6$.}
\item{3.}{$a=-2$, which gives $P(z)=\Phi_2(z)^2$. In this case we
have
$$q_i=(-\textstyle{1\over 2},\textstyle{1\over 2})\  .$$
\item{}{and $H$ has a non--trivial Jordan block. Indeed, this
is precisely the solution \cpmono\ we have discussed in detail in
sect.4.3.
{}From that analysis we see that this model corresponds to the $\BP^1$
$\sigma  $--model (or, equivalently, the Ising two--point function).
Again this also fixes the integral part of the charges.}

\vglue 1cm

In cases 2. and 3. the matrix $B=S+S^t$ is the Cartan matrix for
$A_2$ and
$\hat A_1$, respectively. In fact, the model 3. can also be realized
as
the N=2 $\hat A_1$ Toda theory, \ie\ the LG model with superpotential
\eqn\todaone{W(X)=\lambda\left(e^X+e^{-X}\right),}
and the identification $X\sim X+2\pi i$.

The above number--theoretical result should be
compared with the known regularity theorems for
Painlev\'e III (PIII) \piiimath. (For a massive model with two vacua
the
\ttstar\ equations can be always recast in the PIII form, see
\topatop).
 In the $n=2$ case, the
eigenvalues of the $Q$--matrix are \refs{\topatop,\newindex}
\eqn\Qtwo{ Q(z)=\pm{1\over 4} z {d \phantom{u}\over dz}u(z),  }
where $z=m\beta$ and $u(z)$ satisfies special PIII, \ie\ the radial
sinh--Gordon equation
\eqn\piiiU{{d^2u\over dz^2}+{1\over z}{du\over dz}=\sinh u.}
As discussed in sect.4, the boundary condition for \piiiU\
is encoded in the Stokes parameter $a$. In terms of the more usual
boundary datum $r$, defined by the behaviour of $u(z)$ as
$z\rightarrow 0$
\eqn\zasy{\eqalign{u(z) &\simeq r \log {z\over 2}+s+\dots \hskip
3.8cm
 {\rm for}\
 |r|<2\cr
&\simeq \pm 2 \log{z\over 2}\pm \log\left[-\log\left({z\over
4}+\gamma\right)\right]+\dots \qquad {\rm for}\ r=\pm 2,\cr}  }
$a$ is given by \piiimath\
\eqn\aVr{a=2 \sin\left({\pi r\over 4}\right).}
In view of eq.\Qtwo, the datum $r$ is essentially the $U(1)$ charge
at the
UV fixed point. Indeed,
$$q=\lim_{z\rightarrow 0} Q(z)=\pm {1\over 4} r,$$
so, in physical terms, \aVr\ reads
\eqn\mVq{|\mu|= 2 \sin  (\pi |q|).  }
In fact this result can be derived directly by continuously turning
on the soliton number and considering the eigenvalues of $SS^{-t}$,
which gives a nice illustration of what we mean by continuously
deforming the soliton number in order to recapture the integral
part of $q$.
It is known \piiimath\
that PIII has {\it one} regular solution $u(z)$
for each $r$
with $|r|\leq 2$,
and that all regular real solutions (bounded as $z\rightarrow
\infty$)
have the UV asymptotics \zasy\ for some $r$, $|r|\leq 2$. In view of
\aVr, we have a regular solution for all (real) $a$ with $|a|\leq 2$.
Comparing with \chtwo, we see that this is just the condition
$$-2\leq \tr\, H\leq 2,$$
\ie\  $u(z)$ is regular iff the eigenvalues
of the monodromy have norm $1$. So, in this case, {\it
unitarity implies
regularity}. In particular, the
three possible integral values of $|a|$ do correspond to regular
solutions
having the expected UV behaviour.
Thus for $n=2$ all solutions to the Diophantine problem are realized
by physical systems. (In fact even the non--integral {\it unitary}
solutions play a role for {\it non--generic} $n>2$ models, see \eg\
\refs{\topatop,\sigmamodels,\newindex}).

\subsec{Complete Solution for $n=3$}

Consider next $n=3$.
We put
$$S=\left(\matrix{1 & x_1 & x_2\cr 0 & 1 & x_3
\cr 0 & 0 & 1\cr}\right).$$
Two triples $(x_1,x_2,x_3)$ and
$(x^\prime_1,x^\prime_2,x^\prime_3)$
correspond to (massive perturbations of) the same
superconformal model if we can pass from one to the other by
a repeated application of the following transformations:
\itemitem{\it a)}{flipping the sign of {\it two} $x_j$'s;}
\itemitem{\it b)}{a permutation of $(x_1,x_2,x_3)$;}
\itemitem{\it c)}{replacing $x_1$ by $x_2x_3-x_1$
{\it or} $x_2$ by $x_3x_1-x_2$ {\it or}
$x_3$ by $x_1x_2-x_3$.}

\noindent Indeed, {\it a)} just corresponds to a redefinition of the
signs
for the canonical basis, while {\it b)} and {\it c)} can be obtained
by
suitable combinations of {\it i)} rotations of the Stokes axis, and
{\it ii)} deformation of one vacuum $w_i$ across the line connecting
the other two vacua $w_j$, $w_k$ (see sections 2 and 3).

The characteristic polynomial of the monodromy is
$$\eqalign{& P(z)=\det[z-S(S^t)^{-1}]=z^3+\alpha(x_i)\, z^2-
\alpha(x_i)\, z-1\cr
&{\rm where}\quad \alpha(x_i)\equiv x_1^2+x_2^2+x_3^2-x_1 x_2
x_3-3.\cr}$$
The requirement that $P(z)$ has the form \decomposition\ leads to the
following Diophantine equation
\eqn\uno{x_1^2+x_2^2+x_3^2-x_1x_2x_3=b}
where $b$ depends on the particular product of cyclotomic
polynomials.
Explicitly one has

\vglue 4pt
{\settabs 12\columns
\+ &&& \underbar{\raise 2.5pt\hbox{\qquad $P(z)$\qquad }} &&&
\underbar{\raise 2.5pt\hbox{\ $U(1)$\ \rm charges\ } }
&&&\underbar{\raise 2.5pt\hbox{\ $b$}\ } \cr
\+&&& \ $\Phi_1(z)^3$ &&& \hskip 5mm $(-1,0,\,1)$ &&&\ 0  \cr
\+&&&\ $\Phi_1(z)\Phi_2(z)^2$ &&&\hskip 5mm $(-\half,0,\half)$ &&&\
4  \cr
\+&&&\ $\Phi_1(z)\Phi_3(z)$ &&&\hskip 5mm $(-{1\over 3},0, {1\over
3})$ &&&\ 3  \cr
\+&&&\ $\Phi_1(z)\Phi_4(z)$ &&&\hskip 5mm $(-{1\over 4},0,{1\over
4})$
&&&\  2 \cr
\+&&&\ $\Phi_1(z)\Phi_6(z)$ &&&\hskip 5mm $(-{1\over 6},0,{1\over
6})$
&&&\ 1\cr}
\vglue 4pt
Luckily enough, \uno\ is a well--studied Markoff--type Diophantine
equation
\refs{\markoff,\mordell}. All solutions are explicitly known. Let us
summarize the main results for \uno\ \mordell .

\item{i)}{All {\it non--trivial}
solutions\foot{A solution is trivial if at least {\it two}
of the $x_i$ vanish.
The only trivial solutions are (up to permutations)
$(0,0,0)$ for $b=0$; $(\pm 2,0,0)$ for $b=4$; and
$(\pm 1,0,0)$ for $b=1$. Clearly, these solutions correspond to
physically trivial models.}  \uno\ can be obtained from a
fundamental solution\foot{A solution is fundamental if
$0<x_1\leq x_2\leq x_3$ and $x_1+x_2+x_3$ is minimal.} by a
repeated application of the transformations {\it a)}, {\it b)}, and
{\it c).  So the physically distinct solutions are in one--to--one
correspondence with the fundamental ones.}
\item{ii)}{For $b=0$ the only fundamental solution is $(3,3,3)$.
Using
{\it a), b), c)} this
generates an infinite number of `equivalent' solutions.}
\item{iii)}{For $b=4$ the fundamental
solutions are $(1,1,2)$ and $(2,y,y)$ for $y\geq 2$. Each of these
generates an infinite number of non--fundamental solutions.}
\item{iv)}{For $b=3$ there are no fundamental solutions.}
\item{v)}{For $b=2$ the only fundamental solution is $(1,1,1)$. There
is
only a {\it finite} number of solutions generated by this fundamental
one. Up
to
permutations and changes of signs there are just two: $(1,1,1)$ and
$(1,0,1)$.}
\item{vi)}{For $b=1$  there is no fundamental solution.}

The absence of solutions for $b$ odd is an additional
number--theoretical
`selection rule' on the possible UV charges.

Regarding the Jordan structure, one checks that,
except for the trivial solution $(0,0,0)$,
and for $(2,2,2)$, $H$ is non--derogatory
\ie\ it has  Jordan blocks of
dimension  equal to
the multiplicity of the corresponding eigenvalue.
This claim is equivalent to the statement that, unless $H=1$,
the minimal polynomial for $H$ is the characteristic one.
Since the minimal polynomial $P(z)_{\rm min}$ is given by the formula
\lambdamat\
$$P(z)_{\rm min}={\det[z-H]\over \gcd[{\rm Minors}(z-H)]},$$
we see that $P(z)_{\rm min}\not=P(z)$
is possible only if \decomposition\ contains a factor with
$\nu(m)>1$. From the table we see that this can happen only for
$b=0$, or $4$. Then the above claim is obtained by direct inspection.

\vglue 10pt
\noindent\underbar{\raise 3pt\hbox{\it Comparison with
Degenerate Painlev\'e III}}
\vglue 2pt

As in \S 5.3, the above classification should be compared with the
known results \piiideg\
about the regularity of solutions to the {\it degenerate}
PIII (\ie\ radial Bullough--Dodd)
\eqn\Pdeg{{d\phantom{u}\over d\tau}\left(\tau {d\phantom{u}\over
d\tau}u\right)=e^u-e^{-2 u}.}
The \ttstar\ equations for an $n=3$ massive model can always be
recast
in this form provided we have a $\BZ_3$ symmetry. Hence \Pdeg\ is
connected
to the special cases in our classification  with $x_1=x_2=x_3=s$.
Here $s$ is the Stokes parameter for the degenerate PIII, see eq.(13)
of Ref.\piiideg. A solution of degenerate PIII is a solution to our
integral equation provided the other parameters in Ref.\piiideg\
take the value
$$g_1=g_2=0,\qquad g_3=1.$$
Then, for large $\tau$ ($=$ large $\beta$) the
asymptotic behaviour of the solution is
$$\exp[u(\tau)]\simeq 1+{s\over 2}\sqrt{3\over \pi} (3\tau)^{-1/4}
e^{-2\sqrt{3\tau}}+\dots$$
from which it is obvious that $s$ counts the number of solitons
connecting any
two vacua, in agreement with our
general discussion (see also \refs{\topatop,\sigmamodels}).
In terms of $u(\tau)$ the
two non--trivial eigenvalues of the $Q$--index are
\eqn\qM{Q(\tau)=\pm \tau{\partial\phantom{u}\over\partial
\tau}u(\tau).}
To compute the UV charges $q$, we need the asymptotic expansion
of $u(\tau)$ for small $\tau$. One has \piiideg\
\eqn\tauzero{\eqalign{e^{u(\tau)}&\simeq
-{\sigma  ^2 \over 2\tau\, \sin^2\left\{{i\over 2}\left[\sigma
\log\tau+
\log a\right]\right\} }\hskip 2.5cm s>0,\ s\not=3\cr
&\simeq {2\over \tau [\log\tau-(2\log 3+2\gamma)]^2}\hskip 3.2cm
s=3\cr
&\simeq  {2\over i \nu}\tau^{1/2} \sin\left[{i\over 2}\nu\log\tau-
\log b\right] \hskip 2.7cm s<0,\ s\not=-1,\cr
&\simeq -\tau^{1/2}\log\tau+\tau^{1/2}\left(2\log 3-\gamma-{2\over 3}
\log 2\right)\hskip 0.7cm s=-1,\cr} }
where $\sigma  $, $\nu$, $a$, and $b$ are
\eqn\munu{\eqalign{& \mu= {3\over 2\pi i}\log \left({s-1\over 2}
+\sqrt{\left({s-1 \over 2}\right)^2-1}\right) \hskip 0.8cm
 {\rm with}\  |{\rm Re}\,
\mu|<1\cr
& \nu=3- {3\over \pi i}\log \left({s-1\over 2}
+\sqrt{\left({s-1\over 2}\right)^2-1}\right) \qquad {\rm with}\
 |{\rm Re}\,
\nu|<1\cr} }
\eqn\alphabeta{\eqalign{& a=
3^{-2\mu}\, \Gamma(1-\mu/3)\, \Gamma(1-2\mu/3)\, \Gamma(1+\mu/3)^{-1}
\, \Gamma(1+2\mu/3)^{-1}\cr
&b=
3^{\nu}\, \Gamma(1/2+\nu/6)\, \Gamma(\nu/3)\, \Gamma(1/2-\nu/6)^{-1}
\, \Gamma(\nu/3)^{-1}.\cr}  }
{}From \qM\ and \tauzero\ we see that $\lim_{\tau\rightarrow
0}Q(\tau)$
exists and is real only if $\sigma  $ (resp. $\nu$) is real.
In view of \munu\ this is equivalent to
$$\left|{s-1\over 2}
+\sqrt{\left({s-1\over 2}\right)^2-1}\right|=1,$$
this condition is satisfied iff the expression inside the
square--root
is non--positive, \ie\ for
$$-1\leq s\leq 3.$$
Then there are precisely five regular solutions to \Pdeg\ with
integral soliton number $s$, namely $s=3,2,1,0,-1$. These
correspond to the five ${\bf Z}_3$--symmetric solutions we got
for our Diophantine problem\foot{Notice that the solution $(1,1,2)$
is
equivalent to $(-1,-1,-1)$ by perturbation.}. So, `unitarity'
implies
regularity in this case too. From \munu\ one finds
$$\tr\, H\equiv 1+\cos(2\pi q)=3-3s^2+s^3,$$
in agreement with the result of our Diophantine analysis.
As for the $n=2$ case, the regularity theorem for degenerate
Painlev\'e III can be stated as the condition
$$-1\leq \tr\, H\leq 3.$$

\subsec{Identification of the $n=3$ Models}

Now we discuss the physical realizations of the N=2 models
corresponding to the non--trivial
 solutions of the $n=3$ Diophantine problem. For
some models more than one Lagrangian formulation is known.

\vglue 10pt
\noindent\underbar{\raise 3pt\hbox{\it The $(1,1,1)$ Model}}
\vglue 2pt

The solution $(x_1,x_2,x_3)=(1,1,1)$ corresponds to
the $A_3$ minimal model, \ie\
to the LG model with superpotential
$$W(X)={1\over 4}X^4+\rm lower\ order.$$
This identification is confirmed by the value of the UV charges,
see the table in \S 5.3. The two basic solutions in this class,
$(1,1,1)$ and $(1,0,1)$ correspond to the soliton multiplicities
$\mu_{ij}$ for the two inequivalent geometries in $W$--space.
These two geometries are realized, \eg\
by the ${\bf Z}_3$--symmetric model
$W(X)=X^4-X$ and by the ${\bf Z}_2$ invariant one $W(X)=X^4-X^2$,
respectively. In \topatop\
the \ttstar\
equations for these two models have been solved in terms of
PIII transcendents. The first case leads to  {\it degenerate} PIII
\Pdeg,
while the second to {\it special} PIII \piiiU.

\vglue 10pt
\noindent\underbar{\raise 3pt\hbox{\it The $(3,3,3)$ Model}}
\vglue 2pt

The unique class of solutions for $b=0$, \ie\
$(3,3,3)$ is also an old friend --- the ${\bf C}P^2$
$\sigma  $--model. This identification is consistent with the
$U(1)$ charges (see table). Moreover,
from the explicit
solution of the ${\bf C}P^2$ model \massGap\ we know  that
there are precisely 3 solitons (transforming according
the fundamental representation of $SU(3)$) connecting any two vacua.
Thus the mass--spectrum extracted from the $S$--matrix agrees with
the one
predicted by the Diophantine analysis.

This solution corresponds to a sensible physical theory only for
special geometries in $W$-space. Indeed, if we send one of the three
vacua to infinity (in $W$ space) we end up with a model with only
two vacua connected by $3$ solitons: but this is impossible
in view of  the classification
for $n=2$. The usual ${\bf C}P^2$ $\sigma  $--model corresponds to
the three vacua at the vertices of an equilateral triangle in
$W$--space (its size being related to the K\"ahler class of ${\bf
C}P^2$, and its orientation to the $\theta$--angle). Then
 the model must not
make sense if the vacuum triangle is squeezed more than a certain
amount.
Note that there are three chiral
operators in the ${\bf C}P^2$ model, $1,k,k^2$, where $k$ denotes
the K\"ahler class chiral field.  The operator corresponding to $k^2$
has dimension bigger than 1 and is  non-renormalizable.  Addition
of this term to the action is not allowed.
 The corresponding coupling
controls the shape of the vacuum triangle in $W$--space. So
stretching
the vacuum triangle corresponds to adding non--renormalizable
interactions to
the action, leading to a pathological field theory.
Indeed we see here that
if we insist in adding terms which are not renormalizable we should
either sacrifice unitarity, as the Hermitian charges are becoming
complex,
or the decoupling of the infinitely massive states (i.e., somehow
the vacua that we move to infinity should still be contributing
somehow).  At any rate we see that $tt^*$ equations allow us to
address
the question of adding non-renormalizable terms to the action
in a simple way.  These pathologies must manifest themselves as
singularities
in the solutions of the \ttstar\ equations for certain
critical values of $\beta$ (cf. the discussion
in \S  4.4, 4.5).

{}From the result of this section we can also infer the
classification
of the (compact) complex surfaces with Betti numbers $b_1=0$ and
$b_2=1$ having positive first Chern class\foot{This
condition is needed to ensure AF, see \eg\ ref.\sigmamodels.}
 (\ie\ admitting a K\"ahler--Einsten
metric with positive cosmological constant).
Any such manifold will lead to a $\sigma  $--model having $n=3$.
Its monodromy $H$ should have a Jordan block of order $3$.
Our classification says that there is only one such manifold,
namely $\BC P^2$. This is in agreement with the
known classification of
complex surfaces, see \CCS.
Indeed consider a  K\"ahler manifold
of complex dimension $d$ with $c_1>0$ which has
only diagonal hodge structure ($h^{p,q}\not=0$ only if $p=q$),
i.e. with chiral fields which have zero fermion number $(p-q=0)$.
Sigma model on such manifolds
should correspond to a massive $N=2$ theory which should thus be
showing
up in our classification.  Let
$$n=\sum_{p=1}^d h^{p,p}$$
Then in our classification with $n$ vacua these
sigma models will show up with $U(1)$
charges ranging from $-d/2$ to $d/2$ in integer steps,
for which $h^{p,p}$ is the number of charges equal to $-{d\over
2}+p$.
In particular if two manifolds lead to the same solution
in our classification, then they are `mirror' in the sense that
the sigma model on the two are isomorphic (at least
as far as the ground states are concerned).  It would be interesting
to see if there are any examples of mirror phenomena of this type.
At any rate our soliton diagrams give
a new invariant for these K\"ahler manifolds
(up to braiding action discussed before).    We expect that this
$N=2$ view of diagonal K\"ahler manifolds with $c_1>0$ should lead
to their complete classification.

\vglue 10pt
\noindent\underbar{\raise 3pt\hbox{\it The $(1,1,2)$ Model}}
\vglue 2pt

This solution is equivalent to $(-1,-1,-1)$ by perturbation (as
follows from
how the
soliton numbers change under perturbation \solnc ). The last one is
easier to
realize since it is $\BZ_3$ symmetric. In this case the matrix
\eqn\bmat{B=S+S^t,}
is just the Cartan matrix for the {\it affine} $\hat A_2$ Lie
algebra.
By analogy with \todaone\
it is natural to realize the $(-1,-1,-1)$
model as an N=2 Toda theory related to the $\hat A_2$ root system.
In fact, we claim that it is the N=2
Bullough--Dodd model, \ie\
the LG model with superpotential
\eqn\todatwo{W(X)=t\left(e^X+{1\over 2}e^{-2X}\right),}
again with the identification $X\sim X+2\pi i$.
The simpler way to see
this is to compare with
the usual $A_3$ minimal model \ie\ $(1,1,1)$. Clearly
 the only difference between
the two models is the sign of the basic ${\bf Z}_2$ soliton cycle
(we label the vacua  according to the anti--clockwise
order in $W$ space)
\eqn\ztwocy{\mu_{12}\mu_{23}\mu_{31}=\cases{{-1}\qquad {\rm for}\
A_3\cr {\hskip 3mm 1} \qquad {\rm for}\ \hat A_2.\cr}  }

Let $f_{ij}$ be the Fermi number of the soliton connecting the
$i$--th vacuum to the $j$--th one. Then \ztwocy\ holds provided that
\newindex\
\eqn\tVt{\exp\big[i\pi(f_{12}+f_{23}+f_{31})\big]\Big|_{\hat A_2}=
- \exp\big[i\pi(f_{12}+f_{23}+f_{31})\big]\Big|_{A_3}.}
Indeed  $f_{ij}=f_i-f_j$ where (here $X_k=e^{2\pi i k/3}$, $k=0,1,2$,
 are the classical vacua)
$$f_k=-{1\over 2\pi}{\rm Im}\, \log
W^{\prime\prime}(X_k)
=\cases{\raise 4pt\hbox{$-\textstyle{1\over 3}k$ $ \quad \hat
A_2$}\cr {\hskip 3mm
 \textstyle{1\over 3}k}\quad A_3\cr}\qquad ({\rm mod}.\ 1),$$
which gives \tVt.

This identification is also consistent with the UV behaviour. The
Diophantine analysis shows that the UV central charge is $\hat c=1$.
Since the UV limit of \todatwo\ is just (massless) free field theory,
this is the correct result.

If one is not satisfied with the above argument, we can
do much better, \ie\
we can solve explicitly the \ttstar\
equation for \todatwo\ in terms of (degenerate) Painlev\'e
transcendents. To do this, we
 take the natural vacuum basis generated by spectral
flow \ie\
$$|1\rangle,\qquad |e^X\rangle,\qquad |e^{2X}\rangle.$$
Because of the ${\bf Z}_3$ symmetry $X\rightarrow X+2\pi i/3$, in
this
basis the ground state metric $g$ is diagonal. The diagonal entries
of
$g$ are further restricted by the reality constraint \topatop.
This gives\foot{Notice that these reality conditions are quite
different from those of the $A_3$ minimal model. In fact this is the
{\it only} difference between the two models,   \ie\ they correspond
to two
inequivalent foldings of the $\hat A_2$ Toda equation.}
$$\eqalign{ \langle \bar 1|1\rangle\,
\langle\overline{e^X}|e^X\rangle &={1\over |t|^2},\cr
\langle\overline{e^{2X}}|e^{2X}\rangle &={1\over |t|}.\cr}$$
Using this and setting
\eqn\Wut{\log\, \langle\bar 1|1\rangle= u(\tau)-\half\log |t|^2,
\qquad
\tau= {9\over 4}|t|^2,}
the \ttstar\ equations reduce to \Pdeg\ with $Q(\tau)$ given
by \qM.
In view of our discussion at the end of \S 2.3, to prove that
\todatwo\ is the $(-1,-1,-1)$ model it is enough to show
that $u(\tau)$ in
\Wut\ is the solution to \Pdeg\ with boundary data
$g_1=g_2=0$, $g_3=1$, and $s=-1$. This follows from `regularity'
\topatop. Regularity
requires \sigmamodels\ that $\langle \bar 1|1\rangle$ is
regular as $t\rightarrow 0$, possibly up to logarithmic violation
of scaling (as predicted
by the non--trivial
Jordan structure of the monodromy). Comparing \Wut\
and \tauzero\ we see
that this condition is satisfied only for $s=-1$.

\vglue 10pt
\noindent\underbar{\raise 3pt\hbox{\it The $(2,2,2)$ Model}}
\vglue 2pt

This solution corresponds to the Ising 3--point function \noisi.
Equivalently, we can identify it as the LG model with
superpotential the Weirstrass function \noisi\
$$W(X)=\wp(X),$$
and the identifications
$$X\sim X+n_1\omega_1+n_2\omega_2,\qquad n_i\in \bf N$$
where $\omega_i$ are the two periods of $\wp(X)$.
This model has UV central charge $\hat c=1$ as predicted by
the number--theoretical viewpoint.
Since the Jordan
structure is trivial,
the UV limit of this model
is a viable candidate for a new $\hat c=1$ non--degenerate
superconformal model.  Indeed the fact that the solution for
the metric is non-degenrate at the UV point follows from
the explicit solution of degenerate Painleve III discussed before,
which is thus a confirmation of our general arguments.
Whether or not this is {\it sufficient} to obtain a non-degerate
conformal theory remains to be seen.
 Further details on this model
can be found in
\noisi.  More generally the solution with $n$ vacua with all soliton
numbers equal to $2$ is related to the massive Ising model
$n$ spin correlation functions.  Note that this is the {\it only}
non-trivial theory for which the number of solitons (in absolute
value) does not change by perturbations (see equation \solnc ).

\vglue 10pt
\noindent\underbar{\raise 3pt\hbox{\it The $(2,y,y)$ Models for
$y>2$}}
\vglue 2pt

To our knowledge, no physical realization of these models is known.
On the other hand,
the consistency of these models requires properties which sound so
magical that
one wonders if they exist at all (as sensible QFT's).
 This fact, together with the absence of $\BZ_3$ symmetry, would make
very
difficult to guess an explicit Lagrangian realization for them, even
if they
exist. Anyhow, the positive result is that any
yet--to--be--discovered $n=3$
model should belong to this class, and hence have UV charges
$(-\half,0,\half)$ and
soliton spectra (related by {\it a)}, {\it b)}, and {\it c)}) to
$(2,y,y)$!

We give a discussion of their properties:
Certainly they
cannot be well--behaved for arbitrary  vacuum geometries, since
sending an
appropriate vacuum to infinity we end up with a model containing just
two vacua
connected by $y>2$ solitons, a situation ruled out by the $n=2$
classification.

The UV charges of the three chiral fields are $(0,\half,1)$.
This follows by continuously turning on the soliton number
as discussed before.  However, in this case we find that {\it
there is no way to turn the soliton numbers and go
through phases as the eigenvalues of } $H$.  This
suggests to us that indeed these theories are pathological.
At any rate, if these theories exist, the
$q=1$ field
cannot be a marginal operator, since otherwise all vacuum geometry
will be
allowed, contrary to the above remark. By the same argument it cannot
be an AF
coupling. Then the only possibility is that the leading term in their
$\beta$--function is positive, \ie\ that the coupling is infra--red
stable.
In particular $q=1$ field may be the K\"ahler class
of a dimension 1 complex manifold with negative curvature.  The
fact that there is a non-trivial Jordan block and
that $B$ has a negative eigenvalue in this case support
this picture.
The field with $q=1/2$ may in this set up be related to
a $Z_2$ twist field for an orbifold of this sigma model.

\newsec{The A--D--E Minimal Models Revisited}

If we restrict our general classification to the models with
$\hat c<1$ we should recover the well--known A--D--E classification.
Note that since minimal models by definition have chiral charges
less than $1$, and since the left and right chiral charges
differ by an integer, this implies that for minimal models
the left and right chiral charges are equal.  Moreover
since the charges are all less than one, perturbation with
all of them are relevant and so we should get a massive theory.
Therefore {\it all} the minimal models {\it must} appear in our
classification.
In this section we show
how nicely this particular case fits in our general
framework. From the discussion below it will be evident how
our methods for classification of $N=2$ theories
are the natural generalization of the ones which were
successful for the $\hat c<1$ case.
{}From one point of view, our
discussion of the minimal models is more detailed than the
usual one. In fact as an extra
bonus we get the  classification of the  solitonic spectra
which may appear for a given minimal model perturbed in a generic
way. To get the usual A--D--E result we just have to `forget about'
this
extra information.

\subsec{Positive Inner Products and Root Systems}

Let $B=S+S^t$. We will now show that if $B$ is positive definite,
then
the integral matrix $S$ is automatically
a solution to our Diophantine problem.
In fact from the identity
$$HBH^t=S(S^t)^{-1}(S+S^t)S^{-1}S^t= S+S^t=B,$$
we see that $H$ is orthogonal with respect to the inner product $B$.
If $B$ is positive its orthogonal
group is compact, and hence $H$ is
simple with $|\lambda_j|=1$. Then  $S$ solves our
Diophantine equation.  Note also that if $S$ is close
to the identity matrix then $S+S^t$ is positive definite
and so the eigenvalues of $H$ are always phases.  So in
our argument in previous sections for `building up' the
charges, at least near $t=0$ we are guaranteed that
the charges are real.

We pause a while to digress on the classification of
the
positive definite (integral) matrices $B$.
A first remark is that
$B_{ij}=0$ or $\pm 1$  (for $i\not=j$).  Indeed let (say) $B_{12}=s$.
Then consider the vector $V=(v_1,v_2,0,\dots, 0)$. Then
$VBV^t$, as a quadratic form in $v_1$, $v_2$, is positive definite
iff $|s|<2$. Since $s$ is integral, $s=0$, $\pm 1$.
To any such $B$ we associate a (generalized) Dynkin diagram
by the following rule: the $i$--th and $j$--th vertices are connected
by a solid (resp. dashed) line iff $B_{ij}=-1$ (resp. $B_{ij}=+1$).

Since $\det[B]\not=0$,
we can introduce a basis $\vec e_i$ ($i=1,\dots,n$) of
unit vectors in $\BR^n$
such that
$$(\vec e_i,\vec e_j)=\half B_{ij}.$$
Consider the group $W$ generated by the reflections $R_k$
\eqn\refl{\vec e^{\ \prime}_j
=(\vec e\,  R_k)_j= \vec e_j-B_{jk}\vec e_k.}
$W$, being a discrete subgroup of the compact orthogonal group,
 is finite.
If we take the union of all the images under
$W$ of the vectors $\vec e_i$
we get a finite set of vectors in $\BR^n$
which satisfies the axioms for a (reduced) root system \bourbaki.
In fact it is a simply--laced root system since all elements have
length
$1$. Therefore (assuming irreducibility)
it belongs to the A--D--E series \bourbaki.
There
is a simple rule to get the root system associated to a
given $B$. Since the $\vec e_i$ generate the root lattice,
 $\det B$ is the volume of the fundamental
cell. Then $\det B$ is $n+1$ for $A_n$, $4$ for $D_n$
and $9-n$ for $E_n$.
The general solution to our Diophantine problem with $B$
positive
is obtained as follows. Take a simply--laced Lie algebra of rank $n$,
and choose $n$ linearly independent vectors $\vec e_i$ belonging to
its root system\foot{Notice that two choices differing only for
the order of the elements $\vec e_i$ should be considered as
distinct since they lead to different $S$'s.}. Then put
$$-A_{ij}=\cases{(\vec e_i,\vec e_j)\qquad {\rm for}\ i<j\cr
0 \hskip 1.4cm \rm otherwise.\cr}$$
Let us compute the monodromy $H$ of this solution.
 Consider the matrix
\eqn\coxele{R=-H^t=-S^{-1}S^t.  }
The matrix $R$ satisfies a remarkable identity due to
Coxeter.
One has \coxt
\eqn\coxid{R=R_1R_2R_3\cdots R_n.}
Let us consider first a `standard' solution,
\ie\ the vectors $\vec e_i$
are normal to the walls of a Weyl chamber.
In this case the generalized
Dynkin
diagram reduces to the usual one, and $B$ is just the
Cartan matrix. For this `standard' situation
$R$ is known as the Coxeter element of
the finite reflection group $W$ ($=$ the Weyl group). The Coxeter
element is
independent of
choices (up to conjugation) as long as the Dynkin diagram
contains only {\it solid} lines (which in particular means that
it is a tree).
The order $h$ of $R$
is called the Coxeter number of the associated
Lie algebra. Its eigenvalues are of the form $\exp[2\pi i m_j/h]$
where the integers $m_j$ are the {\it exponents} of the corresponding
Lie algebra \coxt. Comparing with \coxele\ we see that the UV charges
for a `standard' solution are
\eqn\mincharges{q_j= {m_j\over h}-{1\over 2}\quad ({\rm mod.}\ 1).}
Of course, this is precisely the answer for the corresponding
A--D--E minimal model.

The next step is to find the solutions which are equivalent to the
`standard' one, in the
sense of corresponding to different perturbations of the same basic
superconformal model.
This is the same as asking which sets of roots
$\vec e_i$ can be obtained
from a given one by
 a continuous
 deformation
of the couplings $w_k$. These are those
 obtained from a standard solution by a repeated application
of the following `moves'.
First of all, we can replace a root $\vec e_i$
by the opposite root $-\vec e_i$ since this is just a redefinition
of the sign of the corresponding canonical vacuum.
Then we can replace the ordered pair of roots $(\vec e_j,\vec
e_{j+1})$
by the pair $(\vec e_{j+1} R_j, \vec e_j)$.
Indeed, the transformation
$$\eqalign{&\vec e_i^{\ \prime}=\vec e_i
\qquad {\rm for}\  i\not=j,j+1\cr
&\vec e_j^{\ \prime}= \vec e_{j+1} R_j \equiv
\vec e_{j+1}-B_{j+1,j}e_j\cr
&\vec e_{j+1}^{\ \prime}=\vec e_j,\cr}$$
induces the following transformation on $B_{ij}$ ($i\not=j,j+1$)
\eqn\Ti{B^\prime_{ij}=B_{i,j+1}-B_{j+1,j}B_{ij}.}
\ie\ the transformation $T_j$ --- the braiding
action taking the $j$-th vacuum in the anti-clockwise
direction replacing the $j+1$-th vacuum (see sects. 2 and 3).
We know that
this transformation can be realized {\it via} a deformation
of the couplings $w_k$
as all the perturbations are relevant and thus
allowed. Finally we can cyclically permute the $\vec e_i$'s or
take them in the inverse order.

In this way many solutions are reduced to the standard ones.
The corresponding N=2 model is known to be realizable as a
Landau--Ginzburg
model \landgins. From these explicit LG realizations, we see that
their UV
limits are just the corresponding minimal models. Since in minimal
models all
formal perturbations are physically allowed,  the
solution obtained by braiding
the standard ones can be realized as a suitable perturbation
of the
corresponding minimal model\foot{Of course, this is just the usual
description
of the deformations of a minimal singularity \singt .}.
 Then these solutions  are
just (massive perturbations of) A--D--E minimal models.
For example the $A$-series can be realized as Chebyshev perturbations
of the $x^n$ minimal model \fenint .

However not all solutions with $B$ positive
 are equivalent to the standard
ones. This reflects the fact that
the notion of irreducibility for the field algebra of a QFT
is a much stronger constraint than the analog notion
for a reflection group.
Consider e.g. the $4\times 4$ matrix
\eqn\badguy{S=\left(\matrix{1 & \sigma_1
+\sigma_2\cr 0 & 1\cr}\right)}
where $\sigma_i$ are the Pauli matrices.
The corresponding $\vec e_i$ generate the $D_4$ root lattice.
However its monodromy $H$ satisfies
\eqn\wrongmon{\det[z-H]=\big(z^2+1\big)^2, }
and thus has nothing to do
with the charges of the $D_4$ minimal model.
In fact, $-H^t=R_1R_2R_3R_4$, whereas the Coxeter element (for this
unorthodox choice of roots) reads
 $R_3R_1R_4R_2$, and these two elements
are not conjugate in $W$. Now the point is that \wrongmon\ cannot
correspond to an {\it irreducible}  regular critical theory.
There are three good physical reasons to discard the
solution \badguy: {\it i)} the minimal value of $q$ is doubly
degenerate
(that is $1$ is not the only chiral primary with $q=0$).
Then \badguy\ is  reducible. {\it ii)} the
solution has $\tr\, H=0$; this cannot
be for an irreducible theory. Indeed the N=2 superconformal algebra
together with modular invariance shows\foot{In fact comparing with
ref.\lgorv\ we see that $\tr\, H=\Tr(-1)^F g$,
where $g=\exp[2\pi i J_0]$. The modular transformation
$\tau\rightarrow -1/\tau$ transforms the character--valued
susy index $\Tr(-1)^Fg$
 into the Witten index for the
sector twisted by $g$. The ground states in this sector are the
harmonic
representatives of a certain cohomology (the group $H(1,0)$ in the
notation of ref.\lgorv). Chiral primaries of minimal $U(1)$ charge
are always non--trivial elements of $H(1,0)$ \lgorv.
Moreover for $\hat c<1$ (at least) positivity implies that these
chiral
primaries exhausts $H(1,0)$.}
 that
$\tr\, H$ is the susy index
 counting {\it with signs} the number of chiral primaries
with $q=0$. The requirement that the only such object is the standard
identity operator gives\foot{Notice that this restriction is
automatic
in LG models. This is an easy consequence of the results of
ref.\lgorv\
as well as a known fact from singularity theory \singt.}
\eqn\goodconst{\tr\, H=1.}
{\it iii)} the solution to the \ttstar\ equations defined by
the Stokes matrix \badguy\ cannot be {\it both}
regular for all $w_k$'s
and irreducible. Indeed, let us send
${\rm Im}\,w_3$, ${\rm Im}\,w_4\rightarrow\infty$
while keeping ${\rm Im}\, w_1={\rm Im}\,w_2=0$. We end up
with a model with just two vacua and {\it no solitons}.
Again this is a reducible situation.
For a general model this argument shows that, in order to
have irreducibility, we need\foot{At
first sight, it may seem that this
argument imply $B_{ij}\not=0$
for all $i$, $j$. However it is not
so since in the limit ${\rm Im}\,w_k\rightarrow\infty$
(for $k\not=i$, $j$) we still get contributions
to the soliton number
$\mu_{ij}$ from the `vacua at infinity' because
the singularity
mechanism discussed in sect.4 spoils the naive
decoupling. Instead  $-\mu_{i,i+1}=B_{i,i+1}$
because of the bound \wbound\ and thus for these particular entries
the argument is correct.}
\eqn\bors{B_{i,i+1}\not=0.}
 Without loss
of generality we can take $B_{i,i+1}=-1$.
This condition is not fulfilled by \badguy.
The same reasoning shows that, if we have $n$ vacua
and take the limit ${\rm Im}\,w_n\rightarrow\infty$,
the `reduced' Stokes matrix obtained by deleting the last
row and the last column should be such as to correspond
to a regular irreducible solution of \ttstar.

Thus we have three necessary criteria for irreducibility.
Now we give an argument to the effect that if $B$ is positive
definite {\it and} satisfies these criteria\foot{In fact
we can forgot about {\it i)} since it is a consequence of the
other two.} then it is equivalent to a standard solution.
The idea is to argue by induction on the number of vacua.
Assume we know that  this is true for $n$ vacua.
Then in the $n+1$
vacuum case we can use criterion
{\it iii)} to put the `reduced' $S$ in a
standard form for the given
`reduced' root system\foot{Of course this
process  screws up the elements $S_{i,n+1}$.}.
Having fixed the `reduced' $S$ in a standard form
 we are reduced to a much
simpler Diophantine problem for the unknowns $a_i=S_{i,n+1}$.
By \bors\ $a_n=-1$, so we have just $n-1$
unknowns which can only take the values $0$ and $\pm 1$.
At this stage we impose the index
restriction $\tr\, H=1$ which greatly reduces the
allowed values for the $a_i$'s (and kills the
`spurious' solutions like \badguy). Finally one shows that the few
surviving possibilities
either lead to
a non positive\foot{Generically they are not even solutions
to our problem. However some
are `affine' solutions, see next subsection.} $B$,
or they are equivalent to standard solutions.
To illustrate this last step of process
we consider the simpler case in
which the `reduced' $S$ is associated to the $A_n$ root
system. In this case one finds
$$\tr\, H=1+\sum_i\big(a_i-a_i^2)-\sum_{i<j}a_ia_j.$$
Let $m$ ($0\leq m\leq n-1$) be the number of $a_i$'s
with $|a_i|=1$ and $r$ ($0\leq r\leq m$) the number of
those with $a_i=1$. Then $\tr\, H$ becomes
$$1-(m-r)-\half\big[(m-2r)^2+(m-2r)\big].$$
The last two terms in this
expression are non--positive. Hence $\tr\, H=1$ iff they
both vanishes. This happens in just two cases: {\it i)}
all $a_i=0$; or {\it ii)} one $a_i=1$ while all the other vanish.
So we have only $n$ cases to check.
The case $a_i=0$ for all $i$'s and the one
with $a_{n-1}=1$ give the standard $A_{n+1}$ solution.
The cases $a_1=1$ or $a_{n-2}=1$ give the usual
$D_{n+1}$ solution. Finally the cases $a_2=1$
or $a_{n-3}=1$ give the $E_{n+1}$ `solution' (it is a $\hat c<1$
model only if $(n+1)\leq 8$; for $(n+1)<6$ it coincides with
a $D$ solution, as one sees from the corresponding
Stokes matrix). The cases with
$a_i=1$ for an $i$ in the range $2<i<n-3$ have never $B$
positive definite. If the `reduced' $S$ is of the $D_n$ or $E_n$ type
the analysis is similar although more involved.

It remains to show that the A--D--E
series exhausts the models with $\hat c<1$,
\ie\ that $\hat c<1$ implies $B$ positive definite.  One argument
was mentioned before, i.e., the fact that for the minimal
model $|q|<1/2$ implies this.
 We will now give another argument:
In a model with $\hat c<1$ all chiral primaries are relevant, and
hence
all deformations of the theory lead to regular solutions of \ttstar.
 Moreover, the property
that all UV charges are less than $1$ should survive
 perturbation.

If $\hat c<1$ then all Ramond $U(1)$ charges satisfy $|q|<\half$.
Hence
$-1$ is not an eigenvalue of the monodromy $H=S(S^{-1})^t$. Then
\eqn\nonzero{0\not=\det[-1-H]=(-1)^n\det[S^t+S].}
Assume that a model with $\hat c<1$ has $B$ non--positive definite.
By \nonzero, $B$ has (at least) a negative eigenvalue. Consider the
pseudo--reflection group $W$ generated by the corresponding $R_k$'s
\refl. Repeating word--for--word the previous
argument, we see that the elements
of this group can be  realized as
{\it formal} perturbations of the model. But in a minimal model all
formal
perturbations should
be good deformations of the theory. So all the
Stokes matrices generated by these reflections should
correspond to regular solutions of \ttstar\ for all $w_k$'s. But now
$W$ is infinite \bourbaki\ and hence in the equivalence
class we can find arbitrarily big Stokes parameters
$A_{ij}$.
But this is absurd. Indeed the solution of \ttstar\ cannot be
regular for all $w_k$'s for $A_{ij}$ very large as we can see by
considering suitable geometries in $W$--space and taking some vacua
to infinity. Then minimality requires $B$ to be positive.

\subsec{`Affine' Models and Their Lagrangian Realizations}

\vglue 6pt
\noindent\underbar{\raise 3pt\hbox{\it $\hat c=1$ Degenerate Models}}
\vglue 2pt

In our context the A--D--E classification has a natural `affine'
generalization. The main purpose of this subsection is
to provide explicit examples of such
 `affine' models. We make no attempt at completeness.

Suppose our (integral) Stokes matrix $S$ is such that the
associated symmetric form $B=S+S^t$ is {\it singular} (i.e. $\det
B=0$),
while all its (proper) principal minors are {\it positive definite}.
We claim that any such $S$ is
also a solution to our Diophantine problem.
Indeed, since $B$ is singular, there is a vector $v$ such that
$Bv=0$.
This vector is unique (up to normalization) since, by assumption,
${\rm rank}\, B=n-1$. In particular $v$ is {\it real}. Then $v$ is
the
{\it unique} eigenvector of $H^t=S^{-1}S^t$ associated to the
eigenvalue
$\lambda_0=-1$; indeed,
\eqn\usone{H^tv=S^{-1}(B-S)v=-v.}
Consider next the quadratic form $\widehat{B}$ induced by $B$ on the
quotient space $\BR^n/\BR v$. By assumption, $\widehat{B}$ is
positive
definite. From the argument of \S.7.2, we see that the induced
monodromy
$\widehat{H^t}$ acts orthogonally on  $\BR^n/\BR v$.
This, together with \usone, shows that the eigenvalues of $H^t$
satisfy $|\lambda|=1$, as claimed. However the Jordan structure of
$H^t$ is non--trivial. In fact $H^t$ has just one eigenvector
associated to
this eigenvalue, whereas $(-1)$ is a root of the characteristic
polynomial
of {\it even} multiplicity\foot{Because $\det H=1$, and if $\lambda$
is
an eigenvalue of $H$ so is $\lambda^{-1}$.}. Since $\widehat{H^t}$ is
simple,
$H^t$ has just a $2\times 2$ block associated to the eigenvalue
$(-1)$.

{}From our general discussion in sect.5  we know that these solutions
lead to {\it degenerate} superconformal
models with $\hat c_{\rm uv}=1$. Conversely all such degenerate
models are
associated to an $S$ having the above properties.
If we assume that the off--diagonal entries of $S$ are non--positive,
then (up to
permutations)
$B$ is nothing else than the Cartan matrix for a simply laced affine
Lie algebra, i.e. $\widehat{A_{n-1}}$, $\widehat{D_{n-1}}$,
or $\widehat{E_6}$, $\widehat{E_7}$, and $\widehat{E_8}$.
The general solution is obtained as follows. We consider a
(finite) simply laced root system of rank $(n-1)$ and take $n$ roots
$\vec e_i$ ($i=0,1,\dots, n-1$) such that $\vec e_i$
($i=1,\dots,n-1$)
span $\BR^{n-1}$ whereas
$$\vec e_0=\sum_{i=1}^{n-1}k_i \vec e_i,\qquad k_i\ \rm
non-vanishing\ integers.$$
Then
$$B_{ij}=2( \vec e_i,\vec e_j)\qquad i,j=0,\dots,n-1.$$
The Cartan matrix corresponds to the special solution with
$\vec e_i$ ($i\not=0$) the simple roots and $\vec e_0$ {\it minus}
the highest root.
Many  solutions can be obtained one from the other by `formal'
perturbations (\ie\ braiding transformations). Since the chiral
fields
$\phi_i$ are either soft perturbations or asymptotically free
renormalizable interactions, we expect that all the `formal'
perturbations make perfect physical sense.

\vglue 6pt
\noindent\underbar{\raise 3pt\hbox{\it LG Models with
Exponential Interactions}}
\vglue 2pt

In sect.6 we saw that the $\hat A_1$  model can be
realized as the N=2 Sinh--Gordon  provided we make
the identification $X\sim X+2\pi i$. Other `affine' models are
obtained by
changing this identification to
$$X\sim X+2\pi n i.$$
In this way we get a $\hat A_{2n-1}$ model.
 The easiest way to see this is to solve the corresponding \ttstar\
equations in terms of Painlev\'e transcendents. As in Ref.\topatop\
 we introduce the
transformation $T$ which shifts $X$ by $2\pi i$. One has $T^n=1$.
Then we consider the $\theta$--vacua, i.e. the ground states such
that
$$T|a,\theta\rangle= e^{i\theta}|a,\theta\rangle,\qquad
a=1,2.$$
For a fixed value of $\theta$ there are just two ground states, and
hence
the \ttstar\ equations take the PIII form (see ref.\topatop\ for
details).
Then the ground state metric
 is (here $z=m\beta$, with $m$ the mass of the basic
soliton)
$$g(z,\theta)={4\over z}\exp[i\theta\sigma_3/4]
\exp[\sigma_1 L(z,\theta)]\exp[-i\theta\sigma_3/4],$$
where $L(z,\theta)$ is the regular solution to PIII with
$$r(\theta)=2\left(1-{\theta\over \pi}\right),
\qquad (0\leq \theta<2\pi).$$
Returning to the canonical basis, the ground state metric becomes
(here
$|k\pi\rangle$ denotes the canonical vacuum associated to the
critical
point $X_k=k\pi i$)
\eqn\sineorbifold{\eqalign{\langle\overline{k\pi}&|j\pi\rangle=\cr
&={i^{(j-k)}\over 2n}\sum_{s=0}^{n-1} e^{i\pi(j-k)s/n}
\left[e^{L(z,2\pi s/n)}+(-1)^{(j-k)} e^{-L(z,2\pi s/n)}\right]\cr
&\qquad {\rm where}\ (k,j=0,1,\dots, 2n-1).\cr} }
Since as $z\rightarrow \infty$,
$$\exp[L(z,\theta)]\simeq 1-2\cos(\theta/2) {1\over \pi}K_0(z),$$
we have the IR asymptotics,
$$\langle\overline{k\pi}|j\pi\rangle\simeq \delta_{i,j}^{(2n)}-
i^{(j-k)}\Big(\delta_{k,j+1}^{(2n)}
+\delta_{j,k+1}^{(2n)}\Big){K_0(z)\over
\pi},$$
where
$$\delta^{(m)}_{i,j}=\cases{1\quad {\rm if}\ i\equiv j\ {\rm mod}.\,
 m\cr
0\qquad \rm otherwise.\cr}$$
Then the soliton matrix $\mu_{ij}$ reads
\eqn\muorbifold{\mu_{jk}= i^{(j-k-1)} \Big(\delta_{k,j+1}^{(2n)}
+\delta_{j,k+1}^{(2n)}\Big).  }
The $U(1)$ charges in a given $\theta$--sector are equal to
$\pm r(\theta)/4$ (see \topatop). Thus
\eqn\uvCharges{\{{\rm UV}\ U(1)\ {\rm charges}\}
=\pm {s\over n}\mp {1\over 2}\qquad
(s=0,1,\dots, n-1). }
In the present case there are just two critical values, and so
we have $S=1-A$ with
$$A_{ij}=\cases{\mu_{ij}\quad {\rm if}\ i\ \rm even\cr
0\qquad \rm otherwise\cr}$$
After a relabeling of the basis (and a suitable sign redefinition)
the Stokes matrix reads
$$S=\left(\matrix{ 1 & 1+R\cr
0 & 1\cr}\right),$$
where the $n\times n$ matrix $R$ corresponds to a
cyclic permutation. In particular, $R^n=1$, and $R^t=R^{-1}$.
Then the characteristic polynomial of $H$ is
$$\eqalign{P(-z)&=\det[z S^t+S]=\det[(z+1)^2-z (1+R^{-1})
(1+R)]\cr &=\big[\det(z-R)\big]^2= (z^n-1)^2,\cr}$$
in agreement with \uvCharges.

In sect.6 we also saw that the N=2 Bullough--Dodd model, \ie\ the LG
model with superpotential
$$W(X)={2t\over 3}\Big(e^X+\half e^{-2X}\Big),$$
and field identification $X\sim X+2\pi i$,
 leads to a model related to the
$\widehat{A_2}$ root system. Our general discussion above implies
that
the models obtained by the more general identification $X\sim X+2\pi
ni$
also correspond to degenerate $\hat c=1$ theories. We
expect that the corresponding solution to our Diophantine problem
is related to the $\widehat{A_{3n-1}}$ root system.
However this time it is
not possible to check this expectation by writing explicitly
the ground--state metric in
terms of known transcendents. Luckily there is one special case,
namely $n=2$, in which $g$ can be still written in terms of
Painlev\'e
transcendents.

We define the ground states
$$|m\rangle=\sum_{k=0}^5 e^{\pi i m k/3}\left|{2\pi k\over
3}\right\rangle,$$
where $|2\pi k/3\rangle$ is the `point basis' vacuum at $X_k=2\pi
k/3$
($k=0,1,\dots, 5$). Then the ${\bf Z}_6$--symmetry implies
$\langle \bar m|l\rangle=0$ for $m\not=l$. Using the reality
constraint
the \ttstar\ equations decompose into two decoupled degenerate PIIIs.
Then the ground state metric reads
$$\eqalign{& 2|t| \langle\bar 2|2\rangle= 2|t|\langle\bar
5|5\rangle=1\cr
& 2|t|\langle \bar 4|4\rangle= \big( 2|t|\langle\bar
0|0\rangle\big)^{-1}=
e^{u_1(\tau)}\cr
& 2|t|\langle \bar 1|1\rangle= \big( 2|t|\langle\bar
3|3\rangle\big)^{-1}=
e^{u_2(\tau)},\cr}$$
where $u_i(\tau)$ are (regular) solutions to eq.\Pdeg.
In terms of the canonical basis, the ground state metric reads
$$\eqalign{g_{k\bar j}&= {1\over 6}\Big[1+(-1)^{(k-j)}\Big]+
{1\over 6}\Big[e^{\pi i(k-j)/3} e^{-u_1}+ e^{-\pi i(k-j)/3}
e^{u_1}\Big]+\cr
&\qquad + {1\over 6}\Big[e^{2\pi i(k-j)/3} e^{u_2}+ e^{-2\pi
i(k-j)/3}
e^{u_2}\Big].\cr}$$
Then the soliton matrix is
$$\mu_{kj}={1\over \sqrt{3}} s_1 \sin\left[{\pi\over 3}(k-j)\right]
-{1\over \sqrt{3}} s_2 \sin\left[{2\pi\over 3}(k-j)\right],$$
where $s_i$ are the Stokes parameters specifying the boundary
conditions
for $u_i(\tau)$ (see \S.6.2). Consistency with the $m=1$ case (which
can be
identified with a subsector of the present model) fixes $s_2=-1$.
Then the requirement that $\mu_{ij}$ are integers implies
$$s_1=1\quad {\rm mod}.\, 2.$$
Using regularity (\S.6.2) we get $s_1=-1$, $1$, or $3$.
But
$3$ is not possible because it gives $\hat c$ too big. On the other
hand, $-1$
is ruled out on the basis of the uniqueness of the chiral field with
the smallest $U(1)$ charge. Thus $s_1=1$.
Then
$$\mu_{ij}=\delta_{i,j+1}^{(6)}-\delta_{i,j-1}^{(6)},$$
which is the expected result. (Notice that, although the N=2
soliton multiplicities $|\mu_{ij}|$ are the same as in the
Sinh--Gordon with $n=3$, the signs assignments are physically
inequivalent).
The UV charges are easily computed from the above explicit solutions.
They are
$$(-\half, -\textstyle{1\over 4}, 0,0,\textstyle{1\over 4}, \half).$$
Needless to say, this is in agreement with the values obtained from
the monodromy $H$.

On general grounds one expects that similar phenomena also happens
for more general exponential interactions, in particular the ${\bf
Z}_n$
invariant one
\eqn\expzn{W(X)= e^X+\textstyle{1\over n-1} e^{-(n-1)X}.}
However explicit computations are not as elementary as in the above
cases.

\vglue 6pt
\noindent\underbar{\raise 3pt\hbox{\it ${\bf C}P^1$ Orbifolds}}
\vglue 2pt

We saw in \S 6.1 that $\widehat{A_1}$ model has a
second Lagrangian realization\foot{The
structure of the exact $S$--matrices for the ${\bf C}P^1$
sigma model and the N=2 sine--Gordon suggests \fenint\ that
these two models are equivalent as QFT's for
a special choice of the D--terms.}
 as the $\sigma$--model with target space
${\bf C}P^1$. Then our general arguments show that any
(sensible) orbifold of this sigma model should  also be
a {\it degenerate} $\hat c=1$ and hence related to
a simply laced root system as above.
It is tempting to make the following conjecture on the nature of
this correspondence.
An orbifold is obtained by modding out a discrete subgroup $G$ of
the (double cover of) the $\BC P^1$
isometry group $SU(2)$. As is well known, these subgroups
are again classified by A--D--E. Then it is natural to expect that
the $G$ orbifold is related to the root system associated to the
subgroup $G$. More precisely,
the correspondence between  a subgroup and a
root system is obtained by considering the eigenvectors of the
of the $\widehat{A}$--$\widehat{D}$--$\widehat{E}$
Cartan matrices: They are the columns of the character
table for the corresponding group $G$ \ref\duval{J. McKay,
{\it Cartan matrices, Finite groups of quaternions, and
Klenian singularities}, Proc. Am. Math. Soc. (1981) 153.}.
In particular,
\eqn\correspondence{\widehat{A_r}\ \leftrightarrow\ {\bf Z}_{r+1}.}
It is also natural to expect that not all
subgroups can appear, since
the center ${\bf Z}_2$
of $SU(2)$ must belong to $G$
because the physical states are automatically invariant
under this subgroup. (That is the original $\sigma$--model, having
the isometry group $SO(3)=SU(2)/{\bf Z}_2$, corresponds to $G={\bf
Z}_2$
not $G=1$).

That these massive orbifolds are {\it bona fide} quantum field
theories
was shown in ref.\sigmamodels. There the special case $G={\bf
Z}_{2n}$
(corresponding to the orbifold $\BC P^1/\BZ_n$)
was studied in great detail,
and the corresponding \ttstar\
 equations were explicitly solved in terms of
PIII transcendents. The result of \sigmamodels\
 implies that
the ground state metric for the $\BZ_n$--orbifold
 is just that for the
N=2 Sinh--Gordon with the identification\foot{Physically this
is due to the fact that (for a special choice of the D--terms)
both models are orbifolds of the {\it same} QFT.} $X\sim X+2\pi ni$,
see eq.\sineorbifold.
Again one can compute the soliton spectrum out of this \ttstar\
solution.
The computations are a word--for--word
repetition of those leading to eq.\muorbifold.
Then eq.\muorbifold\
gives the mass--spectrum of the ${\bf C}P^1/{\bf Z}_n$
orbifold model. In agreement with the guess \correspondence\
$\mu_{ij}$, as a solution to our classification program, is indeed
related to the $\widehat{A_{2n-1}}$ root system (e.g.
$2\delta_{ij}-|\mu_{ij}|$
is the $\widehat{A_{2n-1}}$ Cartan matrix).  Non-abelian
orbifolds of ${\bf C}P^1$ have been recently considered
in \ref\Ezas{E. Zaslow,{\it Topological Orbifold Models and
Quantum Cohomology Rings}, Harvard preprint HUTP-92/A069.}\
with results in the direction of connecting them to affine $D$ and
$E$ series.

\newsec{More on Sigma Models}

The primary aim of this section is to supply examples which
cannot be realized as LG models. Here we focus
mainly on $\sigma$--models over symmetric spaces (with $c_1>0$).
The main issue is to compute their (solitonic) mass spectra as we did
in \S 7.2 for the ${\bf C}P^1$ orbifolds.
Typically these $\sigma$--models are confining theories
whose physics is quite similar to that of $4d$ gauge theories.
So the possibility of
getting (part of) their exact mass spectrum
by back--of--an--envelope computation is
a very dramatic consequence of our methods.

It is more convenient to start with a
more general problem \ie\ the
 classification of the massive models with
 a $\BZ_n$--symmetry
acting transitively on the $n$ ground states.
Indeed,
besides the cases of small $n$ and small $\hat c$, there is a third
situation in which a complete
classification is possible, namely in
presence of a `big'
symmetry: One looks for all
models having a symmetry group $\CG$ which is
big enough to restrict the $\mu_{ij}$'s
in a significant way. In Ref.\newindex\
it was shown that the $\BZ_n$ invariant models organize themselves
in a `family' and that it is somewhat easier to study all the
models in  the family at once than one at the time.

\subsec{The Classification of ${\bf Z}_n$--Invariant Models}

If our system has a cyclic symmetry, then the matrix
$C$ should
transform according to some irreducible representation of $\BZ_n$
\ie\ like $\zeta_r(a)=\exp[2\pi ia r/n]$ for some $r\in\BZ_n$.
We set $m=n/(n,r)$. Then, without loss of generality, we can
assume that the $m$ distinct
`critical values' $w_k$ are given by
\eqn\critvalues{w_k=\exp[2\pi i k/m]\qquad k=0,\dots, m-1.}
All these values have multiplicity $(n,r)$.
For convenience, we label the vacua with two
indices $(k,a)$ ($k=0,1,\dots, m-1$, $a=1,2,\dots, (n,r)$)
in such a way that $k+m a$ is increased by $1$ under a basic
 ${\bf Z}_n$ rotation while $k$ labels the corresponding
critical value as in \critvalues. In this basis $S$ will not be
upper triangular.

The $\BZ_n$ symmetry restricts the $(n,r)\times
(n,r)$ matrices $(\mu_{kh})_{ab}$.
First of all, one has
$$\mu_{ij}=\mu(i-j)\qquad {\rm with}\qquad \mu(k+m)=J\mu(k),$$
where the $(n,r)\times (n,r)$ matrix $J$ is
$$J_{ab}=\delta_{a,b-1}+\varepsilon \delta_{a,1}\delta_{b,(n,r)}.$$
$\varepsilon$ is fixed by the following identity
$$\mu(k+n)=J^{(n,r)}\mu(k)=\varepsilon \mu(k).$$
At first sight it may seem that this together with
the $\BZ_n$ symmetry predicts
$\varepsilon=1$. However it is not so. The point is that
 the canonical basis is well defined {\it up to sign}, and it is
quite possible that acting $n$ times with the basic $\BZ_n$
transformation we end up to with the {\it opposite} sign. In this
case $\varepsilon=-1$ (in
fact most models work this way).  The sign
assignments for the canonical vacua are specified by
the phases of the topological metric
 $\eta$. Again
$\eta$ should belong to a definite
representation of $\BZ_n$. If it transforms as
$\eta\mapsto e^{-2\pi i q/n}\eta$ then $\varepsilon=(-1)^q$.
Thus $J^{(n,r)}=(-1)^q$.
On the other hand $\mu$ is antisymmetric, so $\mu(k)=-\mu(-k)^t$,
or
\eqn\mures{\mu(m-k)=-J\mu(k)^t.}
Finally, the symmetry implies that $\mu(k)$ commutes with $J$.
But $J$ is non--derogatory and thus
$$\mu(k)=\sum_{l=1}^{(n,r)} \varrho(k,l)J^l,$$
for some (integral) coefficients $\varrho(k,l)$.
In view of \mures\ we have
$$\varrho(m-k,l)=-\varrho(k,(n,r)-l+1),$$
(we used that $J^t=J^{-1}$).

The angles $\phi_{kh}$ of sect.4  are given by
$$e^{i\phi_{kh}}=\cases{-i\exp[\pi i (k+h)/m]\quad {\rm for}\ k>h\cr
+i\exp[\pi i (k+h)/m]\quad {\rm for}\ k<h.\cr}$$
The (reflected) soliton rays belonging to the right half--plane are
at angles
$$\psi_s={\pi\over 2}-{s\over m}\qquad s=0,\dots,m-1.$$
As we have discussed in \S 4.5 the matrices
$\mu^{[\cdot]}$ associated with a given ray
commute. Let $\mu^{\langle s\rangle}$ be the sum of all the matrices
 associated with the $s$--th ray. From the above formulae
one gets ($i,j=0,\dots,m-1$)
$$\Big(\mu^{\langle
s\rangle}\Big)_{ia,jb}=\delta_{i+j,m-s}\lambda_{ia,jb}-
\delta_{i+j,2m-s}\lambda_{jb,ia},$$
where
$$\lambda_{ia,jb}=\cases{\big(\mu_{ij}\big)_{ab}\quad {\rm if}\
i>j\cr
0\qquad \rm otherwise.\cr}$$
{}From the definition it is easy to get the identity
\eqn\strangeid{\mu^{\langle s+1\rangle} \mu^{\langle s\rangle}=0,}
then the Stokes matrix reads
$$\eqalign{S&=(1-\mu^{\langle 2l+1\rangle}-\mu^{\langle 2l\rangle})
(1-\mu^{\langle 2l-1\rangle}-\mu^{\langle 2l-2\rangle})\cdots
(1-\mu^{\langle 1\rangle}-\mu^{\langle 0\rangle})\quad
m=2l+2\cr
&=(1-\mu^{\langle 2l+2\rangle})
(1-\mu^{\langle 2l+1\rangle}-\mu^{\langle 2l\rangle}) \cdots\cdots
\cdots\cdots
(1-\mu^{\langle 1\rangle}-\mu^{\langle 0\rangle})\quad m=2l+3.\cr}$$
Let $R$ be the orthogonal $n\times n$ matrix
$$R=\left(\matrix{
0 & 0 & 0 & \cdots & 0 & J^{-1}\cr
1 & 0 & 0 & \cdots & 0 & 0\cr
0 & 1 & 0 & \cdots & 0 & 0\cr
0 & 0 & 1 & \cdots & 0 & 0\cr
.. & .. & .. & \cdots & .. & ..\cr
.. & .. & .. & \cdots & .. & ..\cr
0 & 0 & 0 & \cdots & 1 & 0\cr}\right),$$
which satisfies $R^m={\bf 1}\otimes J^{-1}$.
Then the cyclic symmetry of $\lambda_{ij}$ yields
$$\mu^{\langle s+2\rangle}= R^{-1}\mu^{\langle s\rangle}R.$$
So, if (say) $m=2l+2$
$$S= R^{-l}[(1-M)R]^{l+1}R^{-1},$$
where $M=\mu^{\langle 1\rangle}+\mu^{\langle 0\rangle}$.
This procedure can be continued through the other half--plane.
Finally we get for the monodromy
$$H=R^{-(m-1)}[(1-M)R]^mR^{-1}\equiv ({\bf 1}\otimes
J) R [(1-M)R]^m R^{-1}.$$
So, up to similarity, $({\bf 1}\otimes J^{-1})H$ is just
$[(1-M)R]^m$.
Since $({\bf 1}\otimes J^{-1})$ commutes with $H$, the monodromy
 eigenvalues
are phases  iff the matrix $(1-M)R$ satisfies the same condition,
\ie\ if $\det[z-(1-M)R]$ is a product of cyclotomic polynomials.

Let us assume for the moment that $(n,r)=1$, so $n=m$. In this case
$J=(-1)^q$. Then if
$$\CQ(z)\equiv \det[z-(1-M)R]=\prod_{s\in \BN} \Phi_s(z)^{\nu(s)},$$
the characteristic polynomial $P(z)$ of $H$ reads
\eqn\pqz{P\big((-1)^q
z\big)=\prod_{s\in\BN}\Big[\Phi_{s/(n,s)}(z)\Big]^{\nu(s)
\alpha_n(s)},}
where
$$\alpha_n(s)=\phi(s)\left/\phi\left({s\over (s,n)}\right)\right. .$$
One finds
\eqn\pozz{\eqalign{\CQ(z)&\equiv\det[z-(1-M)R]=\cr
&= z^n+(-1)^{q+1}+\sum_{k=1}^{(n-1)/2}\mu(k)
\big(z^{n-k}+(-1)^{q+1}z^k\big)
\quad (n\ {\rm odd})\cr
&= z^n+(-1)^{q+1}+(-1)^{q+1}
\mu({\textstyle{n\over 2}})z^{n/2}+
\sum_{k=1}^{n/2-1}\mu(k) \big(z^{n-k}+(-1)^{q+1}z^k\big)
 (n\ {\rm even})\cr} }
\ie\ the coefficients of the polynomial $\CQ(z)$ are precisely the
soliton
numbers $\mu(k)\equiv \mu_{i,i+k}$ (with signs as specified by
eq.\pozz).
Then for, say, $q$ odd the solution of our Diophantine problem take a
very
elegant form: {\it a set of soliton numbers $\mu_{i,j}$ is a
$\BZ_n$--symmetric solution of our problem if and only if the
polynomial}
\eqn\soldio{z^n+1+\sum_k \mu_{i,i+k}\, z^k, }
{\it is a product of cyclotomic polynomials.}
In particular we have the bound
\eqn\mubound{|\mu_{i,i+k}|\leq {n\choose k}.}
Let us give a few simple examples.

1. The basic example is the perturbed
$A_n$ minimal model, \ie\ the LG theory with superpotential
$W(X)=X^{n+1}-t X$. In this case $q=-1$ and the soliton numbers
$\mu(k)$ are all equal to $1$. Then
$$\CQ(z)={z^{n+1}-1\over z-1}=\prod_{d|(n+1)\atop
d\not=1}\Phi_d(z).$$
Using the rule \pqz\
$$P(z)=\prod_{d|(n+1)\atop d\not=1}\Phi_d(-z),$$
which gives the usual result for the UV charges of the $A_n$ model.

2. A second example is the Ising $n$--point function\foot{For a
discussion
of the associated `hyperelliptic' LG models see ref.\noisi.}
with the spins located at the vertices of a regular $n$--gon. In this
case
the
soliton numbers $\mu(k)$ are equal $2$ for all $k$'s. Then
$$\CQ(z)=(z+1){(z^n-1)\over (z-1)}=\Phi_2(z)\prod_{d|n\atop
d\not=1}\Phi_d(z),$$
and \pqz\ gives
$$P(z)=\cases{\Phi_1(-z)^n \hskip 3.1cm {\rm for}\ n\ {\rm even}\cr
\Phi_1(-z)^{(n-1)}\Phi_2(-z) \hskip 1.4cm {\rm for}\ n\ {\rm
odd}.\cr}$$
Needless to say, the corresponding $U(1)$ charges agree with the
physical
picture of the Ising correlators. These two examples (and the trivial
case
$\mu(k)=0$) exhaust the solutions with all soliton numbers equal for
$n\geq
4$.

3. $\mu(k)$ is $1$ (resp. $-1$) for $k=1$ and $0$ otherwise. Then
$$\eqalign{\CQ(z)&=(z+1)(z^{n-1}+1)=\Phi_2(z)\prod_{d|(n-1)}\Phi_{2d}(
z)
\qquad \big(\mu(1)=+1\big)\cr
&=(z-1)(z^{n-1}-1)=\Phi_1(z)\prod_{d|(n-1)}\Phi_{d}(z)
\hskip 1.1cm \big(\mu(1)=-1\big).\cr}$$
Then
$$\eqalign{P\big((-1)^q z\big)&=\Phi_1(z)
\prod_{d|(n-1)}\Phi_d(z)\qquad
\mu(1)=-1\ {\rm or}\ n\ {\rm even}\cr
&=\Phi_2(z) \prod_{d|(n-1)}\Phi_{2d}(z)\qquad
{\rm otherwise}.\cr}$$
The case in the first line (and $q$ odd) leads to the charges
$$q_k={k\over n-1}-{1\over 2}\qquad (k=0,1,\dots, n-1).$$
It is conceivable that these solutions correspond to the models in
 \expzn.
This is the case for $n=2$, $3$.

4. This last example can be generalized to $\mu(k)=+1$ (resp. $-1$)
for $k=k_0$ and zero otherwise. Then
$$\eqalign{\CQ(z)&=\prod_{d|k_0}\Phi_{2d}(z)\prod_{l|(n-k_0)}\Phi_{2l}
(z)
\hskip 1.1cm \big(\mu(k_0)=+1\big)\cr
&=\prod_{d|k_0}\Phi_d(z)\prod_{l|(n-1)}\Phi_{l}(z)
\hskip 1.5cm \big(\mu(k_0)=-1\big).\cr}$$

Let us return to the general case, \ie\ $(n,r)\not=1$. In each
eigenspace
for $J$ the situation is exactly as before. Then, in a sector in
which
$J$ acts by multiplication, the eigenvalues $\lambda$ of $H$ are
\eqn\wlam{\lambda=J z^m,}
where $z$ is a solution to
\eqn\genca{\eqalign{&
z^m-J^{-1}+\sum_{k=1}^{(m-1)/2}\sum_{l=1}^{(n,r)}
\varrho(k,l) J^l \Big(z^{m-k}- J^{-1} z^k\Big)=0\hskip 1.1cm (m\ {\rm
odd})\cr
&z^m-J^{-1}-\sum_{l=1}^{(n,r)}\varrho({\textstyle{m\over 2}},l)
J^{l-1}
z^{m/2}+\cr
&\hskip 2cm +\sum_{k=1}^{m/2-1}\sum_{l=1}^{(n,r)}
\varrho(k,l) J^l \Big(z^{m-k}- J^{-1} z^k\Big)=0\qquad (m\ {\rm
even})\cr}}
and a set $\varrho(k,l)\equiv \mu_{i,i+k+m l}$\
 of soliton numbers gives a
solution to our Diophantine problem iff the roots of the $(n,r)$
polynomials obtained from \genca\ by replacing $J$ with its
eigenvalues
\eqn\Jei{\exp\left[{2\pi i\over (n,r)}k+{\pi i\over
(n,k)}q\right]\qquad
\big(k=1,2,\dots, (n,r)\big),}
are phases. As an example, take the $\BC P^1/\BZ_h$ orbifolds in
\S.7.2.
There $n=2h$, $r=h$, $m=2$, and $q=-h$. Moreover,
$$\mu(1)={\bf 1}-J.$$
Then \genca\ becomes
$$(z-J^{-1})(z+{\bf 1})=0,$$
so $z^2=J^{-2}$ {\it or} ${\bf 1}$. Then eq.\wlam\
gives
$$\lambda=\cases{J^{-1}\cr J,\cr}$$
which, in view of \Jei, is what we got by a direct computation in
\uvCharges.

\subsec{The $\BC P^{n-1}$ $\sigma$--Model}

The $\BC P^{n-1}$ $\sigma$--model has
 Witten index $n$ as it is obvious from its
Hodge diamond $h^{p,q}=\delta_{p,q}$ ($p,q=0,\dots, n-1$).
They are AF and hence the UV limit is described purely in terms
of classical
geometry. The UV $U(1)$ charges are equal to the
degree of the corresponding harmonic form shifted by minus one
half the complex dimension. Then
$$q_j=j-\half (n-1)\qquad j=0,\dots, n-1,$$
and $\exp[2\pi i q_j]=(-1)^{(n-1)}$. From the arguments in \S.5.2
we know that $H$ consists of a single Jordan block associated
to this eigenvalue.

Since the Chern class is $n$ times the hyperplane class, a
chiral rotation by $2\pi/n$ is anomaly--free and we have
a discrete $\BZ_n$ symmetry  (spontaneously
broken by the vacuum). Then the above discussion applies.
The same anomaly argument shows that $q=-1$
and that $r=1$, \ie\ the `critical values' $w_k$ are at the vertices
of a regular $n$--gon.

{}From the above geometrical considerations we see that
the characteristic polynomial of $H$ is
$$P(-z)=\cases{(z+1)^n=\Phi_2(z)^n \qquad n\ {\rm odd}\cr
(z-1)^n=\Phi_1(z)^n\qquad n\ {\rm even}.\cr}$$
Using \pqz\ we get
$$\CQ(z)=\Phi_2(z)^n=(z+1)^n=\sum_{k=0}^n {n \choose k} z^k,$$
Comparing with \soldio\ we get the (solitonic) mass spectrum
\eqn\cpnspec{\mu_{i,i+k}={n\choose k}\qquad (k=1,\dots, n-1),}
which saturates the bound \mubound.
The value of the masses of each kind of solitons can be easily
computed from the vacuum geometry in $W$--space getting \sigmamodels
$$m_{i,i+k}= 4n|t|^{1/n} \sin(\pi k/n),$$
where the coupling $t$ is defined by the chiral ring relation
$X^n=t$ (here $X$ is the chiral primary associated to the hyperplane
class).

The result \cpnspec\ can be understood as follows. Since nothing
depends on the $D$--terms, we can take them to corresponds to
the usual symmetric metric on $\BC P^{n-1}$ (\ie\ the Fubini metric).
Then the isometry group $SU(n)$ is realized as a symmetry of the
mass spectrum. Then the $k$--solitons belong to the $k$--fold
antisymmetric product of the defining $SU(n)$ representation.

With this symmetric choice of the $D$--term the model becomes
exactly solvable and the $S$ matrix has been computed \massGap.
The mass spectrum extracted from the exact solutions is
just \cpnspec. (Notice that the solitons give the full
particle spectrum for these theories.
This is typical  in solvable
models). We stress that the $\BC P^{n-1}$ models are confining
theories with
a very subtle IR structure, see Ref.\sigmamodels\ for a discussion.

\subsec{Grassmanian $\sigma$--models}

Next we consider the $\sigma$--models with target space the
Grassmanian
$$G(N,M)={U(N+M)\over U(N)\otimes U(M)}.$$
$G(N,M)$ is an $NM$ dimensional complex manifold. Its Poincar\'e
polynomial reads
$$\CP_{t,\bar t}\big(G(N,M)\big)=\sum_{p,q} h_{p,q} t^p \bar t^q=
\prod_{k=1}^N {[1-(t\bar t)^{M+k}]\over [1-(t\bar t)^k]}.$$
Since the corresponding $\sigma$--model is AF, the classical
cohomology
fixes the UV behaviour. Hence 
$$\Tr_R\Big[t^{J_0}\bar t^{\bar J_0}\Big]\Bigg|_{\rm uv\ limit}=
(t\bar t)^{-NM/2} \CP_{t,\bar t}\big(G(N,M)\big).$$
By construction, $\tr\, H^m$ is the limit of this quantity as
$(t\bar t)\rightarrow e^{2\pi im}$. Then
$$\tr\, H^m=(-1)^{mNM}\, {(M+N)!\over N!\,  M!}.$$
Then the Witten index is ${M+N \choose N}$ and the characteristic
polynomial of $H$ reads
$$P(z)=\big(z-(-1)^{NM}\big)^{(M+N)!/M! N!}.$$

However, the situation is much subtler than in the $\BC P^{n-1}$
case. First of all, in this case we have not a $\BZ_{(N+M)!/(N! M!)}$
symmetry as above. Even worse, the general theory
discussed in this paper does not apply as it stands.
 In fact we deduced our
main formulae under the `genericity' assumption that no three
vacua are aligned in $W$--space. Usually we can choose a suitable
arbitrarily small perturbation such that any alignment is
destroyed.
However there are special `rigid' cases in which the alignment
cannot be undone --- since all the formal perturbations which would
do the job correspond to non--renormalizable interactions which just
make no sense in the quantum case. The Grassmanian $\sigma$--models
are such a `rigid' case. This is not at all a surprise. It is just
the physical counterpart of the fact that the $G(N,M)$ are
rigid as complex manifolds --- \ie\ the moduli space is just a point.
This rigidity phenomenon may, in principle, lead to a
non--completeness
of our classification scheme. However it is not a real problem.
In fact, on one hand we can extend our theory to these `aligned'
situations just by taking into account a few more terms in the
IR expansions of sect.4. On the other, the rigidly aligned models
have a tendency of being so magical that they can be discussed by
direct means, as we do below for the Grassmanian case.

{}From a direct path integral analysis (summarized in Appendix A)
one learns that
\eqn\idigr{G(N,M)\buildrel \bullet\over=
 \Big(\BC P^{N+M-1}\Big)^N\Big/\hskip -5pt\Big/ S_N,}
where $\buildrel \bullet\over =$
 means equivalence\foot{More precisely,
equivalence up to a deformation of the $D$--term.}
as QFT's for the
corresponding $\sigma$--model. The RHS of \idigr\
is a tensor product of $N$ copies of the $\BC P^{N+M-1}$
$\sigma$--model reduced by the action of the replica symmetry
$S_N$. The double slash in \idigr\ is there to remind the reader
that {\it it is not} the $S_N$ orbifold. Rather (topologically
speaking) it is the set of
{\it maximal} $S_N$ orbits, \ie\ orbits whose elements are {\it not}
fixed by any non--trivial subgroup of $S_N$. This construction is
what was called the `change of variable trick' in ref.\topatop\
(unfortunately this name is appropriate only for the LG case).
Morally speaking, \idigr\ is the QFT counterpart of the standard
description \refs{\chiralring,\gepner,\intri}
of the quantum cohomology ring of $G(N,M)$ in terms
of $N$ copies of the (perturbed) $A_{N+M}$ minimal model.

Let us recall how the `change of variable trick' works.
One has a map $f$
$$\eqalign{& f\colon \CR\hookrightarrow \CR_\ast\cr
& \phi_i\mapsto f(\phi_i)\det[\partial f]\in \CR_\ast,\cr}$$
which identifies isomorphically (as $\CR$--modules)
$\CR$ with its image. Then the \ttstar\ metric for $\CR$ is the pull
back
{\it via} $f$ of that for $\CR_\ast$. This construction
differs in many
respects from an orbifold. In particular it changes the central
charge $\hat c$. One has \topatop
\eqn\cenhat{\hat c=\hat c_\ast-2 q_\ast(J),}
where $q_\ast(J)$ is the $U(1)$ charge of the Jacobian
$J=\det[\partial
f]$ computed in the $\ast$ theory. In the present case the $\ast$
theory
is just $N$ copies of the $\BC P^{N+M-1}$ $\sigma$--model.

Let $X_\alpha$ ($\alpha=1,\dots,N$) be the chiral primary
associated to the hyperplane class of the $\alpha$--th copy
of $\BC P^{N+M-1}$. Then the map $f$ reads
$$f_i(X_\alpha)=\sigma_i(X_\alpha)\qquad i=1,2,\dots, N,$$
where $\sigma_i$ is the $i$--th elementary symmetric polynomial.
Its Jacobian is
$$J=\Delta(X_\alpha)\equiv\prod_{\alpha>\beta}(X_\alpha-X_\beta).$$
We know from the previous subsection that the UV charge of the
$\BC P^{N+M-1}$ operator $X_\alpha$ is $1$. Then
$$q_\ast(J)=\half N(N-1).$$
Using \cenhat, the UV central charge of the rhs of \idigr\ is
$$\hat c_{\rm uv}=N (N+M-1)-N(N-1)=NM,$$
which is the correct result for $G(N,M)$ (\ie\ its complex
dimension).

Let $|k_\alpha\rangle_{\raise -2pt\hbox{$\alpha$}}$
 be the canonical vacuum for the
$\alpha$--th copy of $\BC P^{N+M-1}$ at the $k_\alpha$--th
critical point \ie\
$$X_\alpha(k_\alpha)
=t^{1/(N+M)}\, \exp[2\pi i k_\alpha/(N+M)]\qquad (k_\alpha =0,1,\dots
, N+M-1).$$
Then the canonical vacua for the $\sigma$--model on
$(\BC P^{N+M-1})^N$ are just
\eqn\canpro{\bigotimes_{\alpha=1}^N |k_\alpha
\rangle_{\raise -2pt\hbox{$\alpha$}},}
To get the canonical vacua for the model in the rhs of \idigr\
out of those in \canpro\ we have to perform three elementary
operations \topatop:

\item{\it i)}{Kill the states \canpro\ which are in the kernel
of the chiral field $J$. Since in the present case $J$ is just
the Vandermonde determinant, this means that we must keep only
the states \canpro\ such that the $N$ numbers
$k_\alpha$ {\it are all distinct}.}
\item{\it ii)}{Project into the appropriate subsector
 projectively--invariant under $S_N$.
This is  done just by summing over all permutations with
signs as prescribed by the Jacobian}
$$\sum_{s\in S_N}\bigotimes_{\alpha=1}^N {\Delta(k_{s(\alpha)})\over
\Delta(k_\alpha)}
\left|k_{s(\alpha)}\right\rangle_{\raise -2pt\hbox{$\alpha$}}=
\pm \sum_{s\in S_N} \sigma(s) \bigotimes_{\alpha=1}^N
\left|k_{s(\alpha)}\right\rangle_{\raise -2pt\hbox{$\alpha$}},$$
where $\sigma(s)$ is the signature of the permutation $s$.
Such a state is determined by the {\it unordered}
$N$--tuple $k_i$. Since the $k_i$'s can take $N+M$ values, in this
way we get ${N+M \choose N}$ states \ie\ as many as the Witten
index for the $G(N,M)$ $\sigma$--model.
\item{\it iii)}{Normalize the states so obtained.
Then the canonical $G(N,M)$ vacua are}
\eqn\canvagr{\eqalign{&\big|\{k_1,k_2,\dots, k_N\}\big\rangle
={1\over \sqrt{N!}} \sum_{s\in S_N}\sigma(s)
\bigotimes_{\alpha=1}^N
\left|k_{s(\alpha)}\right\rangle_{\raise -2pt\hbox{$\alpha$}},\cr
&{\rm with}\  0\leq k_1< k_2 <k_2<\cdots < k_N\leq N+M-1.\cr}}

Then the $G(N,M)$ \ttstar\ metric reads
\eqn\gnmmet{\eqalign{\big\langle \overline{\{h_1,h_2,\dots, h_N\}
}&\big|
\{k_1,k_2,\dots, k_N\}\big\rangle =
{1\over N!} \sum_{s,t\in S_N} \sigma(s)\sigma(t)\prod_{\alpha=1}^N
\langle\, \overline{h_{t(\alpha)}}\,
|k_{s(\alpha)}\rangle_{\raise -2pt\hbox{$\alpha$}}=\cr
&=\det\nolimits_{\{\overline{h_\alpha}\},
\{k_\beta\} }\Big[ \langle \overline{h_\alpha}|
k_\beta\rangle\Big],\cr}}
where $\langle\overline{h}|k\rangle$ is the $(N+M)\times (N+M)$
matrix giving the ground state metric for the $\BC P^{N+M-1}$
$\sigma$--model in a canonical basis, and
$\det_{\{\overline{h_\alpha}\}, \{k_\beta\} }$ means the determinant
of
the $N\times N$ minor obtained by selecting the
rows $(h_1,h_2,\dots, h_N)$ and the columns $(k_1,k_2,\dots,k_N)$.
Of course $G(M,N)=G(N,M)$. But the rhs of \idigr\ is not
manifestly invariant under $N\leftrightarrow M$. Instead the final
answer \gnmmet\ is manifestly `duality' invariant.
We begin to show this in the simpler
case $G(1,M)=\BC P^M$. Consider then $G(M,1)$. Its \ttstar\ metric is
given by the $M\times M$ minors of the usual $\BC P^M$ metric $g_{i\bar
j}$. Let $|\hat k_j\rangle=|\{k_1, \dots, \hat k_j,\dots, k_n\}\rangle$
(where hat means omitted). Rewriting the minors in terms of the
inverse metric $g^{\bar j i}$, one gets
$$\langle \overline{\hat h_j}|\hat k_i\rangle= (-1)^{i+j}\det[g]
\, g^{\bar j i}\equiv (-1)^{i+j} g_{i\bar j},$$
where we used that
 $g$ is orthogonal. This shows duality invariance
for $N=1$ (the signs $(-1)^{i+j}$ can be absorbed in
the definition of the states). The general case $G(M,N)$ is handled
analogously using well known properties of minors.

To get the IR (resp. UV) behaviour of \gnmmet\ we have just to
insert the known asymptotics for the $\BC P^\ast$ case.
For instance, for large $\beta$ we have
$$\eqalign{&\langle \overline{h}|k\rangle \cong
\delta_{kh}-i\, {\rm sign}(k-h)\, {N+M \choose |k-h|}{1\over \pi}
K_0\left(m_{kh}\beta\right)
+\dots,\cr
&{\rm where}\ m_{kh}=4(N+M)|t|^{1/(N+M)}
\sin\Big({\pi |k-h|\over N+M}\Big),\cr}$$
which inserted into \gnmmet\ gives the $G(N,M)$ mass spectrum.

\subsec{Applications to `Polytopic' Models}

One of the nicest aspects of the \ttstar\ equations is that, once you
solved a model you easily generalize your result to a whole
family of  models having the same vacuum geometry in $W$--space.
This strategy was exploited in \newindex\ for the `$\BZ_n$--models'
\ie\ theories whose critical values
form  the vertices of a regular $n$--gon. In the same way, the
solution \gnmmet\ generalizes to a family of models with a certain
`polytopic' vacuum geometry \ref\lewarner{W. Lerche and N.P.
Warner, Nucl. Phys. B358 (1991) 571.}. The general
 model in the family is  obtained
by replacing in \idigr\ the $\BC P^{N+M-1}$ $\sigma$--model by
another member of the same $(N+M)$--gon family. The \ttstar\ metric
is still
given by \gnmmet\ but with $\langle \overline{h_\alpha}|k_\beta\rangle$
replaced by the metric for the given $\BZ_{N+M}$ model.

The simplest model in this family is the Kazama--Suzuki Grassmanian
 coset at level 1
\ref\kazama{Y. Kazama and H. Suzuki, Phys. Lett. B216 (1989) 112;
Nucl. Phys. B321 (1989) 232.} perturbed
by the most relevant operator. One has \refs{\chiralring,\intri}
$${U(N+M)_1\over U(N)\otimes U(M)}\buildrel \bullet\over=
 \Big(A_{N+M}\Big)^N\Big/\hskip -5pt\Big/ S_N,$$
where $A_{N+M}$ denotes the minimal model deformed by the most
relevant operator. Then the mass--spectrum for this model is obtained
by inserting example 1 of \S.8.1 in the rhs of \gnmmet.
The result has the properties expected on various grounds (see \eg\
\lewarner).

\newsec{Conclusions}
We have initiated a program to classify
massive $N=2$ supersymmetric theories in two dimensions.
This classification is up to variation in $D$-terms, and
may be viewed, by considering the UV limit, as a classification
program for $N=2$ SCFT's (which admit massive deformation).
The central object in this classification program is a
generalization of `Dynkin diagram' each node of which represents
a non-degenerate $N=2$ vacuum, and the number of lines between
the nodes just counts the number of solitons
(which saturate the Bogomolnyi bound) between the vacua\foot{
As we discussed before, there is an additional sign which is
important.}.   We saw that perturbations of the theory change
the soliton number and modify the Dynkin diagram by the action
of Braid group (which is generated by the generalized
`Weyl reflections').
We discussed what are the restrictions on these generalized Dynkin
diagrams in particular by the condition of reality of $U(1)$
charges of Ramond ground states, which is computable from the
Dynkin  diagram.  We classified all massive $N=2$ theories with
up to three vacua.  We also rederived the classification of
$N=2$ minimal models.
We saw that the Dynkin diagram corresponding
to the minimal $N=2$ models turns out to be just the usual A--D--E
Dynkin diagram.

As a sub-classification we can use these methods to classify (up to
mirror
symmetry) K\"ahler manifolds
with diagonal hodge numbers with $c_1>0$.
We discussed how this works in a particular example
(which leads to a known mathematical theorem).  It
would be very interesting to continue this line of thought
and obtain a complete classification of such K\"ahler manifolds.

We can also use the above models to construct new string
vacua.  All we have to do is to make sure that $\hat c={\rm integer}$
and use an orbifold method \lgorv.

The most important open question is `reconstruction'.  In other
words for each of the generalized Dynkin diagrams which are allowed
for us can we construct a quantum field theory with that solitonic
spectra?  Some of the examples we discussed in the main text suggests
that this may be possible in the form of `generalized' Toda models
constructed from the corresponding generalized Dynkin diagrams.
This may also suggest that there is always an integrable deformation
of the $N=2$ theory, with a particular choice of $D$- {\it and}
$F$-terms.
This would be very interesting to develop further.  In particular
it would be interesting to see if these models are related to
(supersymmetric version of)
 RSOS-like models which have our Dynkin diagram as target space.

Another direction worth investigating is the study of $tt^*$
equations directly in the conformal case (in
the case of three-folds this is known as
{\it special geometry} \specgeom ).  One generically studies the
moduli space of these theories, which is the analog of $w_i$ for us
here.
In our case the natural degeneration point of moduli space
are the UV and IR limits, whereas in the conformal case we will
have a number of degenerate points (or
submanifolds) on moduli space.  The solution
to $tt^*$ equation will undergo a monodromy around each of these
degeneration points.  Then what we should do is {\it to classify
all possible representations of the monodromy group} that
are consistent with the
existence of global regular solutions to $tt^*$.  This would mean
that we begin
to classify
all the Calabi-Yau manifolds at once by studying all the possible
consistent monodromies on the degeneration points of their moduli
spaces.  This would be the massless analog of the classification
program we have initiated in the massive case here.  We intend
to return to this idea to classify $N=2$ SCFT's
(and thus Calabi-Yau manifolds up to mirror symmetry) in future work.

\vglue 1cm

We would like to thank B. Dubrovin for valuable discussions.  C.V.
also wishes to thank the hospitality of ICTP where this
work was initiated.

\vglue 1cm

The research of C.V. is supported in part by Packard fellowship and
NSF grants PHY-87-14654 and PHY-89-57162.

\appendix{A}{Exact Path--Integral Computations for
 $\sigma$--Models}

In this appendix we want to show eq.\idigr\ by  exact path--integral
computations. Since the `change of variable trick' has a simple
meaning in the LG case, it will be helpful if we could write down
 `LG' models\foot{We put LG in quotes because
it is not really a Landau--Ginzburg model.
For the purposes of the present appendix the naive interpretation
of the effective theory as a LG model is good enough and we
shall stick to this naive viewpoint.}
 which are {\it exactly}
equivalent (as QFT's) to our $\sigma$--models.
 For the $\BC P^{n-1}$ case this was done long ago by
the authors of ref.\ref\divecchia{A. D'Adda, A.C. Davis,
P. Di Vecchia and P. Salomonson, Nucl. Phys. B222 (1983) 45.}.
In this appendix we extend their result to the Grassmanian case.
As an aside, this will give us a rigorous path--integral proof of
the quantum cohomology ring\foot{Similar results
have been obtained from a more mathematical
standpoint by D. Franco and C. Reina (to appear).}
 for Grassmanians as predicted on
general grounds in refs.\refs{\chiralring,\gepner,\intri}.
However here --- as well as in \divecchia\
--- one looks for an equivalence at  the full QFT
level, not just for its
topological sector.

In the old days when the authors of \divecchia\ obtained their result
little was known about N=2 field theories and computations were
quite hard. In those days
\divecchia\ was quite an analytic triumph.
Luckily enough, nowadays N=2 theory is so developed that even
more sophisticated models can be analyzed without real effort.

Let us begin by a (modern) review of their work.
Then we shall generalize to the Grassmanians $G(N,M)$.

\vglue 8pt
\noindent\underbar{\raise 3pt \hbox{\it The $\BC P^{n-1}$ Model}}
\vglue 2pt

The starting point \divecchia\ is the `homogeneous coordinates'
formulation of the model. The Lagrangian reads
\eqn\homogeneous{\int d^4\theta\,\left[ \sum_{i=1}^n \overline S_i
e^{-V}
S_i+ {A\over 2\pi} V\right],}
where $S_i$ are chiral superfields which map into the homogeneous
coordinates
on $\BC P^{n-1}$ and $V$ is a Legendre multiplier real superfield.
In \homogeneous\ we have denoted the coupling constant by $A$ because
it
has the geometrical interpretation of the area of the basic
$2$--cycle
generating the homology of $\BC P^{n-1}$.
The field $V$
 gauges the $\BC^\times$ acting diagonally on $\BC^n$, so the
physical
degrees of freedom are
$\BC^n/\BC^\times \cong \BC P^{n-1}$.
Explicitly, eliminating $V$ using its equations of motion one gets
$${A\over 2\pi}
\int d^4\theta\, \log\left[\sum_{i=1}^n \overline S_i S_i\right],$$
that is the usual formulation of the $\sigma$--model (for the Fubini
metric). From its equations
of motion we see that $V$ is nothing else than the
susy version of the  pull back of the
$U(1)$ part of the Fubini spin--connection. Then its field--strength
superfield
\eqn\hyperpla{n\,  X=D^+\bar D^- V,}
(which is a $(c,a)$ field in the notation of \chiralring)
 is the susy analog of the (pull--back
of the) trace part of the $\BC P^{n-1}$ Riemann tensor, \ie\ it is
the
$(c,a)$ primary operator
associated with the first Chern class. Since $c_1(\BC
P^{n-1})=n$, the observable
$X$ in \hyperpla\ is the basic chiral primary associated
with the hyperplane class. This is most easily seen by looking at
the last (\ie\ auxiliary) component of the
superfield $X$. Up to a normalization
coefficient this is
\eqn\auxfield{(D+i\ast \phi^\ast R),}
 where $D$ is the
auxiliary
field of the real superfield $V$, and $\phi^\ast R$ is the pull--back
to the world--sheet of the Ricci form.

Now the idea \divecchia\ is to perform the Gaussian integral over
the $\bar S_i$'s {\it exactly}. Denoting the result by
$\exp\{-S[V]\}$,
this gives an equivalent formulation of the quantum model in terms of
the
(super)field $V$ with action  $S[V]$. We stress that this procedure
is
exact. By gauge--invariance the action should depend on the
field--strength superfield only. Then it should have the general form
\eqn\effact{S[X,\bar X]=\int d^2\theta\, W(X)+\int d^2\theta\,
 \bar W(\bar X)+
D-\rm term.}
A priori we are not guaranteed that $S[X,\bar X]$ is {\it local}.
However any non--locality is in the $D$--term\foot{Why?
Because if you do the same analysis in the TFT case you do not have
any problem with non--locality.}. Since for the purposes of
this paper we can change the $D$--term at will, we can forget
about any problem the action \effact\ may have.
By the same token, we do not need to compute every detail of the
rhs of \effact\ either. Computing $W(X)$ is good enough. In
order to extract
$W(X)$ from $S[X,\bar X]$ notice that the
$D$--terms either contain higher powers of the auxiliary field $D$
or derivatives of $D$. Instead the term
linear in $D$ (at vanishing momentum) reads
$$D\left({\partial W\over \partial X}+{\partial\bar W\over
\partial\bar
 X}\right),$$
so to extract $W$ it is enough to get the term linear in $D$ in
\effact.
By the same argument,
 we can as well assume that all fields are constant
(and the fermions vanish). Then the computation reduces to that
of the determinants of differential operators {\it with constant
coefficients}. Despite this dramatic simplification,
the computation is still exact!

Expanding out the action \homogeneous\ in components,
we get
\eqn\zemos{\exp\Big[-S[X,\bar X]\Big]\Big|_{\rm vanishing\atop
momentum}
=e^{-  A \int d^2\theta X/2\pi}
\left({{\rm Det}\left[\dsl+\left(\matrix{0 &
X\cr \bar X & 0\cr}\right)\right]\over {\rm Det}\Big[-
\partial^2+(D+X\bar
X)\Big]}\right)^n,}
(the exponential prefactor is the classical value of the action
at $S_i=0$; see the last term in \homogeneous).
Taking the derivative of the rhs with respect $D$ and setting $D=0$,
we get
\eqn\tass{\left({\partial W\over \partial X}+{\partial\bar W\over
\partial\bar
 X}\right)= {A\over 2\pi}+ n\, \Tr\left[{1\over -\partial^2+X\bar
X}\right].}
The trace in the rhs is easily evaluated by $\zeta$--regularization
$$\eqalign{\Tr[(-\partial^2+X\bar X)^{-s}]&=
\int {d^2p\over (2\pi)^2} {1\over (p^2+X\bar X)^s}
= {1\over 2\pi}{1\over (1-s)} (X\bar X)^{1-s}
\cr &={1\over 2\pi (1-s)}+
{1\over 2\pi}\log X+ {1\over 2\pi}\log\bar X+ O(1-s).\cr}$$
As $s\rightarrow 1$ this has a pole;
comparing with \zemos\ we see that the only
effect of this infinity is to renormalize the coupling $A$.
We can just forget about this infinity provided
we replace $A$ by its running counterpart $A(\mu)$.
After having subtracted
the infinity, take $s\rightarrow 1$.
Notice that the rhs of \tass\ is a harmonic function of $X$ as it
should; this is a nice consistency check.
Integrating \tass\ we get \divecchia\
\eqn\cpnw{2\pi W(X)= X\left(\log X^n-n+  A(\mu)-i\vartheta\right).}
where $\vartheta$ is a real parameter. Comparing with \auxfield\ we see
that $\vartheta$ is the usual instanton angle.
The quantum cohomology ring of $\BC P^{n-1}$ is just
$${\cal R}=\BC[X]/\partial W=\BC[X]\Big/\Big(X^n-e^{-A(\mu)+i\vartheta
}\Big),$$
which is Witten's result \ref\wittcpn{E. Witten, Nucl. Phys. B340
(1990)
281.}.

\vglue 8pt
\noindent\underbar{\raise 3pt \hbox{\it Grassmanian
$\sigma$--Models}}
\vglue 2pt

Now we generalize the above approach to the Grassmanians $G(N,M)$.
Again we have a `homogeneous' formulation. Now the chiral fields
 $S_{ia}$ have
two indices, a `gauge' $U(N)$ index $i$, and a `flavour'
$SU(N+M)$ index $a$. The Lagrangian reads
$$\int d^4\theta \Big(\sum_a \bar S_a e^{-V}S_a+\alpha\, \tr\,
V\Big),$$
where now $V$ is a $N\times N$ matrix of superfields which
gauge $U(N)$. Also the field--strength superfield $X$ belongs the
adjoint rep. of $U(N)$, and then is gauge {\it covariant}
 rather than invariant
as before. The basic gauge--invariant objects are the Ad--invariant
polynomials in the field--strengths $X$. Their ring is generated
by the superfields
$Y_i$ ($i=1,2,\dots, N$) defined by
$$\det[t-X]=t^N+\sum_{k=1}^N (-1)^k t^{N-k}Y_k.$$
Contrary to the $X$'s, the $Y$'s are {\it bona fide} $(a,c)$
superfields. They generate the (quantum) cohomology
ring  $\cal R$ for $G(N,M)$ (as it
can be shown by going to the classical limit).
A priori computing the determinants
is now quite
a mess, since everything is non--Abelian. Anyhow
we shall use the same strategy as
before \ie\ to use our non--perturbative knowledge of the N=2
theories
to replace the actual
computation with a trivial --- but still exact --- one.

Again we can take all the background
fields constant (and fermions vanishing). Then we make the following
observation: at the TFT level the $X$'s and the $\bar X$'s do not
talk to each other (in fact the $\bar X$'s are just gauge--fixing
devices)
and we can assume, with no loss of generality, that $X$ and $\bar X$
(as matrices) commute. Then $X$ is diagonalizable
\eqn\bachg{X={\rm diag}(X_1,X_2,\dots,X_N).}
Moreover, with probability $1$, all $X_i$ are distinct.
Consider the superfield $Z_a\equiv\partial W(X)/\partial X_a$
($a$ is an adjoint rep. index
for $U(N)$). Obviously it belongs
to the adjoint rep. of $U(N)$.
 By gauge invariance, $W(X)$ is an Ad--invariant function
of the $X$'s. But then, in a Cartan background like \bachg,
also $Z_a$ belongs to the Cartan subalgebra (by invariance
under the corresponding maximal torus). Given that the terms in
$S[X,\bar X]$ which are linear in
the auxiliary fields $D_a$ should have the form
($D_aZ_a$+h.c.), we see that no information is lost if
we restrict  $D_a$
too to the Cartan subalgebra ($=U(1)^N$).
But then the full background is Abelian
and the functional determinants are just the same as in the $\BC
P^{n-1}$
case.

Therefore, in a Cartan background, $W(X)$ is just the sum of $N$
copies
of what we got
for the $\BC P^{N+M-1}$ model. But the Cartan background fixes the
theory completely.
Then
$$2\pi\, W(X_1,X_2,\dots,X_N)=\sum_{k=1}^N X_k\left(\log X^{N+M}_k -
n+A(\mu)-i\vartheta\right),$$
Again we have the relations $X^{N+M}=\rm const.$. However, this
time the good gauge--invariant fields are the Ad--invariant
polynomials
in the $X_i$, which are generated by the elementary symmetric
functions,
that is the fields $Y_i$. Thus the quantum cohomology ring of
$G(N,M)$
is the ring generated by the symmetric functions in $N$
indeterminates
$X_i$ subject to the relations $X^{N+M}_i=1$.

This result is not new (at least
as far as the classical part is concerned) and was obtained (or
found to be very plausible in the quantum case) in
refs.\refs{\chiralring,\gepner,
\intri} from quite different
considerations. However here we have shown a much stronger
result than just computing
 $\cal R$. In fact, we have considered a
topological truncation of a computation (that of $S[X,\bar X]$)
which makes perfect sense in the {\it full} QFT. To put it
differently,
our `topological'
map from ${\cal R}_{\rm Grass.}$ to $\otimes{\cal R}_{\BC P}//S_N$
is induced by a map between the
corresponding QFT's. Of course, we have no explicit form for the
parent QFT map. However, the pure fact that this map exists and
restricts
nicely
to the topological one, has quite dramatic implications. It
shows that both the $tt^\ast$ differential
equation {\it and} their boundary data agree on the two sides
of the `dotted'--equality \idigr.
In view of the theory we have developed in the main body of the paper
this is enough to fix the (solitonic) mass--spectrum of the $G(N,M)$
$\sigma$--model.

\appendix{B}{Subtleties with Collinear Vacua: An Explicit Example}

In sect.4 we saw that special phenomena take place when three vacua
are aligned:

\item{a)}{We have the `half--soliton' mechanism of eq.\halfsoli. When
we
deform slightly the picture by putting
the middle vacuum on
one side of the line connecting the other two vacua the number of
solitons
connecting these two vacua change. For the exactly aligned situation
the large $\beta$ asymptotics looks as if we had
$$\half(\mu_{13}+\mu_{13}^\prime).$$
where $\mu_{13}$ and $\mu_{13}'$ refer to soliton numbers
{\it if} the middle vacuum was perturbed one way or another.
Of course, there is no such a thing as a `half--soliton'. This
discontinuity in the IR asymptotics just signals that the IR
asymptotic series is not uniform (as always). This is clear from
the analysis of sect.4.}
\item{b)}{We have various possibilities for the power of $\beta$
in front of the Boltzmann exponential, see \eg\ eqs.\twoasy\twoasyspe.
Physically this is a consequence of the fact that there are states
with different numbers of solitons and the same energy. These states
have
the same Boltzmann exponential but a different phase--factor.}

The purpose of this appendix is to illustrate these two points in
a concrete example where explicit computations are possible.
Consider the LG model with superpotential
\eqn\model{W(X)={X^6\over 6}- {t X^2 \over 2}.}
In $X$ space the critical points are
$X_0=0$ and $ X_k= i^k t^{k/4}$, ($k=1,\dots,$).
They are at the vertices and center of
a quadrate. The critical values are
$$W_0=0, \qquad W_k=(-1)^{k+1}{1\over 3} t^{3/2}.$$
We have only three distinct critical values, and these three points
are collinear. The naive picture of solitons would suggest that the
`fundamental' solitons are the inverse images of the segments
connecting
the $W_k$'s to  $W_0$, i.e. the half--diagonals of the square in
$X$--space. All other pairs of vacua are connected by multi--soliton
process only.
Is this naive picture
correct? It better be wrong, since it leads to
paradoxes when compared to our general theory.
Luckily we can do exact computations to see what is going on.

The ground--state metric for \model\
was computed in \topatop. There it was also checked
that this is the only regular solution and that it reproduces
the known results in the UV limit.
The non--vanishing elements are \topatop\
\eqn\ground{\eqalign{& \langle \bar 1|1\rangle={2\over |t|}
e^{-u(z)}\cr
& \langle\bar 1|X^4\rangle={t\over |t|}e^{-u(z)};\quad
 \langle\overline{X^4} | 1\rangle={\bar t\over |t|}e^{-u(z)}\cr
& \langle\bar X|X\rangle= {1\over |t|^{1/2}} \exp\left[-\half
u(2z)\right],\cr
&\langle \overline{X^2}|X^2\rangle=1\cr
& \langle\overline{X^3}|X^3\rangle= |t|^{1/2} \exp\left[\half
u(2z)\right],\cr
& \langle\overline{X^4}|X^4\rangle= |t|\cosh\big[u(z)\big],\cr}}
where $u(z)$ is the regular PIII transcendent (cf. \S.6.1)
with $r=-2/3$ and
$$z= {2\over 3} |t|^{3/2}\beta =2|W_k-W_0|\beta.$$
Let $|f_k\rangle$ ($k=0,1,\dots,4$) be the canonical vacuum
associated with each critical point $X_k$. Then in the canonical
basis
\ground\ becomes ($r,s\not=0$)
$$\eqalign{ \langle \bar f_0|f_0\rangle=&
\cosh[u(z)]\cr
\langle \bar f_0|f_r\rangle=& {i\over 2}  \sinh[u(z)]\cr
\langle \bar f_s|f_r\rangle=&
{1\over 4}\Big[i^{(s-r)}\exp\big[-\half u(2z)\big]
+(-i)^{(s-r)}\exp\big[\half u(2z)\big]+\cr
& \ + (-1)^{(s-r)}+\cosh[u(z)]\Big].\cr}$$

Let us study the large $z$ asymptotics of this solution.
One has ($r,s\not=0$)
$$u(z)= -{2\over \pi} K_0(z)+ O(e^{-2z}).$$
Then one has ($r,s\not=0$)
\eqn\sixasy{\eqalign{ \langle \bar f_0|f_0\rangle =&
1+{2\over (\pi)^2}K_0(2z)^2+O(e^{-3z}),\cr
\langle \bar f_0| f_r\rangle= &-i\, {1\over \pi}K_0(z)+
O(e^{-3z}),\cr
 \langle \bar f_s|f_r\rangle=& {1\over 4}
\left\{i^{(s-r)}\Big[1+{1\over \pi}K_0(2z)\Big]
+(-i)^{(s-r)}\Big[1-{1\over \pi} K_0(2z)\Big]+\right.\cr
&\ + \left.[1+(-1)^{(s-r)}]+{2\over (\pi)^2}K_0(z)^2\right\}+
O(e^{-3z}).\cr} }
So,
$$\langle \bar f_i|f_j\rangle\Big|_{z=\infty}=\delta_{ij}.$$
{}From \sixasy\ we see that in the IR expansion
of $\langle \bar f_s|f_r\rangle$ with $r\not=s$ and $r,s\not=0$
 there are two kind of contributions of order
$O(\exp[-2z])$, those of the form $K_0(2z)$ and those of the
form $K_0(z)^2$. The power--law in front of the exponential
is $\beta^{-1/2}$ and $\beta^{-1}$ respectively as expected on
the basis of eqs.\halfsoli\twoasyspe. This is phenomenon b).

To get the phenomenon a), just extract ``$\mu_{ij}$" as the
coefficient of the leading terms with a $\beta^{-1/2}$ power--law.
Then
$$\eqalign{\mu_{0r}&=1\cr
``\mu_{rs}"&={i\over 4}\big[ i^{(s-r)}-(-i)^{(s-r)}\big]
=\cases{-\half\quad {\rm for}\ s=r+1\ \rm mod. \ 4,\cr
\hskip 3mm 0\quad {\rm for}\ s=r+2\ \rm mod.\ 4,\cr
\hskip 3mm \half\quad {\rm for}\ s=r+3\ \rm mod.\ 4.\cr}\cr}$$
This shows how a) phenomenon appears.

\def\BQ{{\bf Q}}

\appendix{C}{Conjectures on the OPE Coefficients}

{}From one point of view our work here may be seen as
a generalization of the connection formula for PIII  as
discussed in \S 6.1. However for PIII the authors of
\refs{\piiimath,\aits} did a better job, since their result
 not
only allows us to compute the UV $U(1)$ charges but also the UV
ground state metric, or equivalently the absolute normalizations
for the OPE coefficients (see \topatop\ for a number of explicit
examples). Then it is natural to ask what we can do in
the direction of computing OPE coefficients for the general case.

In this appendix we show how the number--theoretical nature
of our classification program may reduce the computation of
these normalization factors to the so--called
{\it standard} conjectures of
number theory and algebraic geometry.
Here we present some preliminary thoughts in this direction.
It may seem that there is not much point in building conjectures
over `facts' which are themselves conjectures. However sometimes
conjectures may be deeper than established facts!

In order to formulate our fancies we
should rephrase our Diophantine problem in more abstract terms.

The crucial point is to realize that we have a
lattice $\CL\in \CR$. A chiral primary operator $\CO$ belongs to
$\CL$ iff
$$\CO= a_1 e_1+a_2 e_2+\dots +a_n e_n, \quad a_i \in {\bf Z},$$
where $e_i$ are the idempotents of $\CR$, i.e.
the elements of the `point basis'. Then the integral
elements in the topological Hilbert space $\CH$ are those of the form
$$|\CO\rangle \equiv
a_1 |e_1\rangle +a_2
|e_2\rangle+\dots + a_n |e_n\rangle,
\quad a_i \in {\bf Z},$$
where the map $e_i \rightarrow |e_i\rangle$
is the spectral--flow {\it as
realized by the topological path--integral}.
We denote this $\BZ$--module
as $\CH_\BZ$ and consider the $\BQ$--space $\CH_\BQ=\CH_\BZ
\otimes_\BZ \BQ$.
The most important fact about the lattice $\CH_\BZ$
is that {\it it is preserved by the monodromy}
$H$ as a consequence of the integrality of the number of soliton
species.
We can introduce a natural `Hodge decomposition' of the space $\CH$.
In general it is a {\it mixed} one. To make things as easy as
possible,
here we assume that this additional
complication is not present in the model of
interest. More concretely, we assume that the characteristic
polynomial $P(z)$
has the form
\eqn\asump{P(z)=\prod_{m_i \ \rm distinct} \Phi_{m_i}(z).}
The Hodge decomposition is defined by declaring that the subspace
$H^{p,\hat
c-p}\subset \CH$ consists of the states $|\CO\rangle$
with UV behaviour
$$\langle \overline{\CO}|\CO\rangle \sim \beta^{-2p}\qquad
{\rm as}\ \beta\rightarrow 0.$$
For a $\sigma$--model on a CY space this definition of
`type' $(p,q)$ corresponds to the usual one
(up to `mirror symmetry')
but in general $p$ and $q$ are not even integral (however they are
always in
$\BQ$). As it is well know, the data  $X=\{\CH_\BQ, \oplus_p
H^{p,\hat
c-p}\}$ is a Hodge structure (specified up to isogeny).

Among all Hodge structures there are special ones
having peculiar number
theoretical properties.
Let $\CF$ be an Abelian extension\foot{I.e. a Galois
extension whose Galois group is Abelian.}
of the field $\BQ$, and let $f$ be its
transcendency degree. We say that a
Hodge sub--structure $M\subset X$ has {\it
complex multiplication} by $\CF$ if it
has rank $f$ as a $\BZ$--module and there
is an injection of $\CF$ into ${\rm End}(\CH_\BZ)\otimes_\BZ\BQ$.

In this language the
statements around Eq.\cycloprod\ can be rephrased by saying that
to each cyclotomic factor there corresponds
a Hodge sub--structure
$M_{m_j}$ of rank $\phi(m_j)$ with complex multiplication by the
cyclotomic
field $\BQ(e^{2\pi i/m_j})$. Over $\BQ$ (i.e. modulo isogeny)
the subspace $M_{m_j}$ is
defined by
$$\Phi_{m_j}(H)M_{m_j}=0.$$
The product is defined as follows: The element (here $\zeta_j=e^{2\pi
i/m_j}$
and $z_i\in \BQ$)
$$z_0+z_1 \zeta_j+ z_2 \zeta_j^2+\dots z_{m_j-1} \zeta_j^{m_j-1}
\in \BQ(\zeta)$$
acts on $M_{m_j}$ as the linear operator
\eqn\compmol{z_0+z_1 H+ z_2 H^2+\dots z_{m_j-1}
H^{m_j-1} \in {\rm End}(\CH_\BZ)\otimes_\BZ \BQ.}

The interest of this point of view for physics
stems from the fact
that in presence of complex multiplication there are
standard results (conjectures) for the
corresponding period maps. In the N=2 language
this means that we can predict the
{\it normalized} UV OPE coefficients in terms of
characters for the cyclotomic fields.

This is done as follows.
Consider\foot{As usual $\bar \BQ$ denotes
the algebraic closure of $\BQ$.}  an
operator $\CO_1$
$$\CO_1 = \sum_i \kappa_i e_i, \quad \kappa_i \in {\bar
\BQ},$$
satisfying (to save print we write $m$ for $m_j$)
$$H \CO_1= e^{2\pi i r/m}\CO_1,\qquad (r,m)=1.$$
Such an operator always exists as discussed in the main body of the
paper.
Let $p(1)$ be the `type' of $|\CO_1\rangle$. From \compmol\ we see
that the one--dimensional subspace
spanned by $|\CO_1\rangle$ carries a representation $\varphi_1$ of
the
cyclotomic field $\BQ(\zeta_m)$.
Let $l\in (\BZ/m\BZ)^\times\simeq {\rm Gal}(\BQ(\zeta_m)/\BQ)$
be an element of the Galois group of this cyclotomic extension.
Since $l$ corresponds to an automorphism of $\BQ(\zeta_m)$, $\phi_l=
l\circ \phi_1$ is also a one--dimensional representation of
$\BQ(\zeta_m)$
on $M_m$. Let $|\CO_l\rangle$ be a state spanning the corresponding
representation. This state has also a definite `type' $p(l)$.

For $a\in \BZ$, we write $\langle a\rangle$ for unique number
$0,1,2,\dots, m-1$ congruent to $a$ mod. $m$.
Then we introduce a function $f(a)\colon
\BZ/m\BZ\rightarrow \BQ$ by\foot{This definition does not fix $f(a)$
uniquely but the
ambiguity is immaterial. The existence of $f(a)$ is a consequence of
PCT
together with a lemma by Deligne.}
\eqn\trigsum{p(l)={1\over m}\sum_{a=0}^{m-1} f(a)\langle l a\rangle.}
Then the general period conjecture can be restated as follows.
We fix the moduli $w_i$ in such a way that $w_i-w_j\in \bar\BQ$ for
all
$i$, $j$. Then as $\beta\rightarrow 0$
\eqn\genconj{\langle \overline{\CO_1}|
\CO_1\rangle\Big|_{\beta\sim 0}=
{z\over \beta^{2p(1)}}
\prod_{a=0}^{m-1}\Gamma\left(1-{a\over m}\right)^{[f(a)-f(-a)]},  }
where $z$ is a `trivial kinematical factor' belonging to $\bar\BQ$.

It is tempting to conjecture the validity of this statement in
general.
As evidence for this we discuss the $A_n$ minimal models.

\vskip 8pt
\noindent\underbar{\raise 3pt\hbox{\it Example: The $A_n$ Minimal
Models}}
\vskip 2pt

As an example consider the LG
models $W=X^{n+1}+{\rm lower\ degree}$, where the coefficients are
assumed
(for convenience) to be rational numbers.
To make things even
easier, we assume $n+1$ to be an odd prime\foot{In fact,
this is the only case one needs. Assume that $W=X^{ab}+\dots$. Then
the change of variables $Y=X^a$ reduces to $W=Y^b+\dots$. In this way
we
can always (choosing special submanifolds of moduli space) restrict
to
odd prime powers.} $p$.
Since $H^{2p}=1$,
\compmol\ will give a complex multiplication by $\BQ(\zeta_{2p})$. Of
course this is the same as $\BQ(\zeta_p)$. To rewrite the action in a
canonical $\BQ(\zeta_p)$ form it is sufficient to change sign to $H$,
since
$\Phi_p(-H)=0$.
Then $\CR=\{X^k| k=0,1,\dots,
p-2\}$ and the $U(1)$ charge of the Ramond state $|X^k\rangle$ is
$$q_k={k+1\over p}-{1\over 2}.$$
The `type' of this state is
$$p_k= q_k+{1\over 2}\equiv {k+1\over p},$$
where the extra ${1\over 2}$ arises because of the chiral anomaly
as discussed\foot{The `anomalous'
combination $q_k+{n\over 2}$ is the natural one from the singularity
viewpoint too.} in Ref..
Now we apply the above conjecture to this situation. One has $m=p$.
As
$\CO_1$ we take the operator $X^{k-1}$, which is associated to the
eigenvalue $(\zeta_p)^k$ of $-H$. Under the action of the Galois
group ${\bf F}_p$
this element generates the full ring\foot{The element $-1$ of the
Galois group
corresponds to spectral flow. So complex multiplication can be seen
as
a fancy generalization of spectral flow.} $\CR$. The
corresponding operators $\CO_l$ are
just $X^{\langle lk\rangle-1}$. Then
$$p_k(l)={\langle lk\rangle \over p},$$
and \trigsum\ becomes
$$p_k(l)={1\over p}\sum_{a=0}^{p-1} f(a)\langle a l\rangle \quad
\Rightarrow
\quad f(a)=\cases{1 \ {\rm if}\ a=k\cr 0 \ {\rm if}\ a\not=k.\cr}$$
Thus from \genconj\ we have for the UV OPE coefficients
\eqn\wellkn{\langle \overline{X^{k-1}}|X^{k-1}\rangle=
z_k{\Gamma\left({k\over
p}\right)\over \Gamma\left({p-k\over p}\right)},}
for some {\it algebraic} numbers $z_k$.
\wellkn, with $z_k=1$,
is the well known answer for these coefficients.

\listrefs

\end

\input harvmac
\def\BC{{\bf C}}
\def\ie{{\it i.e.}}

\end

\input harvmac

\def\BZ{{\bf Z}}
\end